\documentclass[a4paper,12pt,times,numbered,print,index]{PhDThesisPSnPDF}

\input{Preamble/preamble}

\title{Tensor Contraction with \texorpdfstring{\\}{}ZX Calculus}   

\subtitle{A Graph-Theoretic Approach}

\author{Yuhao Han}

\dept{Mathematical Institute}

\university{University of Oxford}
\crest{\includegraphics[width=0.2\textwidth]{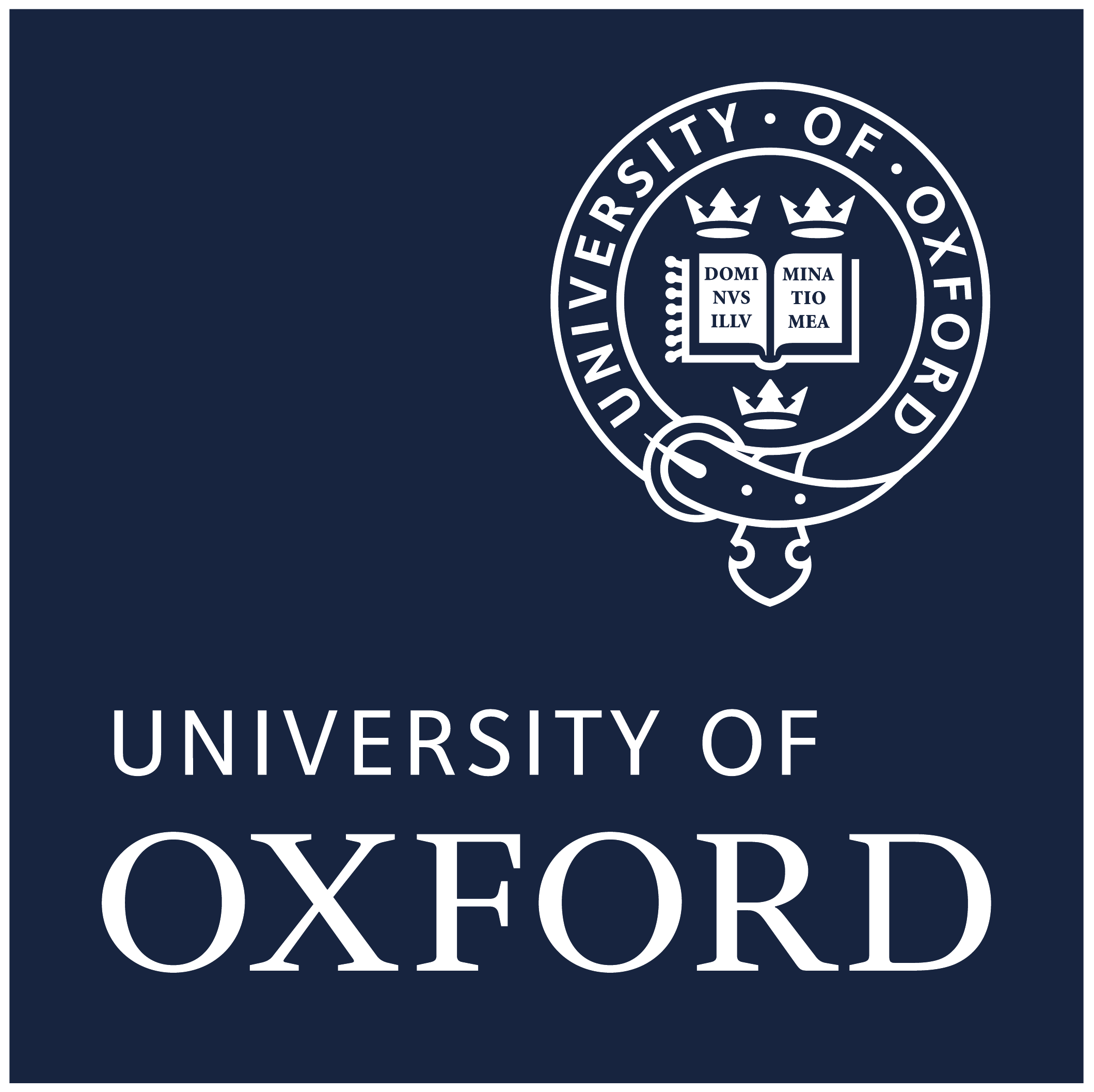}}

\degreetitle{MSc in Mathematics and Foundations of Computer Science}

\college{Reuben College}

\degreedate{September 2026} 

\subject{Tensor Contraction with ZX Calculus} 

\ifdefineAbstract
\fi

\ifdefineChapter
\fi

\begin{document}

\frontmatter

\maketitle


\begin{dedication} 

I would like to dedicate this thesis to my loving parents \dots

\end{dedication}


\begin{declaration}

The author hereby declares that except where specific reference is made, the contents of this dissertation are original and have not been 
submitted in whole or in part for consideration for any other degree or 
qualification in this, or any other university. 
This dissertation is the independent work of the author under the supervision of the author's supervisors, and contains nothing which is the outcome of work done in collaboration with others, except as specified otherwise. 
The author has made his best efforts to ensure the correctness of the results presented, yet there can be no guarantee that the results are free from errors.
\end{declaration}

\begin{spacing}{1.1}
\begin{acknowledgements}
	Crystal dew on the morning grass \\
	Warm lights out of the evening glass\\
	Lo, the larks sing and the owls bark\\ 
	We have come a good way through!\\
\\
	We alone cannot do this\\ 
	Without constant support behind us \\ 
	My parents, for most, for their constant support\\ 
	And always being my last resort. \\ 
	Dr Quanlong Wang, our unsolicited tutor, \\ 
	Professor Aleks Kissinger, his voluntary mentor, \\ 
	And Professor Bob Coecke, his kind referrals. \\
\\
	The city of academics, Oxford \\ 
	And the college of newborn, Reuben\\ 
	Such a pleasure to study here! \\ 
	Delicious food, welcoming staff, \\ 
	And shelf after shelf of books \\ 
	Under the ancient name of Bodleian Library.\\
	And as importantly, the University Park \\ 
	Home of larks, owls, purple beeches,\\
	And both kinds of chestnuts.\\
\\
	James, Ketong, Tianrun, my friends, \\ 
	Who sailed the same boat with me. \\
	Gwen, Huari, Jason and Xiaotong, \\
	Who ate in the same bowl as me! \\ 
	And, of course, Lulu, my dear \\ 
	The penultimate line is reserved for you \\ 
	And the ultimate for our lovely memories!\\
\end{acknowledgements} 
\end{spacing}

\begin{abstract}
Tensors are higher dimensional generalisations of vectors and matrices.
Tensor networks are a diagrammatic representation of complicated relations between tensors, and tensor contraction is the evaluation of a tensor network analogous to matrix multiplication.
Tensor networks and tensor contraction are ubiquitous in computer science, physics, and applied mathematics.

Most existing tensor contraction algorithms are focused on finding an optimal contraction order of the tensors' bonds.
However, finding the optimal order is an NP-hard problem.
It is equivalent to another possibly more straightforward but still NP-hard problem of finding the treewidth of the line graph of the tensor network.

While in general finding the optimal bond-by-bond contraction order is hard, in many practical cases there is no need to find an optimal order.
This observation leads to the core innovation of this thesis, which is a new tensor contraction paradigm that uses graph theoretic operations to produce an equivalent tensor network whose line graph has lower treewidth than the original tensor network, before applying the bond-by-bond contraction algorithm.
We call these graph theoretical operations \textit{rewrite preprocessing}, and we name the algorithms that apply rewrite preprocessing before regular bond-by-bond contraction \textit{Tensor Contraction with Graph Operation}.

There are many possible ways to perform rewrite preprocessing; in this thesis, we focus on the graph operations of ZX calculus, a diagrammatic language for quantum circuits.
Ample numerical simulations have been performed to demonstrate the effectiveness of our proposed method.

\end{abstract}


\tableofcontents

\listoffigures



\printnomenclature

\mainmatter

\chapter{Introduction}

\section{Background of Tensor Networks}

Tensors, to put it plainly, are arrays of numbers that can be regarded as higher dimensional matrices, and tensor contractions are operations between tensors that are analogous to matrix multiplication.

The word \emph{tensor} was coined by Woldemar Voigt in his study of the elastic and optical properties of crystals in 1898 \cite{voigt1898}, and its use was further expanded by Ricci-Curbastro and Levi-Civita under the name \emph{absolute differential calculus} \cite{ricci1900}.
They defined a tensor, in a simplified way, as a multilinear map on a product of vector spaces and their duals to $\C$.
This definition later evolved to the current well accepted notion of tensor product in multilinear algebra defined via the universal property of a bilinear map \cite{lang2002algebra}.

The usefulness of tensors originates from their basis-independent property, which is especially important in physics.
A physical law must be independent of the coordinate system, and a tensor equation is the most economical way of guaranteeing that it does not.
The moment of inertia of a rigid body is a rank-two tensor rather than a single number, because how a body resists being spun depends on the axis about which one spins it.
The stress inside a deformed solid is likewise rank-two, for it takes the orientation of a surface and returns the force transmitted across it, while the elastic moduli relating stress to strain form a rank-four tensor.
General relativity carries the idea to its conclusion: gravity is not a force but the curvature of spacetime, and Einstein's field equations \cite{einstein1916} equate a curvature tensor with the energy-momentum tensor of matter, a statement whose very grammar asserts that no coordinate system is privileged.
Quantum mechanics is no exception, for the state space of a composite system is by postulate the tensor product of the state spaces of its parts, the algebraic root of both entanglement and the exponential growth of Hilbert space \cite{Nielsen_Chuang_2010}.

In mathematics, tensor product is defined via the universal property of a bilinear map, as we will show in chapter \ref{ch:tensors}.
This universal property makes it a natural construction in category theory.
In differential geometry a tensor field assigns a tensor to every point of a manifold, and metrics, differential forms, connections, and curvature are all tensor fields; the formalism is what makes calculus possible on a space that admits no preferred coordinates.

It was Roger Penrose who first used diagrammatic notation, called tensor networks, to represent complicated tensor structures \cite{penrose1971applications}.
His intuitive notation immediately promoted the popularity of tensors.
Since then, tensor networks have found fruitful applications in many areas of science, and above all in the theoretical and numerical study of quantum many-body systems and of quantum computation \cite{orus2014}.
A chain of $n$ spin-half particles inhabits a Hilbert space of dimension $2^n$, so that merely storing the state of fifty of them exceeds any conceivable computer; but the states that nature actually realises are far from generic, and their modest entanglement allows them to be written as the contraction of a network of small tensors.
This is the insight behind matrix product states and the density matrix renormalisation group \cite{white1992, schollwock2011}, and behind projected entangled pair states in two and higher dimensions \cite{verstraete2004peps}.
Closer to the subject of this dissertation, the output amplitude of a quantum circuit is by definition the full contraction of a tensor network \cite{Markov_2008}, so that the classical simulation of a quantum computer is precisely the problem of contracting a network efficiently.
The claim of quantum advantage made for Google's Sycamore processor \cite{GoogleSycamore} was contested, and largely deflated, by exactly such contraction algorithms \cite{panzhang2022sycamore}.

Outside physics, tensors are the currency of high dimensional numerical computation and of data analysis.
Low-rank decompositions of tensors, pioneered a century ago and long standard in psychometrics, form a classical toolkit \cite{koldabader2009}; the tensor-train format in particular is how numerical analysts defeat the curse of dimensionality \cite{oseledets2011tt}.
Maybe most famously, tensors are the core and most indispensable data structure of machine learning, where the dominant numerical frameworks are built around them and indeed named after them, and where tensor network methods borrowed from physics serve as models in their own right \cite{stoudenmire2016}.

What all of these applications share is a single computational bottleneck: the cost of evaluating a tensor network.
The cost depends on the order in which its indices are summed away (this operation is called contraction of an edge), and finding the best order is NP-hard \cite{NPcomplete_tensor_network_contraction}.
Finding the optimal contraction order is equivalent to finding a special graph theoretical property of the line graph of the tensor network, called treewidth.

\section{Related Work: Finding Good Contraction Orders}\label{sec:related-work}

Because finding the optimal contraction order is NP-hard, every practical contraction algorithm settles for a good order found by heuristic means.
It is worth surveying these heuristics, both because they are the algorithms that our preprocessing is designed to complement, and because they arrange themselves naturally into the same graph-theoretic categories as chapter \ref{ch:treewidth}.

The first family descends from the treewidth literature itself.
Markov and Shi \cite{Markov_2008} introduced tree-decomposition-based contraction of quantum circuits, the construction we restate as theorem \ref{thm:treewidth_tensor_contraction}; within graph theory, good tree decompositions have long been produced by elimination heuristics such as minimum degree and minimum fill-in \cite{rose1970triangulated, rose1976elimination}, by maximum cardinality search, and by combinations of these wrapped in randomisation \cite{bodlaender1993tourist}.
In the language of this dissertation these are contraction order heuristics in disguise: simplicial elimination of a vertex of the line graph is exactly the contraction of the corresponding bond (lemma \ref{lem:contracting-makes-neighbours}), so an elimination ordering of $G^*$ and a contraction order of $G$ are the same object, and each heuristic above is directly an ordering heuristic for tensor contraction.

The second family works directly on contraction trees, in the spirit of the dynamic programming solution of the matrix chain problem \cite{hu1982matrixchain}.
Its most widely used representative is \texttt{opt\_einsum} \cite{smith2018opteinsum}, which since NumPy 1.12 has supplied the path optimiser of \texttt{numpy.einsum}: it repeatedly performs the pairwise contraction of least incremental cost, and for small networks it can branch and bound over partial orders to certify optimality.

The third family replaces greedy construction by randomised search.
Gray's hyper-optimised contraction \cite{gray2021hyperoptimised} re-runs greedy ordering thousands of times with randomly perturbed costs, applies simulated annealing and genetic algorithms over the space of contraction trees, and hands very large or sparse networks to graph partitioners; this is the current state of the art for arbitrary tensor networks.

A fourth family exploits the structure of quantum circuits specifically.
One approach divides the space-time volume of the circuit into patches that are contracted almost independently before being stitched back together \cite{panzhang2022sycamore}; another sweeps through the circuit line by line while keeping only a matrix product state boundary, so that the cost is governed by entanglement rather than by treewidth and degrades gracefully on noisy devices \cite{zhou2020limits}.

The family closest to this dissertation rewrites the diagram before an order is ever chosen.
The standard rewrite rules of ZX calculus, applied automatically by \texttt{PyZX} \cite{kissinger2019pyzx}, are routinely used to simplify circuits prior to contraction, and Cam and Martiel \cite{cam2023speedingquantumcircuitssimulation}, whose work inspired this dissertation, likewise reduce the contraction cost of a circuit's tensor network by graphical rewriting.
What is missing from all of these, and what this dissertation contributes, is an explicit optimisation target together with a safety guarantee: we take the treewidth of the line graph as the quantity to be reduced, and we prove (chapter \ref{ch:zx-algorithms}) that our rewrite operations --- K3 remove edge and unfuse and K3 complementation --- can never increase the cost of the best bond-by-bond contraction.

While existing algorithms search for a good contraction order on a fixed network, this dissertation adopts a different approach: transform the network itself into an equivalent tensor network whose line graph has lower treewidth before applying the bond-by-bond contraction algorithm.
Our key observation is that in many circumstances this reduction can be achieved by graph operations alone.
The preprocessing is complementary to every heuristic surveyed above, any of which can be run afterwards on the rewritten, strictly easier network.
The idea is explored in detail in chapter \ref{ch:zx-algorithms}.

\section{Outline}

We start this thesis with a thorough introduction to tensors and tensor networks in chapter \ref{ch:tensors}.
Chapter \ref{ch:treewidth} introduces the concept of treewidth and how treewidth can be used to determine the complexity of tensor contraction.
Chapter \ref{ch:zx_calculus} describes a popular graphical language, ZX calculus, for a special type of tensor networks called quantum circuits. 
ZX calculus enables us to perform diagrammatic rewrites, which are transformations of a tensor network into another equivalent one. Some of these transformations may reduce treewidth and therefore reduce the complexity of tensor contraction.
Chapter \ref{ch:zx-algorithms} explores and proves how each transformation affects treewidth. We also propose an algorithm to exploit these transformations to reduce the complexity of tensor contraction for quantum circuits. 
These transformations are named as \textit{rewrite preprocessing}.
In the last chapter \ref{ch:generic_diagram}, we discuss different diagrammatic calculi that have the potential to be used for generic tensor contraction.

\section{Statement of Originality}

Most of the results in this thesis except those in chapter \ref{ch:zx-algorithms} are well established in the literature and they are cited properly.
All of the explanations, proofs, diagrams, demonstrations, and discussions are my own work and interpretations, unless otherwise stated.

All of the proofs in chapter \ref{ch:treewidth} are my own, although they are classical results in graph theory.
They drew much inspiration from \textit{Introduction to Graph Theory} by Douglas B. West \cite{west2001introduction} and several other papers including
\cite{dirac1961rigid,fulkerson1965incidence,rose1970triangulated,gavril1974intersection}.

The sections on ZX and ZXW calculus in chapters \ref{ch:zx_calculus} and \ref{ch:generic_diagram} are mostly based on the book by Kissinger and van de Wetering \cite{KissingerWetering2024Book}, as well as several related papers including those by Wang et al.~\cite{wang2025penrosetensordiagramszx, Wang_complete}.

The tensor contraction algorithms with rewrite preprocessing are novel, although they were inspired by the paper by Cam and Martiel \cite{cam2023speedingquantumcircuitssimulation}.

Another important part of this thesis is the computer programs I wrote to implement the algorithms,
which I named \texttt{cartographer}. It is responsible for most of the computer simulation and is available at \url{https://github.com/harryhanYuhao/cartographer}.

The figures in this thesis draw inspiration from the illustrations in \cite{KissingerWetering2024Book} and \cite{waintal2026competequantumcomputerslecture}.
Most of them are drawn by myself using \texttt{cartographer} and the open-source software \texttt{zxw-graphboard}\footnote{zxw-graphboard is available at \url{https://github.com/harryhanYuhao/graphboard}.}.

I use the open-source LLM DeepSeek, and Zhipu GLM, as well as the proprietary LLM ChatGPT to help with proofreading, literature search, and some coding tasks.
No other parts of this thesis were involved with any AI tools.
No other AI models are used.

\subsection{Format and Diagram}

The thesis is typeset using \LaTeX{} and the CUED Template created by \href{https://github.com/kks32/phd-thesis-template} {Krishna Kumar}.
The figures are created using TikZ and the open-source graph-board program at \url{https://github.com/harryhanYuhao/graphboard}.

\chapter{Tensors} \label{ch:tensors}

\section{Tensors and Basic Operations}

We present the following straightforward definition of tensors, which is a rephrasing of Penrose's original definition \cite{penrose1971applications}.

\begin{definition}\label{def:tensor}
Given $n\in \N$, let $[n]$ denote the set $\{0, 1, \cdots, n-1\}$. 
A tensor of rank $k \in \N$ is a function $T : [n_0] \times [n_1] \times \cdots \times [n_{k-1}] \rightarrow \C$.
\end{definition}

By convention, we shall write the value $T(i_0, i_1, \cdots, i_{k-1})$ as $T_{i_0 i_1 \cdots i_{k-1}}$, where each $i$ is called an index.
Of course, the output of $T$ can be $\R$ or any other ring. 
We take it to be $\C$ in this dissertation for convenience.
The value $n_j$ is called the dimension of the $j$-th index, which will always be clear from the context.

Following the convention of Einstein summation, a tensor $T$ is denoted by a capital letter with its indices as free variables shown in the subscript.
The class of all tensors is denoted by the calligraphic letter $\T_{ijk \cdots}$.

A tensor is, plainly and unostentatiously, an arrangement of numbers with specified orderings by the indices.
For example, a rank-one tensor is a vector, a rank-two tensor is a matrix, and a rank-three tensor is an arrangement of matrices.

\begin{figure}[ht]
\centering
\input{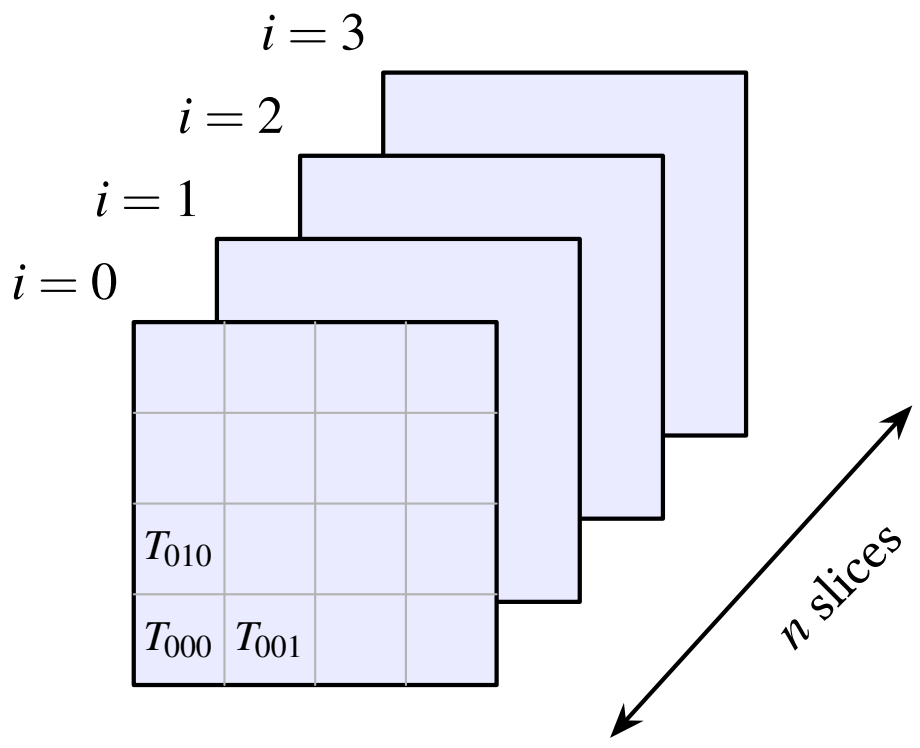}
\caption{A rank-three tensor $T_{ijk}$ regarded as a stack of $n$ matrices.}
\label{fig:rank3-stack}
\end{figure}

The following tensors denote a vector and a matrix, while figure \ref{fig:rank3-stack} depicts the rank-three tensor as a stack of $n$ such matrix slices.
\begin{equation*}
	T_i =
	\begin{pmatrix}
		T_0 \\ T_1 \\ \vdots \\ T_{n-1}
	\end{pmatrix},
	\qquad
	T_{ij} =
	\begin{pmatrix}
		T_{00} & T_{01} & \cdots & T_{0,n-1} \\
		T_{10} & T_{11} & \cdots & T_{1,n-1} \\
		\vdots & \vdots & \ddots & \vdots \\
		T_{n-1,0} & T_{n-1,1} & \cdots & T_{n-1,n-1}
	\end{pmatrix}.
\end{equation*}

There are at least seven natural operations for tensors.
\begin{observation}
	\begin{itemize} 
		\item \textbf{Addition} $\T_{ijk \cdots} + \T_{ijk \cdots} \rightarrow \T$. Let $T_{ijk \cdots}$ and $S_{ijk \cdots}$ be two tensors of the same rank. Then their sum is defined by $(T + S)_{ijk \cdots} = T_{ijk \cdots} + S_{ijk \cdots}$.
		\item \textbf{Scalar multiplication} $\C \times \T_{ijk \cdots} \rightarrow  \T_{ijk \cdots}$. Let $T_{ijk \cdots}$ be a tensor and $c \in \C$. Define $(cT)_{ijk \cdots} = c (T_{ijk \cdots})$.
		\item \textbf{Permutation.} $\T_{ijk \cdots} \rightarrow \T_{ijk \cdots}$. Let $T_{ijk\cdots}$ be a tensor of rank $m$. For any permutation $\sigma \in S_m$, the permuted tensor is defined by $(\sigma \cdot T)_{i_{\sigma(1)} i_{\sigma(2)} \cdots i_{\sigma(m)}} = T_{i_1 i_2 \cdots i_m}$.
		\item \textbf{Flatten} $\T_{ijk \cdots} \rightarrow \T_{jk \cdots}.$ Let $T_{ijk \cdots}$ be a tensor. 
		Any two of the indices of $T$ can be flattened into one to create a new tensor.
		Say we would like to flatten the indices $i$ and $j$, of dimensions $n_1$ and $n_2$. Let $\alpha$ be an index of dimension $n_1n_2$, and we define the new tensor $S_{\alpha k \cdots}$ as $S_{(n_2 i + j)k \cdots} = T_{ijk\cdots}$.
		\item \textbf{Unflatten}, which is exactly the opposite operation of flatten.
		\item \textbf{Outer product} $\T_{ijk \cdots} \times \T_{\alpha \beta \gamma \cdots} \rightarrow \T_{ijk \cdots \alpha \beta \gamma \cdots}.$ Let $T_{ijk \cdots}$ and $S_{lmn \cdots}$ be any two tensors.
		Their tensor product is defined by $(T \otimes S)_{ijklmn\cdots} = T_{ijk\cdots} S_{lmn\cdots}$. Outer product is sometimes called \textbf{tensor product} or \textbf{Kronecker product}.
		\item \textbf{Contraction.}  $\T_{ijk \cdots} \times \T_{ilm \cdots} \rightarrow \T_{jk \cdots lm \cdots}.$ Let $T_{ijk\cdots}$ and $S_{ilm \cdots}$ be tensors of rank $m$ and $n$. Their contraction over the index $i$ is a tensor of rank $m+n - 2$ defined by $(T \cdot S)_{jk\cdots lm\cdots} = \sum_{i=0}^{d_i - 1} T_{ijk\cdots} S_{ilm\cdots}$, where $d_i$ is the dimension of the index $i$.
		If there are many different tensors with the corresponding contraction indices specified, their full contraction gives a new tensor defined as
		\begin{equation}
			\sum_{i_1=0}^{n_1-1} \sum_{i_2=0}^{n_2-1} \cdots \sum_{i_{k-1}=0}^{n_{k-1}-1} T^{(1)}_{i_1 j_1 \cdots} T^{(2)}_{i_1 i_2 j_2 \cdots} \cdots T^{(k)}_{i_{k-1} j_k \cdots}
		\end{equation}\label{eq:contraction}
		where $i_1$ to $i_{k-1}$ are the indices to be contracted.
	\end{itemize}
\end{observation}

Addition and scalar multiplication of tensors are clear enough. 
Permutation of indices is a generalisation of the matrix transpose. 
For example, consider a rank-two tensor $T$, and define $S_{ij} = T_{ji}$.

\begin{equation*}
	T_{ij} =
	\begin{pmatrix}
		T_{00} & T_{01} & \cdots  \\
		T_{10} & T_{11} & \cdots \\ 
		\vdots & \vdots & \ddots \\
	\end{pmatrix}, \qquad
	S_{ij} =
	\begin{pmatrix}
		T_{00} & T_{10} & \cdots  \\
		T_{01} & T_{11} & \cdots \\ 
		\vdots & \vdots & \ddots \\
	\end{pmatrix}
\end{equation*}

\begin{figure}[ht]
\centering
\input{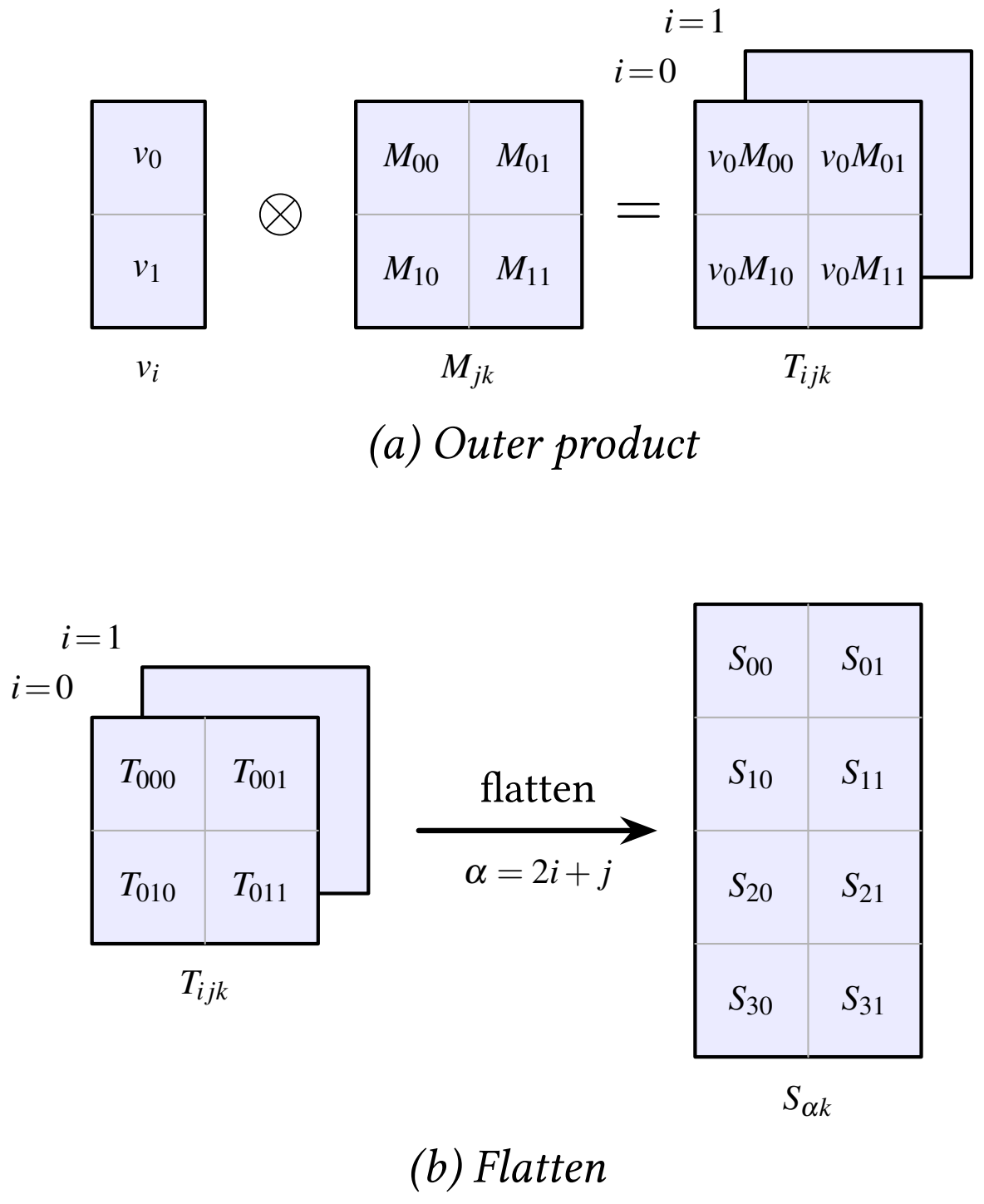}
\caption{(a) The outer product of a vector $v_i$ and a matrix $M_{jk}$ yields the rank-three tensor $T_{ijk}$. (b) Flattening the indices $i$ and $j$ of $T_{ijk}$ into a single index $\alpha = n_2\, i + j$ yields the matrix $S_{\alpha k}$.}
\label{fig:outer-flatten}
\end{figure}

Outer product and flattening are demonstrated in figure \ref{fig:outer-flatten}.

Usually, outer product and flattening are used together to convert tensor products of two arbitrary tensors into a single matrix, for example
\begin{equation}
	A \otimes \vec{v} = 
	\begin{pmatrix}
		a_{00} & a_{01}\\
		a_{10} & a_{11}
	\end{pmatrix}
	\begin{pmatrix}
		v_0 \\ v_1	
	\end{pmatrix}
	= 
	\begin{pmatrix}
		a_{00} v_0 & a_{01} v_0 \\
		a_{10} v_0 & a_{11} v_0 \\
		a_{00} v_1 & a_{01} v_1 \\
		a_{10} v_1 & a_{11} v_1
	\end{pmatrix}
\end{equation}

In general, if we have two matrices $M$ and $N$, their outer product will be 
\begin{equation}
	M \otimes N = 
	\begin{pmatrix}
		M_{00} N & M_{01} N & \cdots \\
		M_{10} N & M_{11} N & \cdots \\
		\vdots & \vdots & \ddots
	\end{pmatrix}
\end{equation}

Contraction is the most complicated operation. 
For the matrices, $M_{ij}$ and $N_{jk}$, their contraction over the index $j$ is a matrix $P_{ik}$ defined by $P_{ik} = \sum_{j=0}^{n-1} M_{ij} N_{jk}$, which is exactly the matrix multiplication.

Given a set of tensors and the corresponding contraction indices, the contraction order does not matter, because we can freely change the order of and add parentheses to nested summations.

\begin{equation}
	\begin{split}
		\sum_{i_1=0}^{n_1-1} \sum_{i_2=0}^{n_2-1} \cdots \sum_{i_{k-1}=0}^{n_{k-1}-1} T^{(1)}_{i_1 j_1 \cdots} T^{(2)}_{i_1 i_2 j_2 \cdots} \cdots T^{(k)}_{i_{k-1} j_k \cdots}
		&=
		\sum_{i_2=0}^{n_2-1} \sum_{i_1=0}^{n_1-1} \cdots \sum_{i_{k-1}=0}^{n_{k-1}-1} T^{(2)}_{i_1 i_2 j_2 \cdots} T^{(1)}_{i_1 j_1 \cdots} \cdots T^{(k)}_{i_{k-1} j_k \cdots} \\
		&= \sum_{i_1=0}^{n_1-1} \sum_{i_2=0}^{n_2-1} \cdots \sum_{i_{k-1}=0}^{n_{k-1}-1} T^{(1)}_{i_1 j_1 \cdots} \left( T^{(2)}_{i_1 i_2 j_2 \cdots} \cdots T^{(k)}_{i_{k-1} j_k \cdots} \right) \\
		&= \sum_{i_1=0}^{n_1-1} \sum_{i_2=0}^{n_2-1} \cdots \sum_{i_{k-1}=0}^{n_{k-1}-1} \left( T^{(1)}_{i_1 j_1 \cdots} T^{(2)}_{i_1 i_2 j_2 \cdots} \right) \cdots T^{(k)}_{i_{k-1} j_k \cdots} \\
		&= \cdots
	\end{split}
\end{equation}\label{eq:contraction-order}

This is an important observation and we will make it a lemma. 
\begin{lemma}[Contraction Order Does Not Matter]
	Given a set of tensors and the corresponding contraction indices, the contraction order does not matter.
\end{lemma} \label{lem:contraction-order}

\subsection{Abstract Tensor Product}\label{sec:abstract-tensor-product}

In multilinear algebra, the definition of tensor product is thus. 

\begin{figure}[ht]
	\centering 
    \input{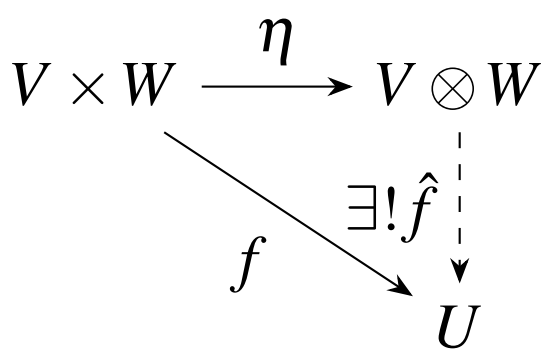}
	\caption{Commutative triangle for tensor product.}\label{fig:tensor-product-universal}
\end{figure}

\begin{definition}[Tensor Product of Vector Spaces]
	Let $V, W$ be vector spaces over a field $\F$. Their tensor product, denoted as $V \otimes W$, is the unique vector space over $\F$ that satisfies the following universal property as demonstrated in figure \ref{fig:tensor-product-universal}:
	there exists a bilinear map $\eta: V\times W \rightarrow V \otimes W $, and for every bilinear map $f : V \times W \rightarrow U$ to any vector space $U$, there exists a unique linear map $\hat{f} : V \otimes W \rightarrow U$ such that $f = \hat{f} \circ \eta$. 
\end{definition}

Equivalently, the tensor product of two finite-dimensional vector spaces $V$ and $W$, with basis $v_1, \cdots, v_n$ and $w_1, \cdots, w_m$ is the $n\cdot m$ dimensional vector space $V \otimes W$ with basis $v_i \otimes w_j, 1\leq i \leq n, 1 \leq j \leq m,$ that satisfies the following bilinearity conditions:
\begin{equation*}
	\begin{aligned}
		(v_i + v_j) \otimes w_k &= v_i \otimes w_k + v_j \otimes w_k, \\
		v_i \otimes (w_j + w_k) &= v_i \otimes w_j + v_i \otimes w_k, \\
		c (v_i \otimes w_j) &= (c v_i) \otimes w_j = v_i \otimes (c w_j), \quad c \in \F.
	\end{aligned}
\end{equation*}
where $i, j, k$ are the appropriate indices of the basis vectors.

I hope the original definitions of tensor product justify our nomenclature of tensor in definition \ref{def:tensor}.
It is clear that 
\begin{enumerate}
	\item A rank-one tensor, $T_i$, is a vector in some vector space $V$,
		where the dimension of the index $i$ is the dimension of $V$.
	\item In general, a rank-$k$ tensor $T_{i_1 i_2 \cdots}$ has additions and scalar multiplications, so the set of rank-$k$ tensors with fixed dimension for each of their indices, denoted by $\T_{i_1, i_2 \cdots}$, is a vector space.
	This vector space has a basis $\{e_{j_1, j_2, \cdots, j_k}, \cdots\}$, where $e_{j_1, j_2, \cdots,}(i_1, i_2, \cdots) = 1$ if and only if $i_1 = j_1, i_2 = j_2, \cdots$, and $0$ otherwise.
	This set of bases is precisely the basis of the tensor product of vector spaces $V_0, V_1, \cdots, V_k$.
	\item Flatten is the canonical isomorphism between the vector space $V_1 \otimes V_2 \otimes \cdots \otimes V_k \rightarrow (V_1 \otimes V_2) \otimes \cdots \otimes V_k$. (And unflatten is its inverse.)
	\item Outer product $(\T_1, \T_2) \rightarrow \T_1 \otimes \T_2$ is the canonical bilinear map guaranteed by the universal property of tensor product.
\end{enumerate}

In linear algebra, there is an intrinsic notion of input and output. 
For example, in vector space $V$ a vector is an element of $V$. Its dual is a function $V \rightarrow \C$, and a matrix is a function $V \rightarrow V$.
In physics, the inputs and outputs of a tensor are differentiated by the upper and lower indices. 
For example, $\T^k_l$ denotes the class of tensors $ \underbrace{V \otimes V \otimes \cdots \otimes V}_{k} \otimes \underbrace{V^* \otimes V^* \otimes \cdots \otimes V^*}_{l} \cong \underbrace{V \otimes \cdots \otimes V}_{l} \rightarrow \underbrace{V \otimes \cdots \otimes V}_{k}$,
where $V$ is a vector space and $V^*$ is its dual.
While differentiating inputs and outputs seems an intuitive idea, it is unwieldy in our algorithmic investigation of tensor product. 
Therefore, we will not differentiate input from output unless otherwise specified.

\section{Tensor Networks}

Suppose we have two tensors $T_{ijk}$ and $S_{kl}$, where indices $i$ and $j$ have dimensions $n_1$ and $n_2$ while both of the indices $k$ and $l$ have dimension $n_1$.
Suppose that we would like to perform the following operation: `contracting the first index of $T$ with the second index of $S$, and contracting the last index of the resulting tensor with the first index of another copy of $T$.'
The precise statement of specifying operations on tensors is too verbose and painful to articulate.
The tensor network, a graphical representation of tensors and their operations, is used to make these statements concise and clear.

\begin{definition}[Tensor Network]
	A tensor network is a generic graph structure consisting of vertices and edges that denotes the tensor resulting from the various operations on a set of tensors, following the three rules below.

\textbf{Rule 1}. In a tensor network, a rank-$k$ tensor is represented as a node with $k$ edges.
Usually, the shape, color, and labels of the nodes are used to differentiate different tensors.
For example, a vector $v$, a matrix $M$, and a rank-three tensor $T$ are represented as in figure \ref{fig:simple_tensors}.

\begin{figure}[ht]
	\centering
	\input{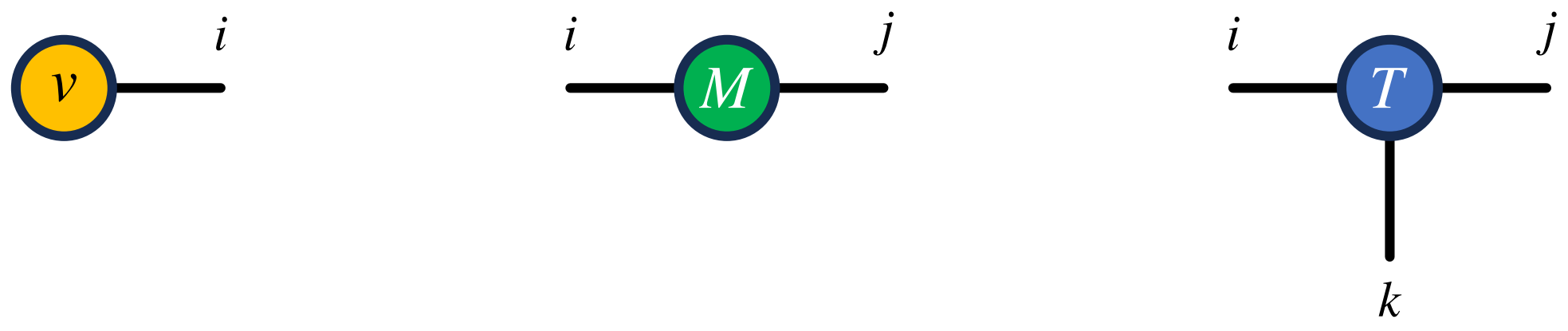}
	\caption{Rank 1, 2, and 3 tensors}
	\label{fig:simple_tensors}
\end{figure}

\textbf{Rule 2}. A diagram with two tensors placed side by side is interpreted as the outer product of the two tensors, as shown in figure \ref{fig:tensor-product}.

\textbf{Rule 3}. An edge connecting two nodes represents the contraction of the two tensors over the index represented by that edge.
\begin{figure}[ht]
	\centering
	\input{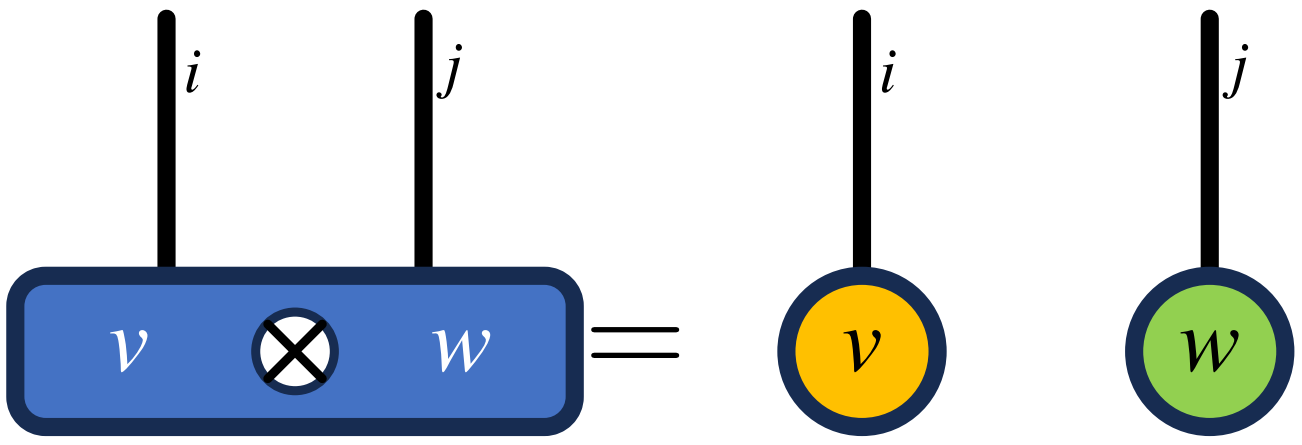}
	\caption{Outer Product in Tensor Network}
	\label{fig:tensor-product}
\end{figure}

\end{definition}

We may also define tensor networks as a special kind of subgraph such that
\begin{enumerate}
		\item it allows loops and multiple edges,
		\item its vertices are of different types,
		\item the order of the edges of a vertex matters.
\end{enumerate}

The value of a tensor can be evaluated by plugging in the values of the indices.
In a tensor network, this is analogous to assigning a value to each of the edges and the graph can then be evaluated to some number.

\begin{figure}[ht]
	\centering
	\input{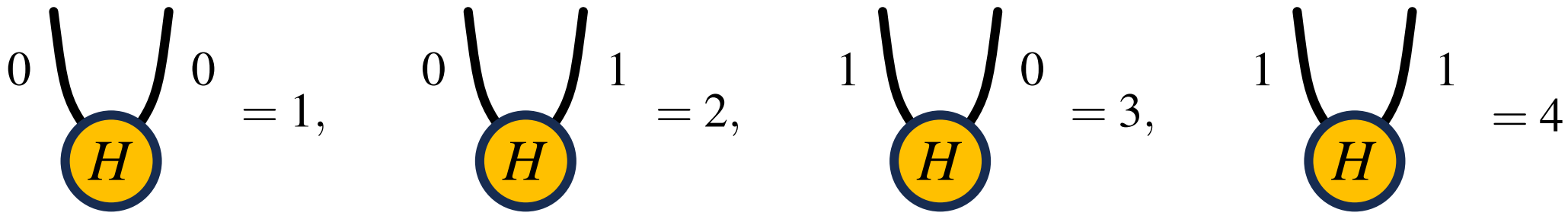}
	\caption{Evaluating a tensor network}
	\label{fig:tensor-net-eval}
\end{figure}

For example, consider the tensor $H = \begin{pmatrix}
	1 & 2 \\
	3 & 4
\end{pmatrix} $. In explicit tensor notation, $H_{00} = 1, H_{01} = 2,  H_{10} = 3$ and $H_{11} = 4$.
In tensor network notation, $H$ is represented as a node with two edges. Labelling the edges with $0$ and $1$ respectively evaluates the tensor as shown in figure \ref{fig:tensor-net-eval}.

\subsection{Order of Edges}

\begin{figure}[ht]
	\centering
	\input{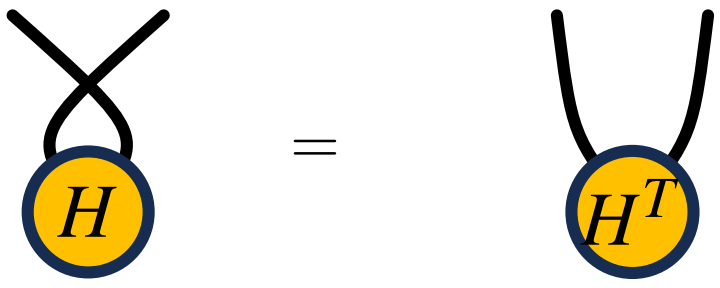}
	\caption{Transpose in Tensor Network}
	\label{fig:tensor-transpose}
\end{figure}

While tensor networks do not differentiate input and output, they do differentiate the order of the bonds of a tensor. 
For example, figure \ref{fig:tensor-transpose} shows the transpose of $H$.

\begin{figure}[ht]
	\centering
	\input{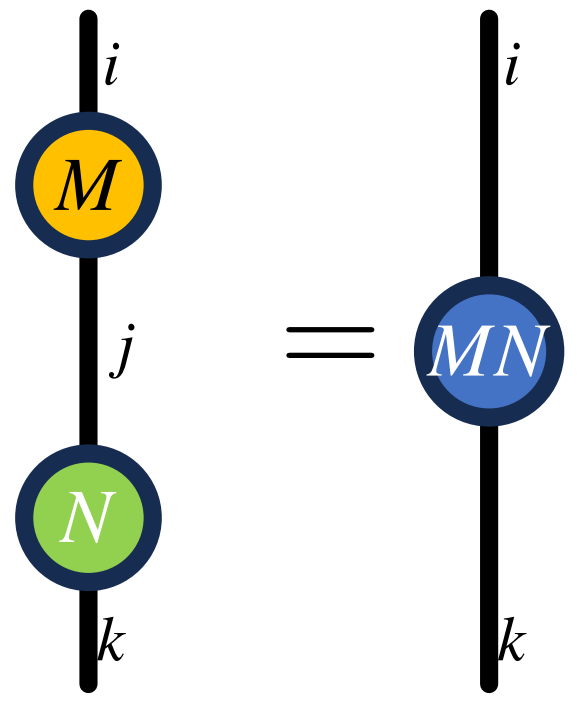}
\caption{Contracting the two rank-two tensors $M_{ij}$ and $N_{jk}$ over their shared index $j$ gives the rank-two tensor $(MN)_{ik} = \sum_j M_{ij} N_{jk}$.}
	\label{fig:contraction-matrix-mult}
\end{figure}

The edges are typically ordered from top to bottom and from left to right. 
Orders can also be specified by labelling the edges with indices of the tensor. 
For example, in figure \ref{fig:contraction-matrix-mult} we demonstrate matrix multiplication, a non-commutative operation, in tensor network notation.

\subsection{Only Topology Matters}

There is a subtlety in the graphical notation.
Each subgraph of a tensor network can itself be regarded as a new tensor resulting from the full contraction of its nodes.
For example, in figure \ref{fig:contraction-matrix-mult}, the two nodes $M$ and $N$ can be regarded as a new node $MN$ with two edges, and in figure \ref{fig:tensor-product}, the two nodes $v$ and $w$ can be regarded as a new node $v \otimes w$ with two edges. 
Given a tensor network, there are many different ways to decompose a tensor network into subgraphs and many different orders for contraction; will all of them give the same result?

Fortunately, lemma \ref{lem:contraction-order} guarantees that the answer is yes, as long as the nodes and their connections are determined.
The tensor network notation precisely captures the nodes and the connections between them, so the tensor network notation is unambiguous and well-defined.
This property is summarised in the phrase `only topology matters'.

\subsection{Symmetric Tensors}

A special class of tensors is worth our attention. 

\begin{definition}[Symmetric Tensor]
	A tensor $T_{ijk\cdots}$ is called symmetric if it is invariant under any permutation of its indices, i.e. for any permutation $\sigma \in S_k$, we have
	$$
		T_{i_1 i_2 \cdots i_k} = T_{i_{\sigma(1)} i_{\sigma(2)} \cdots i_{\sigma(k)}}.
	$$
\end{definition}\label{def:symmetric-tensor}

\begin{figure}[ht]
	\centering
	\input{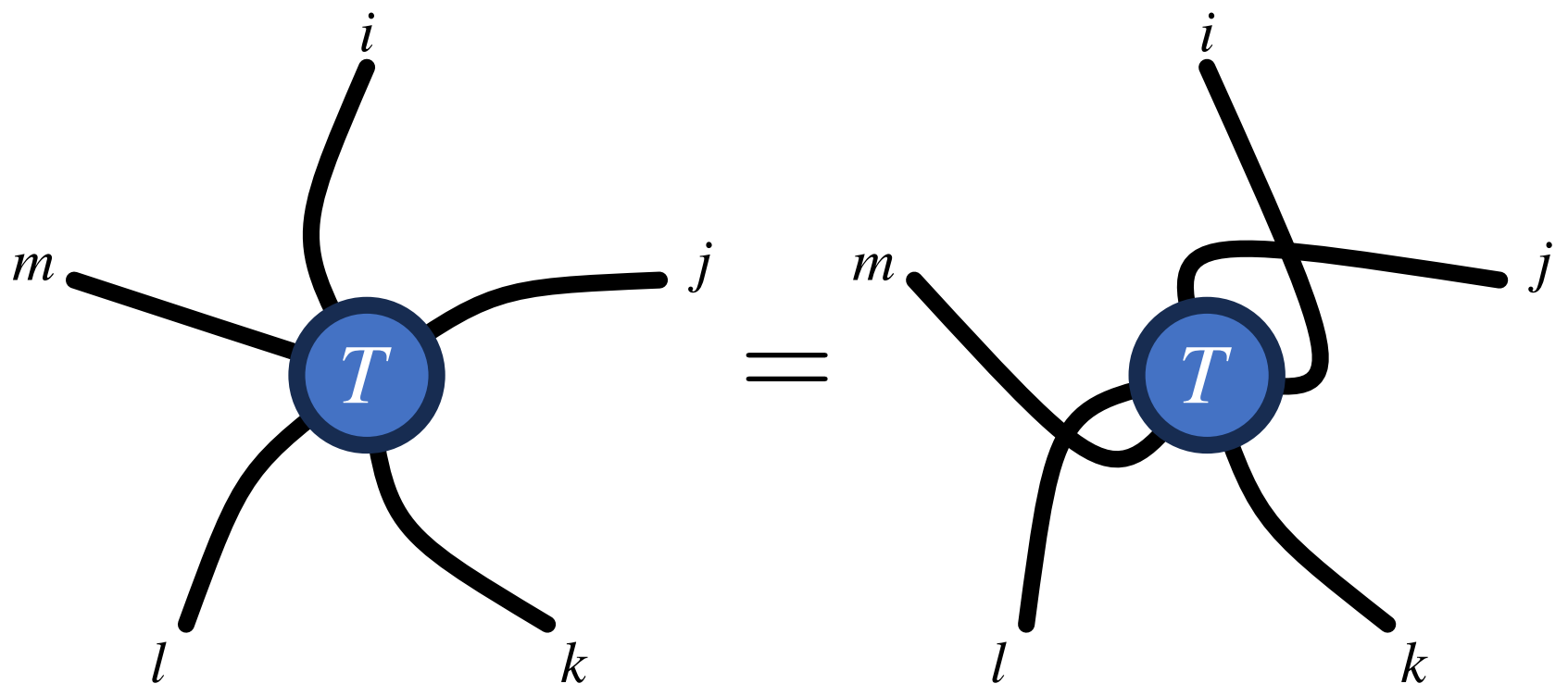}
	\caption{Permuting the edges of a symmetric tensor does not change the tensor.}
	\label{fig:symmetric-tensor}
\end{figure}

In tensor network notation, permuting the edges of a symmetric tensor does not change the tensor, as shown in figure \ref{fig:symmetric-tensor}.

Tensor networks that consist of only symmetric tensors can be regarded as coloured hypergraphs, as the order of the edges of a symmetric tensor does not matter.
Therefore, if there are no dangling legs, the networks of symmetric tensors can indeed be regarded as coloured hypergraphs, and two networks that are isomorphic as hypergraphs are equivalent as tensor networks.

If there are dangling legs, the order and connection to those legs still matter, as contraction between symmetric tensors may give non-symmetric tensors.
In this case, we can regard each dangling leg as a normal edge connected to a special node coloured in its unique colour.

We shall summarise this observation as \define{only connectivity matters}.

\begin{lemma}[Only Connectivity Matters]
	Two tensor networks without dangling legs that are composed of only symmetric tensors are equivalent if they are isomorphic as coloured hypergraphs.
	That is, they equal the same tensor if they have the same nodes and connections between them.

	If they do have dangling legs, the ordering of the dangling legs still matters. 
\end{lemma}\label{lem:only-connectivity-matters}

\section{Complexity of Contraction}

Let us investigate the algorithmic complexity of tensor contraction, which is a problem of paramount importance as it is used in so many applications, such as quantum circuit simulation, machine learning, and almost all kinds of mathematical modelling.

For simplicity, we assume all indices in this section have dimension $d$.

Contraction of two rank-2 tensors, $M_{ij}$ and $N_{jk}$, gives another rank-2 tensor $P_{ik}$.
Each entry of $P$ is computed using the formula
$$
	P_{ik} = \sum_{j=0}^{d-1} M_{ij} N_{jk}	
$$
Computing a single entry thus takes $d$ multiplications and $d-1$ additions.
There are a total of $d^2$ entries to compute, so the total number of operations is $d^2(d + d - 1) = O(d^3)$.

Similarly, contraction of one bond between a rank $m$ and a rank $n$ tensor will result in a rank $m + n - 2$ tensor. 
Each entry of the resulting tensor is computed by summing over $d$ products of entries from the two tensors, so the total number of operations is $O(d^{m+n-1})$.

We may also choose to contract multiple bonds simultaneously. 
In general, contracting $k$ bonds between a rank $m$ and a rank $n$ tensor will result in a rank $m + n - 2k$ tensor.
Each entry of the resulting tensor is computed by summing over $d^k$ products of entries from the two tensors, so the total number of operations is $O(d^{m+n-k})$.

It is thus clear that contracting $k$ bonds simultaneously over a tensor network with $n$ external bonds will take $O(d^{n+k})$ operations.
This is an important observation that is worth writing down.

\begin{lemma}[Complexities of Tensor Contraction by Counting Edges]\label{lem:contraction-complexity}
	Let $T$ be a tensor network with $n$ external bonds and $k$ internal bonds. 
	Contracting all the internal bonds of $T$ will take $O(d^{n+k})$ operations.
\end{lemma}

\begin{figure}[ht]
	\centering
	\input{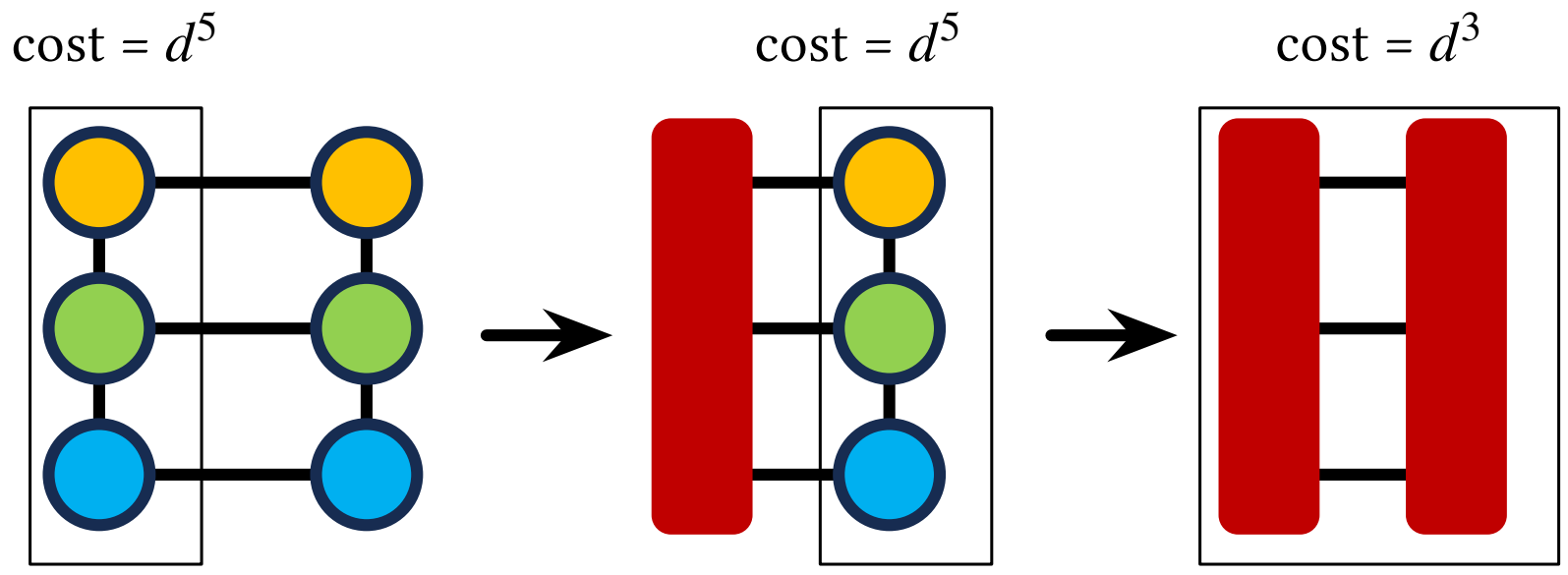}
	\caption{Slow contraction order}
	\label{fig:contract-order-slow}
\end{figure}

\begin{figure}[ht]
	\centering
	\input{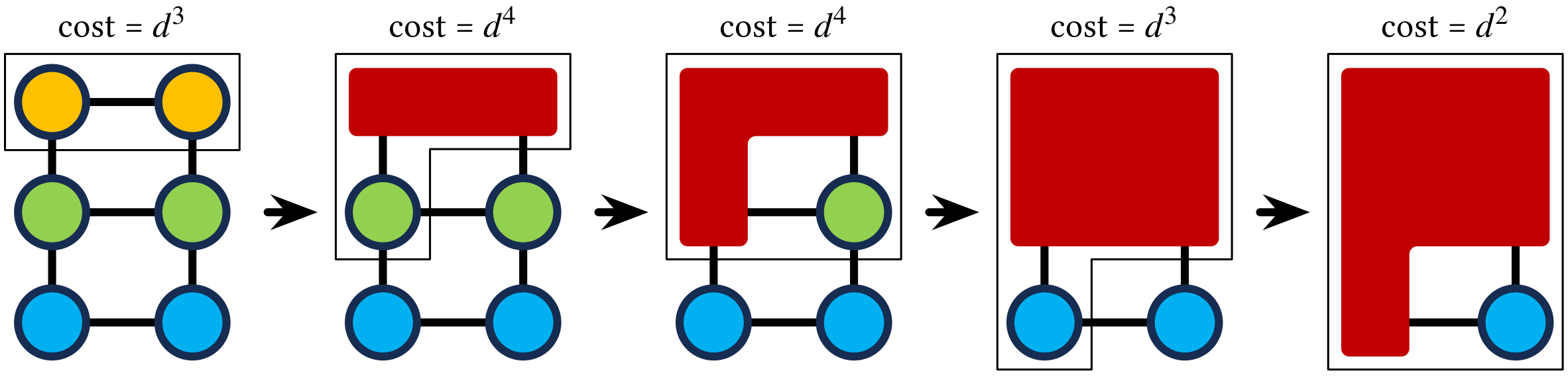}
	\caption{Fast contraction order}
	\label{fig:contract-order-fast}
\end{figure}

It may surprise the reader that different orders of contraction may have different computation costs.
For example, contracting the following tensor fully, which will give a scalar, can be done in two different orders as shown in figures \ref{fig:contract-order-slow} and \ref{fig:contract-order-fast}.
The first contraction order takes $O(d^5)$, while the second contraction order takes $O(d^4)$.


\subsection{Why is There a Difference in Complexity?}

As equation \eqref{eq:contraction-order} shows, different orders of tensor contraction correspond to different orders of the nested summation, which is a result of three basic rules of arithmetic: commutativity, associativity, and distributivity.
It is natural to wonder why such simple arithmetic identities can lead to such a difference in computation cost.
This section examines the question.

Consider the following identity
$$
	(a_1 + a_2 + \cdots + a_n)(b_1 + b_2 + \cdots + b_m) = a_1 b_1 + a_1 b_2 + \cdots + a_n b_m
$$

The left-hand side of the equation takes $n-1$ additions for $a_i$, $m-1$ additions for $b_i$, and one multiplication between the sums, which is a total of $m+n -1$ operations.
The right-hand side, however, takes $nm$ multiplications between $a_i$ and $b_i$ and $nm-1$ additions, amounting to a total of $2nm - 1$ operations. 
This is a big difference all caused by the distributive law.

The impact of the order of operations for matrices is even more pronounced.
For example, consider $\vec{v_1}, \vec{v_2} \in \C^n $ and $M \in \C^{n \times n}$. 
Consider the following operations calculated in two different orders:
$$
(M\vec{v_1} ) \vec{v_2}^T = M(\vec{v_1} \vec{v_2}^T)
$$

The left-hand side takes two matrix-vector multiplications, which is $O(n^2)$ operations, while the right-hand side takes one outer product, costing $O(n^2)$, and one matrix-matrix multiplication, costing $O(n^3)$ operations.

In general, both finding the optimal order and determining if a given order is optimal are NP-hard problems.
There are, nevertheless, many heuristic algorithms to find a good contraction order.
In this thesis, we will discuss in detail one such algorithm, which relies on a graph-theoretic property called treewidth.

\textbf{Note on Strassen's Algorithm}

While this dissertation focuses on tensor contraction, we do not touch the even more basic problem of matrix multiplication. 
It is perhaps surprising that even the simple matrix multiplication defined by computing the sum 
$$
\sum_{k=0}^{n-1} A_{ik} B_{kj}
$$
may have a more efficient algorithm than the naive $O(n^3)$ complexity \cite{dumas2024strassensalgorithmoptimallyaccurate}. 
The first matrix algorithm that beats the naive algorithm was Strassen's algorithm with complexity $O(n^{2.81})$ \cite{strassen1969gaussian}.

Recently, there have been studies on whether the idea of Strassen's algorithm can be generalised to tensor contraction. 
It has been discovered that the complexity of tensor contraction is closely related to another property called tensor rank, which is a generalisation of matrix rank.

Studying algorithms of this nature deserves another independent dissertation.

\chapter{Treewidth and Tensor Contraction} \label{ch:treewidth}

The concept of tree decomposition and treewidth was first introduced by Robertson and Seymour to prove the famous graph minor theorem, which we state here for reference. 

\begin{theorem}[Graph Minor Theorem, Robertson and Seymour \cite{robertson_seymour_1}]
	For any infinite sequence of finite graphs $G_1, G_2, \ldots$, there exist indices $i < j$ such that $G_i$ is a minor of $G_j$.
\end{theorem}

Treewidth was later proved to be very useful in many different areas of graph theory. 
As the core algorithmic idea in this dissertation relies on treewidth, it is worth our effort to introduce treewidth and its properties in detail.

\section{Treewidth and its Elementary Properties}

\begin{definition}[Tree Decomposition]
	Let $G$ be a finite graph and $V$ its vertex set. A \define{tree decomposition} of $G$ is a pair $(T, \{X_t\}_{t \in V(T)})$ where $T$ is a tree and $\{X_t\}_{t \in V(T)}$ is a family of subsets of $V$ (called bags) such that
	\begin{enumerate}
		\item $\bigcup_{t \in V(T)} X_t = V$,
		\item If vertices $u, v \in G$ are adjacent, there exists a $t \in T$  such that $u, v \in X_t$.
		\item Let $a, b$ be vertices of $T$. Since $T$ is a tree, there is a unique path connecting $a$ and $b$, which is denoted as $P$. If $v \in G$ and $v \in X_a, v \in X_b$, then for all $c \in P$, we have $v \in X_c$.
	\end{enumerate}

	Let $X_t$ be the bag such that $ \forall i \in V(T), |X_t| \geq |X_i|$ (That is, $X_t$ is the bag of the largest order). The \define{width} of the tree decomposition $(T, \{X_t\})$ is $|X_t| - 1$.
\end{definition}

\begin{definition}[Treewidth]
	The \define{treewidth} of a graph $G$ is the minimum width over all of its tree decompositions.
\end{definition}

A tree decomposition requires the existence of a bag that contains all adjacent vertices. 
If $u,v \in G$ and $\exists t$ such that $ u, v \in X_t $, however, this does not guarantee that $u$ and $v$ are adjacent in $G$.

\begin{example}[All graphs have a trivial tree decomposition]
	Let $G$ be a graph with vertex set $V$. Then $(T, \{X_t\})$ is a tree decomposition of $G$ where $T$ is a single node and $X_t = V$. The width of this tree decomposition is $|V| - 1$.

	Therefore, the treewidth of any graph $G$ is bounded above by $|V| - 1$.
	By Zorn's lemma, there must exist a tree decomposition of $G$ with minimum width.
\end{example}

Let us look at some examples of tree decompositions.

\begin{figure}[ht]
	\centering
	\input{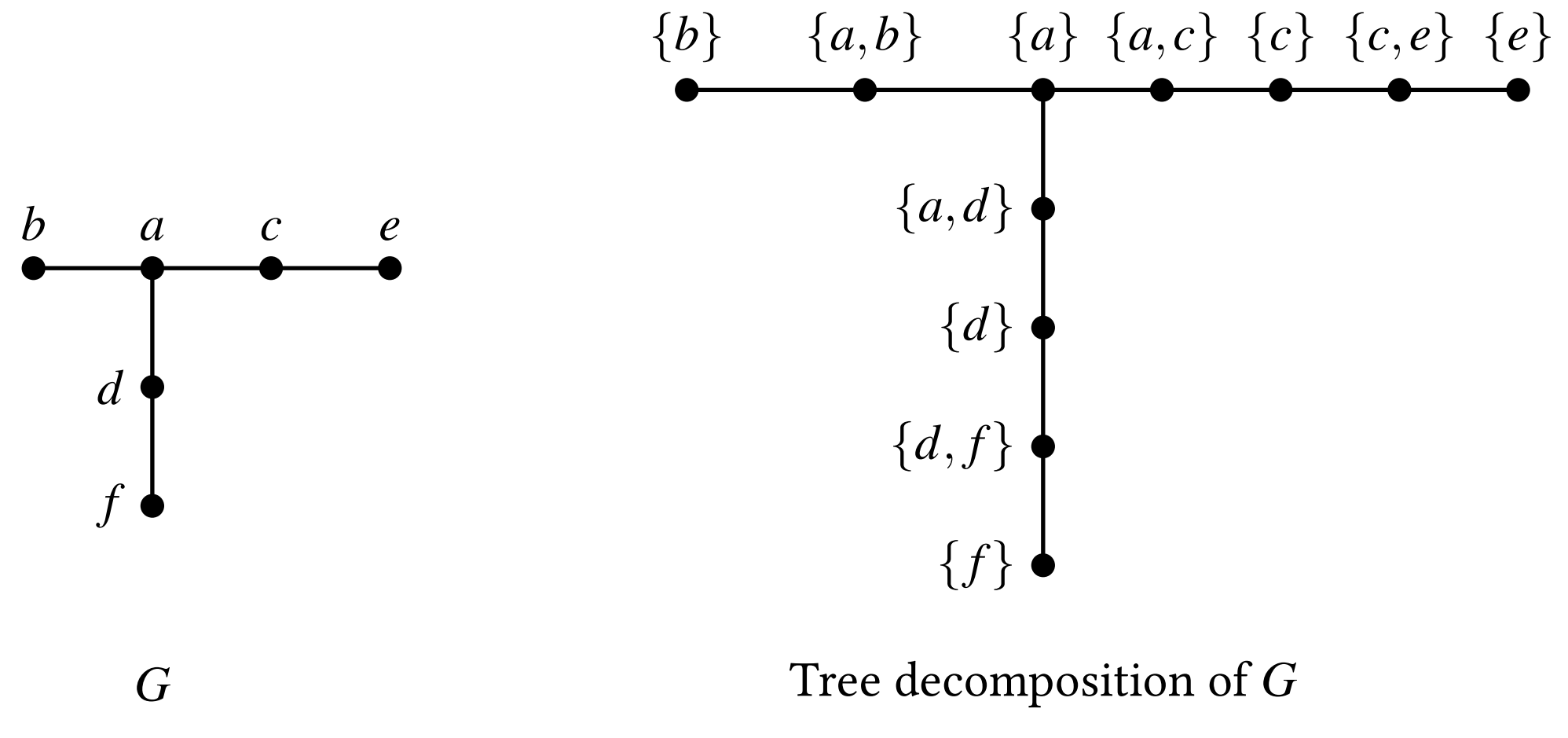}
	\caption{A tree decomposition of a tree.}
	\label{fig:tree_decomposition_of_tree}
\end{figure}

\begin{example}[Tree decomposition of a Tree]
	Let $G$ be a tree. We can construct a tree decomposition of $G$ in the following way.
	\begin{enumerate}
		\item Let each vertex of $G$ be a bag that contains only itself.
		\item On each edge of $G$, add a bag that contains the two endpoints of the edge.
	\end{enumerate}
	An example of this tree decomposition is shown in figure \ref{fig:tree_decomposition_of_tree}. 
	This shows that the treewidth of a tree is $1$. 
	For any tree with at least one edge, there must be a bag that contains two or more vertices. This proves that the treewidth of a non-trivial tree is exactly $1$.
\end{example}

\begin{figure}[ht]
	\centering
	\input{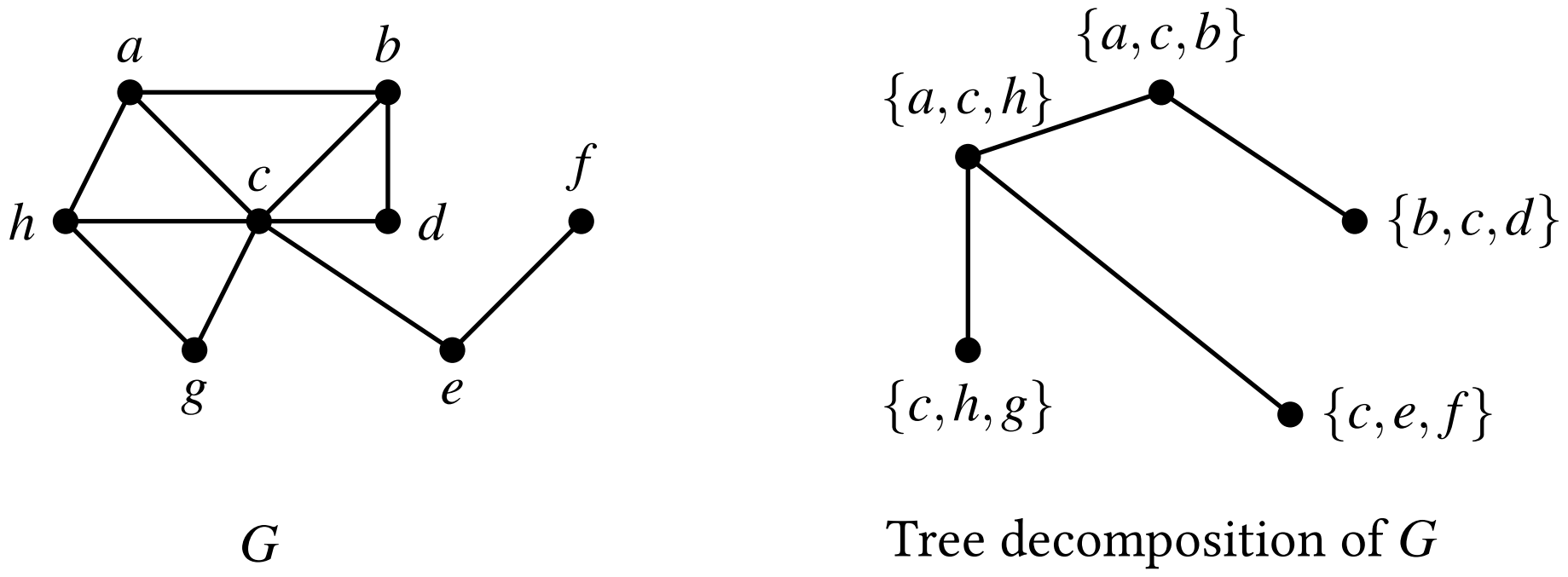}
	\caption{A tree decomposition of a triangulated graph.}
	\label{fig:tree_decomposition_of_random_graph}
\end{figure}

\begin{example}[Tree Decomposition of Some Triangulated Graphs]
	See figure \ref{fig:tree_decomposition_of_random_graph} for an example of a tree decomposition of a triangulated graph.
\end{example}

The following lemma will be useful.

\begin{lemma}
	Let $G$ be a graph and $T$ its tree decomposition with at least three vertices. Let $a \in V(T)$.
	If deleting $a$ from $T$ produces at least two disconnected subgraphs of $T$, which we label as $T_1, T_2$, then the subgraphs of $G$ induced by the vertex sets $\bigcup_{t\in V(T_1)} X_t \setminus X_a$ and $\bigcup_{t\in V(T_2)} X_t \setminus X_a$ are disconnected.
\end{lemma} \label{lem:tree_decomposition_disconnected_subgraph}

\begin{proof}
	It is sufficient to show that no vertex $v \in \bigcup_{t\in V(T_1)} X_t \setminus X_a$ is adjacent to any vertex $u \in \bigcup_{t\in V(T_2)} X_t \setminus X_a$.

	Assume that $v \in \bigcup_{t\in V(T_1)} X_t \setminus X_a$ and $u \in \bigcup_{t\in V(T_2)} X_t \setminus X_a$ and they are adjacent.
	Let $v \in X_s$, $u \in X_r$, and let $P_0$ be the unique path from $s$ to $r$ in $T$. Since $s$ and $r$ lie in different components of $T$ after deleting $a$, the path $P_0$ contains the vertex $a$.

	By property 2 of tree decomposition, there exists $t \in T$ such that $v, u \in X_t$.
	Since $T$ is connected, there exists a path $P_1$ from $s$ to $t$ and path $P_2$ from $r$ to $t$.
	If none of them pass through $a$, then there is a cycle in $T$ consisting of $P_0$, $P_1$, and $P_2$. This contradicts the fact that $T$ is a tree. Therefore, at least one of them must pass through $a$.

	Without loss of generality, let $P_1$ pass through $a$.
	$P_1$ is a path whose endpoints are bags that contain $v$, so by property 3 of tree decomposition the bag $X_a$ contains $v$ as well.
	This is a contradiction to the assumption that $v \notin X_a$.
\end{proof}

\begin{example}[Treewidth of a Complete Graph]
	The treewidth of a complete graph $K_n$ is $n-1$, as all tree decompositions of $K_n$ must contain a bag that contains all vertices of $K_n$.

	To prove this claim, assume $T$ is a tree decomposition of $K_n$ and let $X_i$ be the largest bag of $T$ that does not contain all vertices of $K_n$. 
	Since $X_i$ does not contain all vertices of $K_n$, there exists a vertex $v \in K_n$ such that $v \notin X_i$, which lies in some other bag $X_j$. Again, as $v$ is connected to all other vertices and $X_j$ does not contain all vertices, there must be a third bag in $T$. 

	In a tree with at least three vertices, there must exist a vertex $a$ such that deleting $a$ from $T$ produces at least two disconnected subgraphs of $T$.
	Now we can invoke lemma \ref{lem:tree_decomposition_disconnected_subgraph}, which states that the corresponding subgraphs of $K_n$ are disconnected. This is not possible, as $K_n$ is a complete graph, so all vertices are connected.

	Therefore, all tree decompositions of $K_n$ must contain a bag that contains all vertices of $K_n$, which proves the claim.
\end{example}

If a graph $G$ is a minor of $H$, then the treewidth of $G$ is less than or equal to the treewidth of $H$.
The definition of graph minor is presented here for reference.

\begin{definition}[Graph Minor]
	Let $G$ and $H$ be two graphs. We say that $G$ is a \define{minor} of $H$ if $G$ can be obtained from $H$ by a sequence of the following operations:
	\begin{enumerate}
		\item Deleting an edge,
		\item Deleting a vertex,
		\item Contracting an edge.
	\end{enumerate}
	The process of contracting an edge $e = (u, v)$ is to delete the edge $e$ and merge the two vertices $u$ and $v$ into a single vertex $w$. The new vertex $w$ is adjacent to all the neighbours of $u$ and $v$.
\end{definition}

\begin{lemma}[Graph Minor and Treewidth]\label{lem:minor_treewidth}
	Let $G$ and $H$ be two graphs. If $G$ is a minor of $H$, then the treewidth of $G$ is less than or equal to the treewidth of $H$.
\end{lemma}

The proof of this lemma is presented in section \ref{sec:proofs}.

We can use lemma \ref{lem:minor_treewidth} to prove various lower bounds on the treewidth. 

\begin{corollary}\label{cor:treewidth_delta_lower_bound}
	Let $\delta$ be the minimum vertex degree of a simple graph $G$. The treewidth of $G$ is at least $\delta$.
\end{corollary}

\begin{proof}
  The cases $\delta = 0$ or $\delta = 1$ are trivial, because only one-vertex trivial graph has tree width $0$ and everything else has treewidth at least $1$.

  For graph with $\delta \geq 2$, we only need to realise that 
  \begin{enumerate}
    \item There are at least $\delta + 1$ vertices in $G$
    \item Contracting an edge connected two vertices $a, b$ each with degree equal or more than $2$ will create a new vertex with degree greater than or equal to $\max(a,b)$.
  \end{enumerate}
  Therefore, we can continuously contract edges in $G$ until we obtain a complete graph $K_{\sigma}$, where $\sigma \geq \delta + 1$.
  By lemma \ref{lem:minor_treewidth}, the treewidth of $G$ is at least the treewidth of $K_{\sigma + 1}$, which is greater than $\delta$.
\end{proof}

Using these lemmas, we can find the treewidth of more complicated graphs.

\begin{figure}[ht]
	\centering
	\input{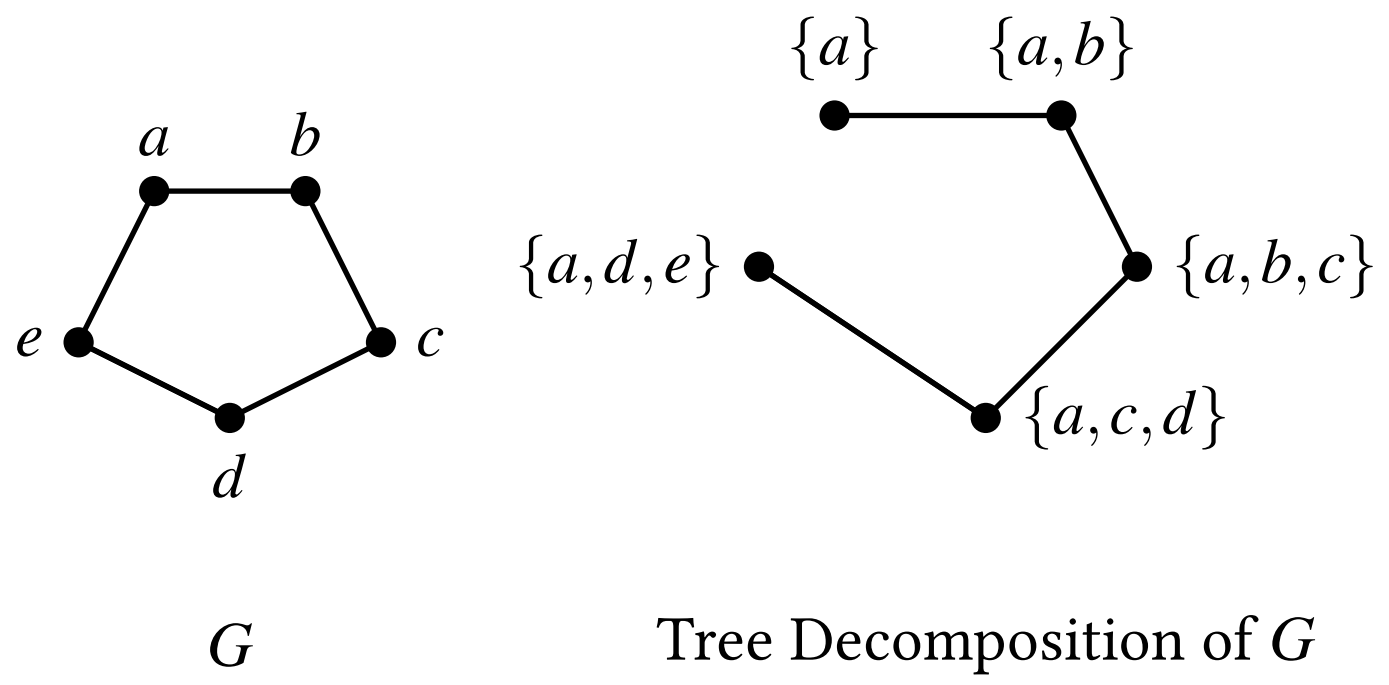}
	\caption{A tree decomposition of a cycle graph.}
	\label{fig:tree_decomposition_of_circle_graph}
\end{figure}

\begin{example}
	The treewidth of a cycle graph is $2$.
	By corollary \ref{cor:treewidth_delta_lower_bound}, the treewidth of a cycle graph is at least $2$ as the minimum vertex degree of a cycle graph is $2$.
	We can construct a tree decomposition of a cycle graph with width $2$ as follows.

	Let vertex $a$ be the start of the cycle. Delete one edge next to $a$ and treat the rest of the graph as a tree, with the maximum bag of size $2$.
	Adding $a$ to all bags of the tree decomposition gives a tree decomposition of the cycle graph with bags of maximum size $3$ and width $2$.

	See figure \ref{fig:tree_decomposition_of_circle_graph} for an example.
\end{example}

\section{Triangulated Graphs and Treewidth}

Treewidth is closely related to triangulated graphs, as we shall explain in this section.

\begin{definition}[Triangulated Graph]
	A graph $G$ is \define{triangulated} if every cycle of length greater than or equal to $4$ has a chord.
	A chord is an edge that connects two non-adjacent vertices in the cycle.
	Triangulated graphs are also called \define{chordal graphs}.
\end{definition}

The triangulation property is a subtle requirement for the connections between vertices.
In one sense , for triangulated graph two vertices are far away from each other if and only if there is no loop connecting them.

\begin{figure}[ht]
	\centering
	\input{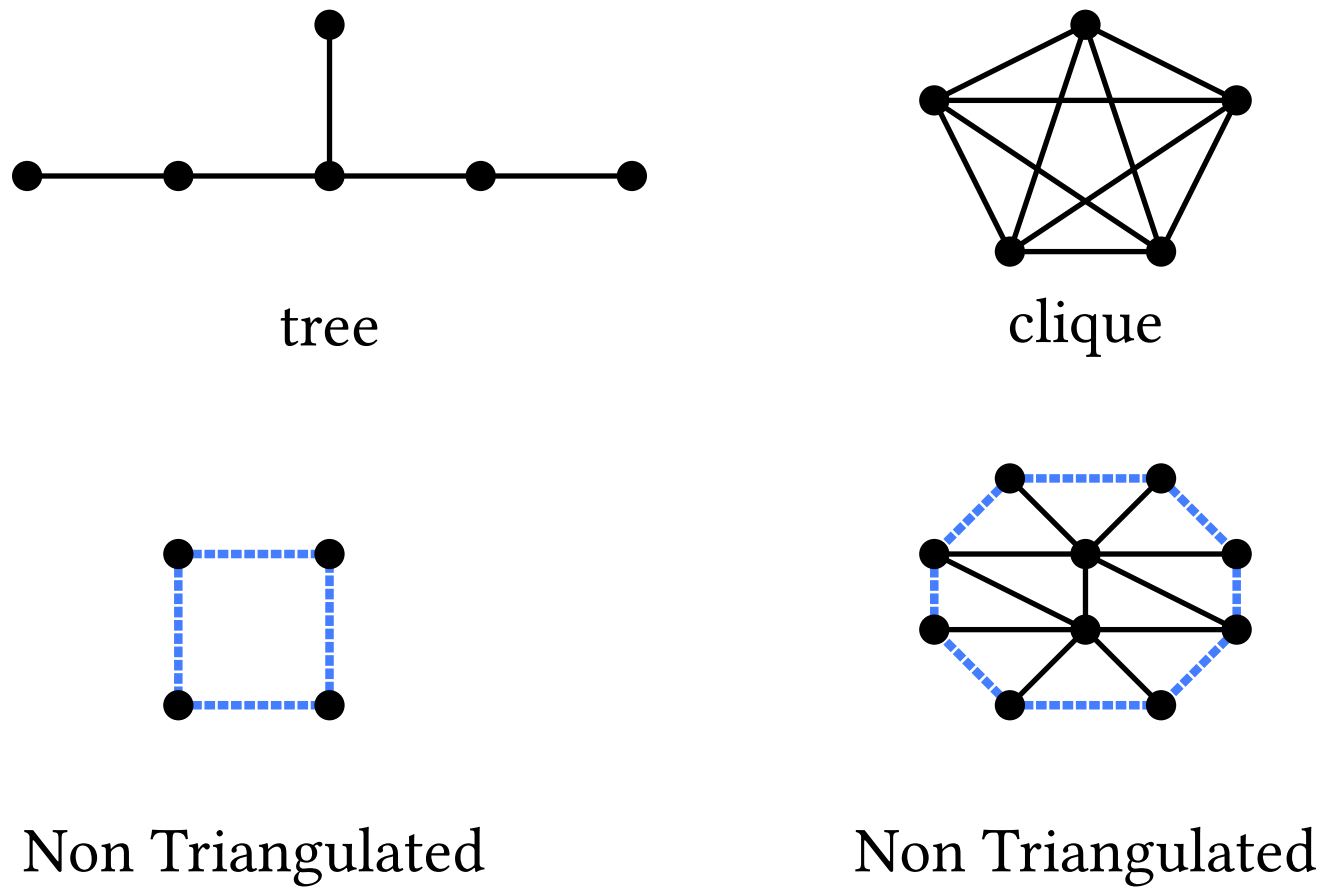}
	\caption{Examples of two triangulated graphs (tree and clique) and two non-triangulated graphs (cycle and union of triangles).
	The chordless cycles of non-triangulated graphs are dashed in blue.}
	\label{fig:triangulated_example}
\end{figure}

Figure \ref{fig:triangulated_example} shows examples of two triangulated graphs. All trees and complete graphs are triangulated. 
Cycles of length greater than three are not. 
Unions of triangles are not guaranteed to be triangulated.
See figure \ref{fig:triangulated_example} for examples of triangulated and non-triangulated graphs.

\begin{lemma}[Removing a vertex does not make triangulated graph non-triangulated]\label{lem:eliminate_vertex_triangulated}
	Let $G$ be a triangulated graph and $v$ any vertex in $G$. Then the graph $G \setminus \{v\}$ is triangulated.
	Note that $G \setminus \{v\}$ is the graph induced by the vertex set $V(G) \setminus \{v\}$.
\end{lemma}

\begin{proof}
	If $G\setminus \{v\}$ is not triangulated, then there exists a cycle of length greater than or equal to $4$ in $G\setminus \{v\}$ that does not have a chord.
	The same cycle exists in $G$ and does not have a chord, meaning $G$ is not triangulated either.
	This concludes the proof by contraposition.
\end{proof}

The following lemma follows by induction. 

\begin{lemma}[Induced Subgraph of Triangulated Graph is Triangulated]
	Let $G$ be a triangulated graph and $H$ an induced subgraph of $G$. Then $H$ is triangulated.
\end{lemma}

\begin{definition}[Simplicial Vertex]
	A vertex $v$ in a graph $G$ is \define{simplicial} if the subgraph induced by the neighbours of $v$ is a complete graph.
\end{definition}

\begin{lemma}[Existence of Simplicial Vertex in Triangulated Graphs]\label{lem:simplicial_vertex}
	Every triangulated graph has at least one simplicial vertex.
\end{lemma}

The proof of this lemma is presented in section \ref{sec:proofs}.

\begin{definition}[Perfect Elimination Ordering]
	A \define{perfect elimination ordering} of a graph $G$ is an ordering of the vertices $[v_1, v_2, \ldots, v_n]$ such that each vertex $v_i$ is simplicial in the subgraph induced by $[v_i, v_{i+1}, \ldots, v_n]$.
\end{definition}

\begin{figure}[ht]
	\centering
	\input{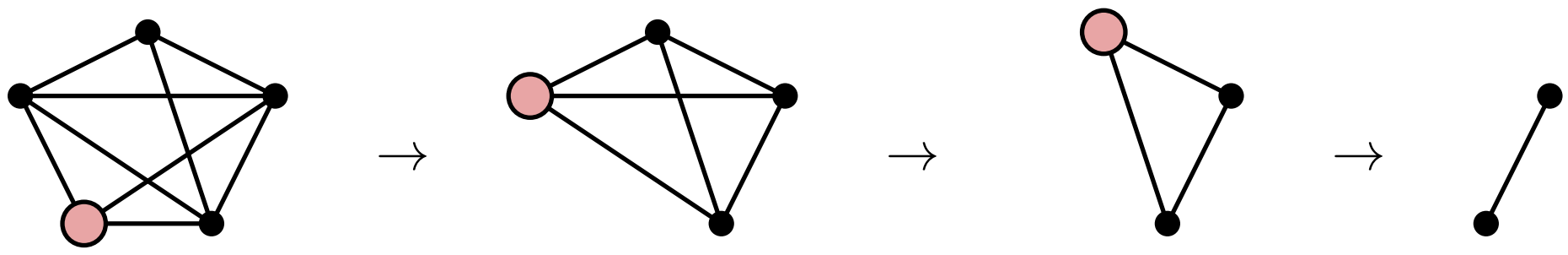}
	\caption{Perfect elimination ordering of a triangulated graph. The simplicial vertices at each step of elimination are coloured in red. }
	\label{fig:perfect_elimination_ordering}
\end{figure}

The reason it is called a perfect elimination is that $v_1$ is a simplicial vertex in $G$, $v_2$ is a simplicial vertex in $G \setminus \{v_1\}$ (that is, the graph $G$ with vertex $v_1$ removed), and so on.
An example of a perfect elimination ordering of a triangulated graph is shown in figure \ref{fig:perfect_elimination_ordering}.

\begin{theorem}[Triangulated Graph and Perfect Elimination Ordering]\label{thm:triangulated_perfect_elimination_ordering}
	A graph $G$ is triangulated if and only if it has a perfect elimination ordering.
\end{theorem}

\begin{proof}
	Proof of the sufficient condition. Assume $G$ has a perfect elimination ordering $[v_1, v_2, \ldots, v_n]$. Let $P$ be a cycle of length greater than 3.
	Among the vertices in $P$, one of them is ordered first in the perfect elimination ordering, say $v_i$.
	$v_i$ connects to at least two other vertices in $P$, which we label as $a, b$. 
	Since $v_i$ is simplicial in the subgraph induced by $[v_i, a, b]$, there is an edge connecting $a, b$, which is our desired chord.

	The necessary condition can be proved by induction on the number of vertices in $G$ using lemma \ref{lem:simplicial_vertex}.
	
	The base case is when $G$ has only one vertex, which is trivially simplicial.

	Let $G$ be a triangulated graph with $n$ vertices. By lemma \ref{lem:simplicial_vertex}, there exists a simplicial vertex $v$ in $G$.
	Elimination of $v$ gives $G \setminus \{v\}$ with $n-1$ vertices. 
	$G\setminus \{v\}$ is triangulated by lemma \ref{lem:eliminate_vertex_triangulated}. So the proof is done by induction.
\end{proof}

\subsection{From Tree Decomposition to Perfect Elimination Order}

Recall that an induced subgraph of a graph $G$ is a graph formed from a subset of the vertices of $G$ and all of the edges connecting pairs of vertices in that subset.
Using the notion of induced subgraph, we have the following important theorem that connects treewidth and triangulated graphs.

\begin{theorem}[Treewidth and Triangulated Graph]\label{thm:treewidth_triangulated_graph}
	For every graph $G$, there exists a triangulated graph $T$ such that $G$ is the induced subgraph of $T$ and $\text{tw}(G) = \text{tw}(T)$.
\end{theorem}

\begin{proof}
	Let $G$ be a graph. It is sufficient to show that for every tree decomposition of $G$, which we label as $T$, there exists a triangulated graph $F$ with the same tree decomposition and such that $G$ is the induced subgraph of $F$.
	We shall give a constructive proof.

	We can construct the triangulated graph $F$ from $G$ in the following way.  Start by initialising $F$ with only the vertices of $G$ and no connections between them.
  Perform the following steps, where in each step we add more edges to $F$ and remove bags in $T$.
  The algorithm terminates when $T$ becomes an empty graph.
  
    \begin{enumerate}
    	\item Let $X_l$ be a leaf in $T$.\footnote{A leaf is a vertex in a tree with degree 1. The parent of the leaf is the only other vertex connected to it. All trees must contain at least two leaves.}
        Let $X_p$ be the parent of $X_l$ in $T$.
		   	We assume that $X_l \setminus X_p$ is a set of one element. If it is not, modify the tree decomposition $T$ in the process described below to make it so. 
		\item Let $v$ be the unique vertex in $X_l \setminus X_p$. We claim that no other bag in $T$ contains $v$. If there is another bag that contains $v$, the path from $X_l$ to that bag must pass through $X_p$, and $X_p$ must contain $v$ by property 3 of tree decomposition. This is a contradiction to the assumption that $v \notin X_p$.
		\item Add enough edges to the graph $F$ such that the subgraph induced by $X_l$ becomes a complete graph.
		\item Remove $X_l$ from $T$ and repeat the process until $T$ is empty. 
    \end{enumerate}

  This process must terminate because in each iteration one bag is eliminated from $T$ and $T$ is finite.
  In the $i$-th iteration, we work on the special vertex $v_i$ that is the only vertex contained in $X_l \setminus X_p$. 
  It is clear that $[v_1, v_2, \cdots]$ is a perfect elimination ordering of $F$; therefore $F$ is triangulated.
	The original tree decomposition $T$ is still a valid tree decomposition of $F$, because the structure and the bags of $T$ are not changed, and each edge added to $F$ is contained in some bag of $T$.

	If $X_l \setminus X_p$ is a set of more than one element, we perform the following modification to the tree decomposition $T$ to make it have one leaf bag that contains exactly one more vertex than its parent.
	\begin{enumerate}
		\item If $X_l \setminus X_p$ is empty (that is, $X_l$ does not contain any vertices beyond those in $X_p$), then we can remove $X_l$ from $T$ and the tree decomposition is still valid.
		\item If $X_l \setminus X_p$ has more than one element, we can split $X_l$ into multiple bags, each containing one more element of $X_l \setminus X_p$. Check figure \ref{fig:demo_leaf_tree_decomposition} for a demonstration of this process.

\begin{figure}[ht]
	\centering
	\input{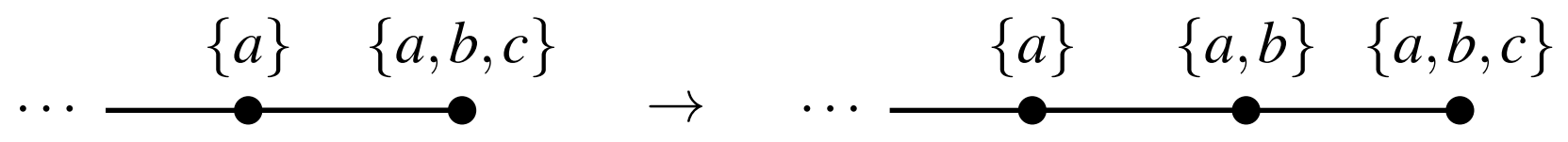}
	\caption{The process of making a leaf in a tree decomposition contain only one more vertex than its parent.}
	\label{fig:demo_leaf_tree_decomposition}
\end{figure}\end{enumerate}
\end{proof}

The triangulated graph $F$ constructed from $G$ in theorem \ref{thm:treewidth_triangulated_graph} is also called an optimal triangulation or an optimal chordal completion of $G$ in the literature \cite{rose1970triangulated}.

This constructive procedure is summarised in Algorithm \ref{alg:tree_decomposition_to_triangulated_graph}.

\begin{algorithm}[ht]
	\caption{Triangulation from a tree decomposition}\label{alg:tree_decomposition_to_triangulated_graph}
	\begin{algorithmic}[1]
		\Require A graph $G = (V, E)$ and a tree decomposition $(T, \{X_t\})$ of $G$.
		\Ensure Return a triangulated graph $F$ with $V(F) = V$ and $G \subseteq F$, and a perfect elimination ordering $\sigma$ of $F$, such that $(T, \{X_t\})$ is a tree decomposition of $F$.
		\State $F \gets (V, \varnothing)$; \quad initialise $\sigma$ as the empty list.
		\While{$T$ has at least two bags}
			\State Let $X_l$ be a leaf bag of $T$ and $X_p$ its parent.
			\If{$X_l \subseteq X_p$}
				\State Delete $X_l$ from $T$. \Comment{redundant leaf}
			\Else
				\State Write $X_l \setminus X_p = \{w_1, \dots, w_k\}$.
				\State Replace $X_l$ by the path $X_p - (X_p \cup \{w_1\}) - \cdots - (X_p \cup \{w_1, \dots, w_k\})$. \Comment{leaf now has one extra vertex}
				\For{$i \gets k, k-1, \dots, 1$}
					\State $B \gets X_p \cup \{w_1, \dots, w_i\}$.
					\State Make $B$ a clique in $F$ (add all edges among its vertices).
					\State Append $w_i$ to $\sigma$; delete $B$ from $T$.
				\EndFor
			\EndIf
		\EndWhile
		\State Let $X_r$ be the single remaining bag of $T$.
		\State Make $X_r$ a clique in $F$; append the vertices of $X_r$ to $\sigma$.
		\State \Return $(F, \sigma)$.
	\end{algorithmic}
\end{algorithm}

\begin{figure}[ht]
    \centering 
    \resizebox{1\textwidth}{!}{
        \input{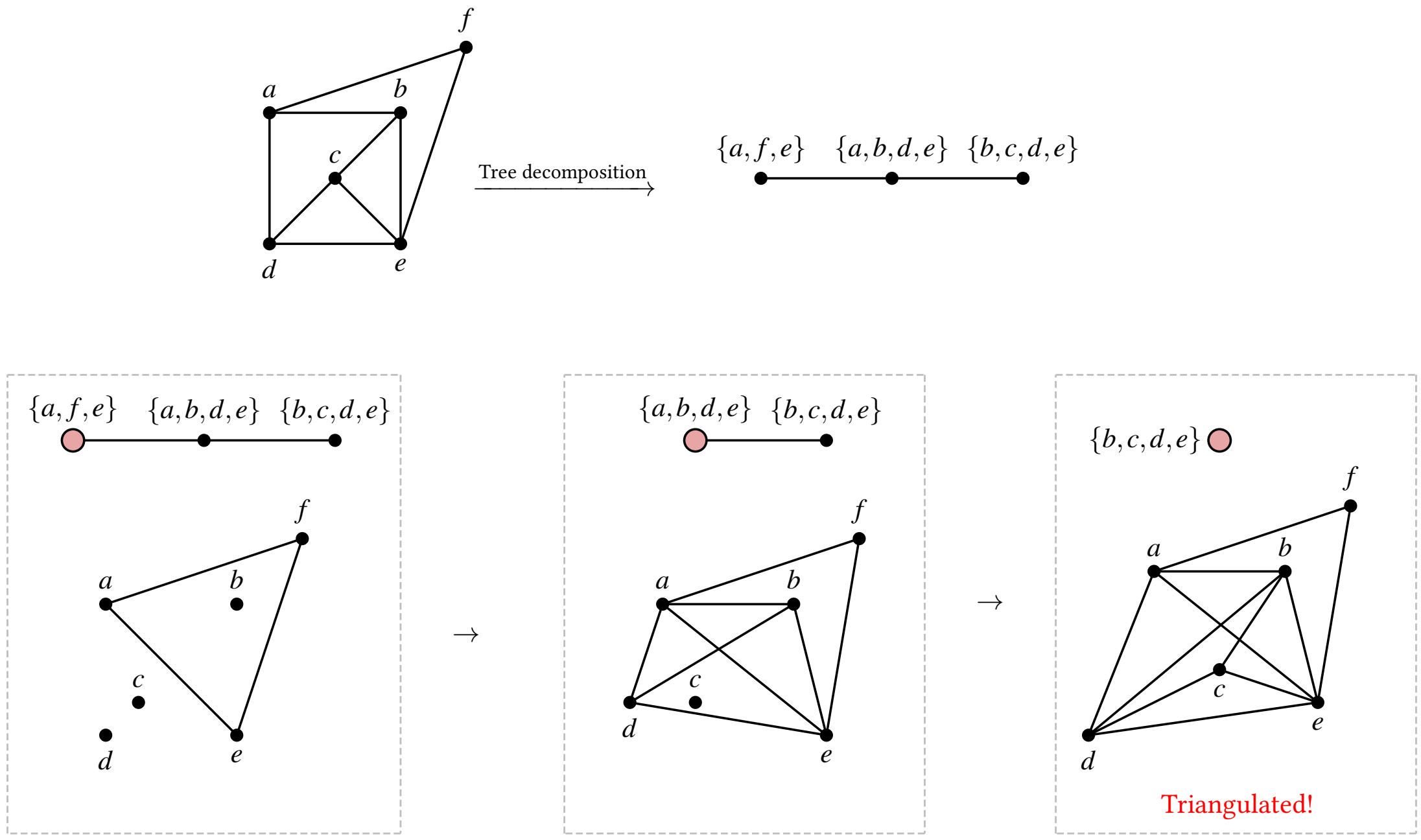}
    }
    \caption{Demonstration of constructing a triangulated graph with the same tree decomposition. The original graph is shown in the top left, and the tree decomposition is shown in the top right. 
    Each step of generating the triangulated graph is shown in the bottom with boxes. The corresponding bags in the tree decomposition are highlighted in red.}
    \label{fig:triangulated_from_tree_decomposition}
\end{figure}

A demonstration of this process is shown in figure \ref{fig:triangulated_from_tree_decomposition}.
Note that this process generates a perfect elimination ordering for the triangulated graph.

There is a different way to describe this process, using the term simplicially eliminated.

\begin{definition}[Simplicial Elimination]\label{def:simplicial_elimination}
    Let $G$ be a graph and $v$ any vertex, not necessarily simplicial, in $G$.
    The graph operation of simplicial elimination of $v$ from $G$ is to add edges between all the neighbours of $v$ in $G$ and then remove $v$ from $G$ to produce a new graph.
\end{definition}

\begin{corollary}[Simplicial Elimination Process to Generate Triangulated Graph]
    Let $G$ be a graph and $[v_1, v_2, \ldots, v_n]$ be an ordering of the vertices generated by the process described in theorem \ref{thm:treewidth_triangulated_graph}.
    Simplicially eliminate $v_1$ from $G$ to produce a new graph $H_1$.
    Inductively simplicially eliminate $v_i$ from $H_{i-1}$ to produce a new graph $H_i$ for $i = 2, \ldots, n$.

    Let $H$ be the graph on the vertices of $G$ obtained by adding to $G$ all the edges that are added during the process of eliminations above.
    The graph $H$ is triangulated and $G$ is an induced subgraph of $H$.
    (That is, the graph $H$ contains all the vertices of $G$ and all the edges of $G$.)
\end{corollary}\label{cor:simplicial_elimination_triangulated}

The order of vertices $[v_1, \cdots, v_n]$ in theorem \ref{thm:treewidth_triangulated_graph} is a simplicial elimination ordering of $G$. 
This ordering is important for our later discussion, and therefore deserves a formal definition.
\begin{definition}[Simplicial Elimination Ordering]\label{def:simplicial_elimination_ordering}
	\footnote{This is my own term.}
	Let $G$ be a graph not necessarily triangulated. Let $T$ be one of its tree decompositions and $[v_1, v_2, \ldots, v_n]$ be an ordering of the vertices generated by the process described in theorem \ref{thm:treewidth_triangulated_graph}.
	The ordering $[v_1, v_2, \ldots, v_n]$ is called a \define{simplicial elimination ordering} of $G$.
\end{definition}

\section{Treewidth is the Complexity of Tensor Contraction}

This section gives the major result of this chapter, theorem \ref{th:ultimate_complexity_tensor_contraction}.

\subsection{Tree decomposition Gives a Contraction Order}

It is possible to derive an upper bound for tensor contraction cost using treewidth.

\begin{theorem}[Treewidth Bound on Tensor Contraction Cost]\label{thm:treewidth_tensor_contraction} 
	Let $G$ be a tensor network with no loose bonds, regarded as a graph. 
	Let all the bonds of $G$ be of dimension $d$.
	It is possible to contract $G$ with a cost of $O(d^{\text{tw}(G^*) + 1})$, where $\text{tw}(G^*)$ is the treewidth of the line graph of $G$.
\end{theorem}

Markov and Shi first discovered this connection \cite{Markov_2008}.
The theorem is constructive, and there exists an algorithm to find such a contraction order, which we break down in the following steps.

\begin{definition}[Line Graph]
	Let $G$ be a graph. The \define{line graph} of $G$, denoted as $G^*$,\footnote{in the literature the line graph of $G$ is also written as $L(G)$.} is a graph such that each vertex of $G^*$ represents an edge of $G$, and two vertices of $G^*$ are adjacent if and only if their corresponding edges share a common endpoint in $G$.
	For an edge $e \in G$, we label the corresponding vertex in $G^*$ as $v_e$.
\end{definition}

\begin{figure}[ht]
	\centering
	\input{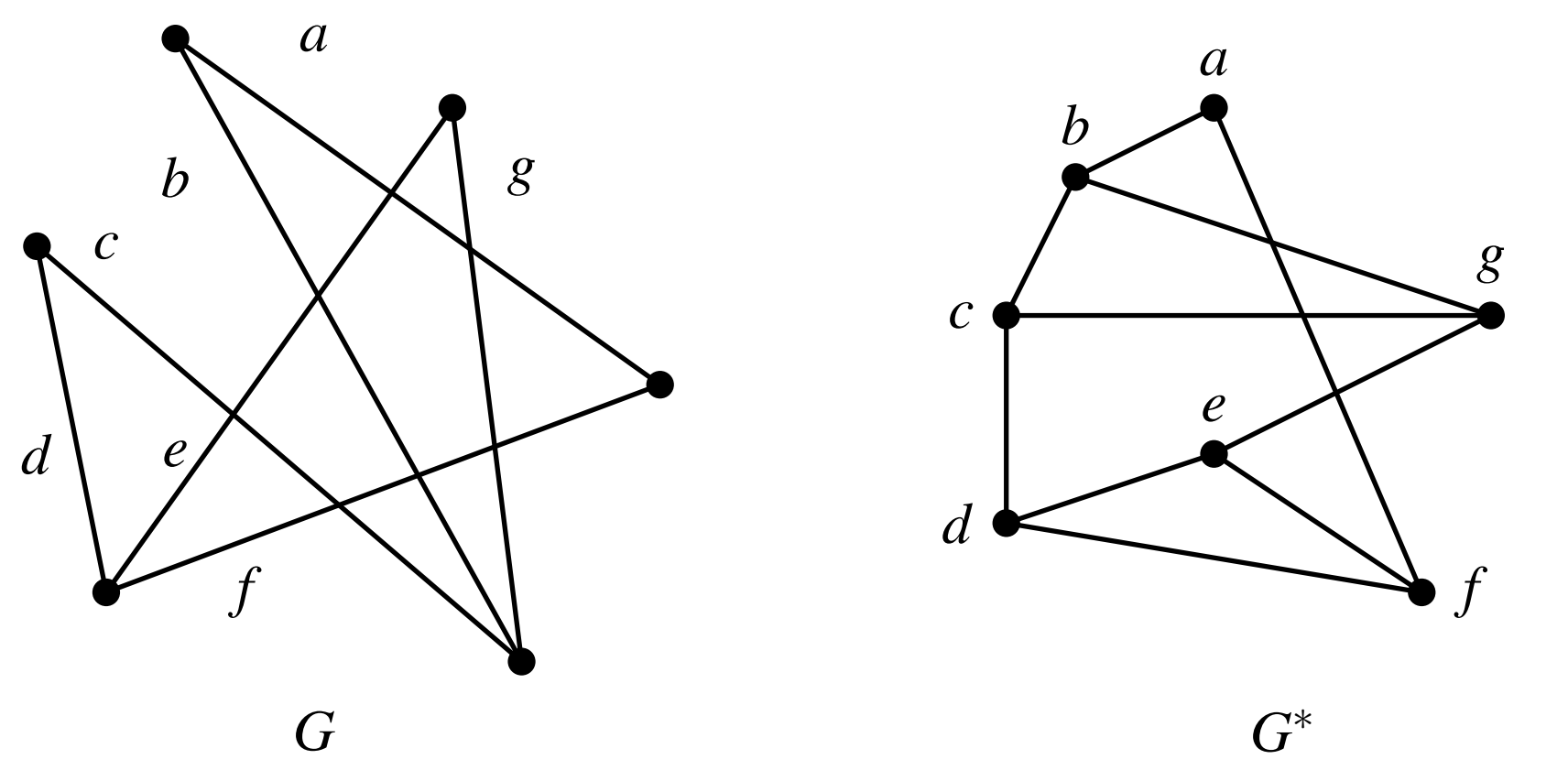}
	\caption{Graph $G$ and its line graph $G^*$. The edges of $G$ and vertices of $G^*$ are labelled with the same letter.}
	\label{fig:line_graph_1}
\end{figure}

\begin{example}
	See figure \ref{fig:line_graph_1} for an example of a graph with 6 vertices and 7 edges and its line graph with 7 vertices and 11 edges.
\end{example}

Let $G$ be a graph and $G^*$ its line graph.
An edge $e$ in $G$ has a corresponding vertex in $G^*$, which we denote as $v_e$.
The number of edge neighbours of $e$ in $G$ precisely equals the degree of vertex $v_e$ in $G^*$.
Therefore, using lemma \ref{lem:contraction-complexity}, we see that the cost of contracting a bond $e$ in $G$ is $O(d^{\text{deg}(v_e)+1})$.
\begin{lemma}[Complexity and Vertex Order]\label{lem:contraction-complexity-vertex-order}
	Contracting an edge $e$ in a graph $G$ has a cost of $O(d^{\text{deg}(v_e) + 1})$, where $v_e$ is the corresponding vertex of $e$ in the line graph $G^*$.
\end{lemma}

\begin{figure}[ht]
	\centering
	\resizebox{0.7\textwidth}{!}{
		\input{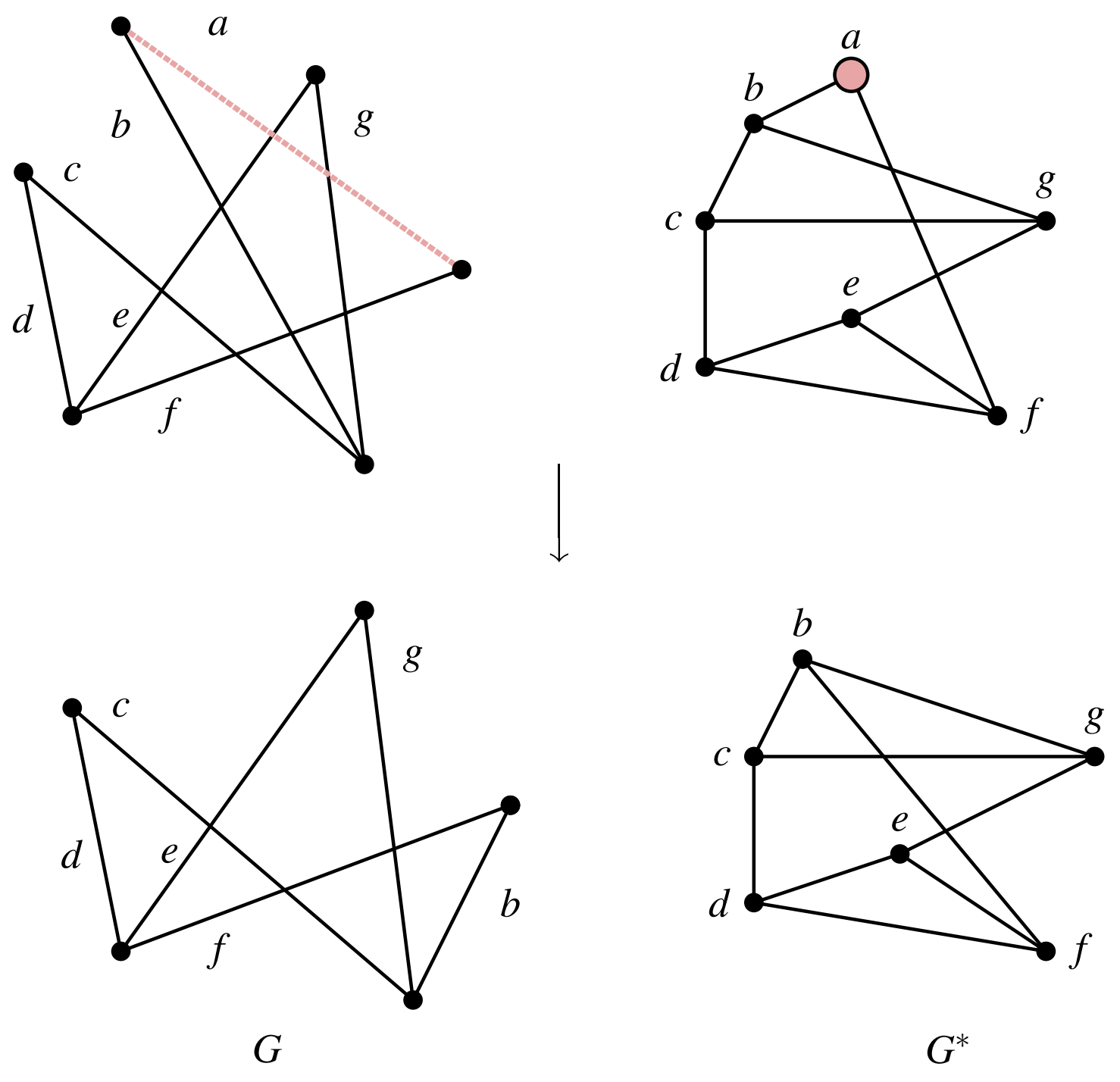}
	}
		\caption{Contracting an edge $a$ (in dotted red) in $G$
				makes all the neighbours of the corresponding vertex $v_a$ in $G^*$ neighbours of each other. This is exactly the simplicial elimination of $v_a$ in $G^*$.
				The cost of this contraction is $O(d^{3})$, as the degree of $v_a$ in $G^*$ is 2.
		}
	\label{fig:line_graph_simplicial_elimination}
\end{figure}

If an edge $e$ is contracted in $G$, all the neighbouring edges of $e$ become neighbours of each other if they are not already.
The corresponding changes in its line graph $G^*$ are exactly simplicial elimination (definition \ref{def:simplicial_elimination}) of the vertex $v_e$ in $G^*$.
This process is demonstrated in figure \ref{fig:line_graph_simplicial_elimination}.

\begin{lemma}[Contracting Edge is Simplicial Elimination]\label{lem:contracting-makes-neighbours}
	Contracting an edge $e$ in a graph $G$ is equivalent to simplicially eliminating the corresponding vertex $v_e$ in the line graph $G^*$.
\end{lemma}

\begin{figure}[ht]
	\centering
	\resizebox{0.9\textwidth}{!}{
		\input{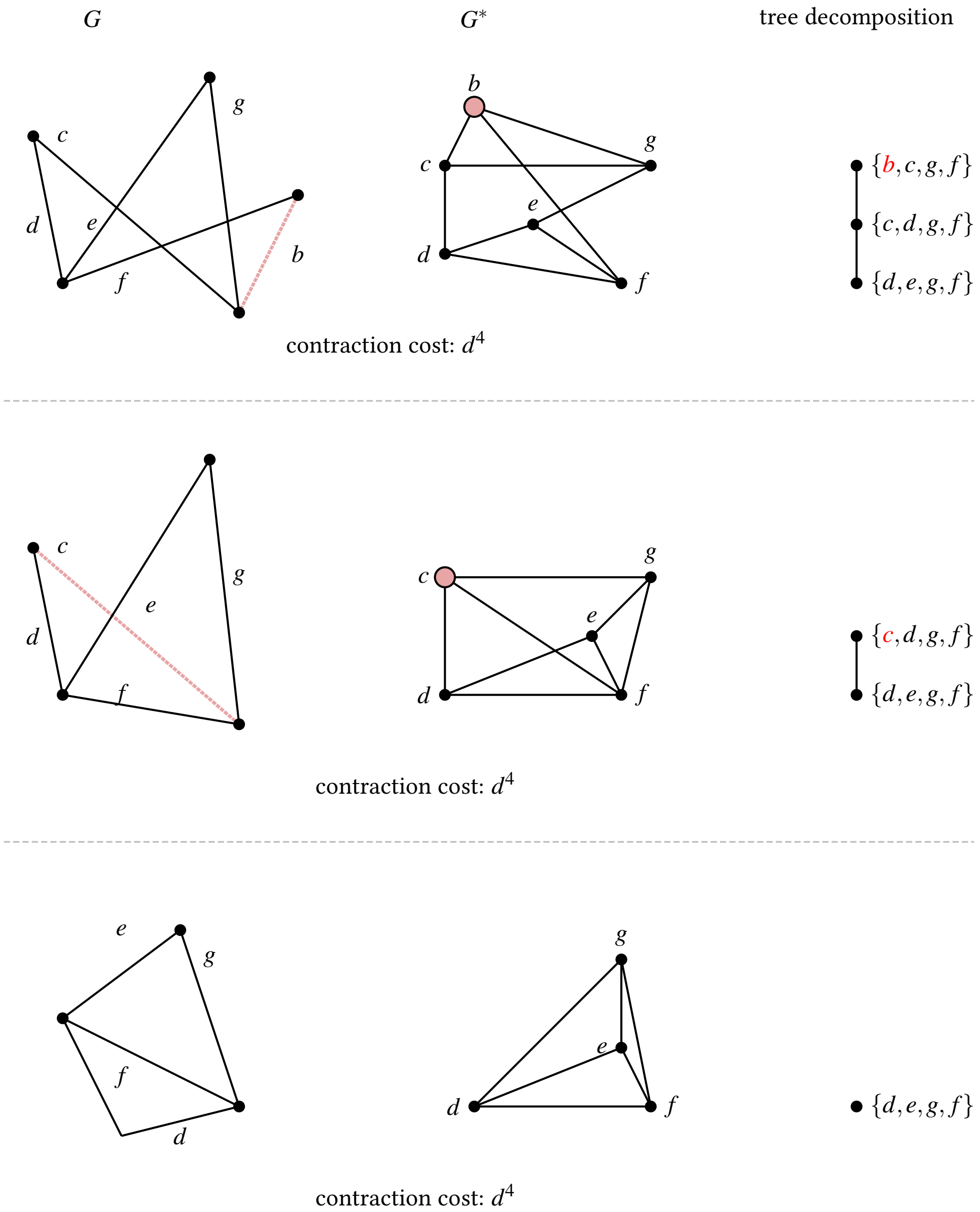}
	}
	\caption{Given the tree decomposition of the line graph of a tensor network, it is possible to find a contraction order with a cost of $O(d^{\text{tw} + 1})$, where tw is the treewidth of the line graph.
		This figure is a continuation of Fig \ref{fig:line_graph_simplicial_elimination}, and demonstrates how to find the total contraction order given the tree decomposition.
	The bond to be contracted, its corresponding vertex in the line graph, and the element in the bag of the tree decomposition are highlighted in red.}
	\label{fig:contraction_order_from_tree_decomposition}
\end{figure}

Therefore, given any tree decomposition of $G^*$, we can find a simplicial elimination ordering $[v_{\alpha}, v_{\beta}, \cdots ]$ guaranteed by theorem \ref{thm:treewidth_triangulated_graph}.
The corresponding ordering of edges in $G$, $[\alpha, \beta, \cdots]$, is our desired contraction order. 
Each contraction will take at most $O(d^{\text{tw}(G^*) + 1})$ time, and the total cost of contracting $G$ is bounded above by $O(n \cdot d^{\text{tw}(G^*) + 1})$, where $n$ is the total number of edges.
The process of finding the contraction order is demonstrated in figure \ref{fig:contraction_order_from_tree_decomposition}.

We shall summarise our important discovery in a theorem.

\begin{lemma}[Treewidth Gives a Contraction Order with Same Complexity]\label{lem:treewidth_contraction_order}
	Let $G$ be a tensor network with no loose bonds, regarded as a graph. 
	Let all the bonds of $G$ be of dimension $d$.
	Given a tree decomposition of $G^*$ (which is the line graph of $G$), it is possible to find a contraction order with a cost of $O(d^{\text{tw}(G^*) + 1})$.
\end{lemma}

\subsection{Tensor Contraction Gives a Tree Decomposition}

Theorem \ref{thm:treewidth_triangulated_graph} gives us important insights into how to generate a perfect elimination ordering $[v_1, v_2, \cdots, v_n]$ from a tree decomposition and how to use that to create a triangulated graph. 
Importantly, the treewidth of the triangulated graph by construction is the maximum number of neighbours any $v_i$ has in the subgraph induced by $[v_i, v_{i+1}, \cdots, v_n]$.

We can indeed go from the other direction as well.
The converse direction produces what is known as the \define{clique tree} (also called a join tree) of a triangulated graph in the literature, which is a tree decomposition in which every bag is a clique. 
See the survey of Blair and Peyton \cite{blair1993chordal}. 

\begin{theorem}[Tree decomposition from Perfect Elimination Ordering]\label{thm:tree_decomposition_from_perfect_elimination_ordering}
	Let $G$ be a triangulated graph and $[v_1, v_2, \cdots, v_n]$ be a perfect elimination ordering of $G$. 
	Define the simplex number
	\footnote{Simplex number is my own appellation. It equals $\omega(G) -1$, where $\omega(G)$ is the clique number of $G$.}
	of vertex $v_i$ as the number of neighbours $v_i$ has in the subgraph induced by $[v_i, v_{i+1}, \cdots, v_n]$.
	It is possible to construct a tree decomposition of $G$ with treewidth equal to the maximum simplex number. (By maximum simplex number, we mean the maximum of all simplex numbers of all vertices in $G$.)
\end{theorem}

\begin{proof}
	We will present a constructive proof using induction on the number of vertices in $G$.
	The idea of the proof is exactly the reverse of the proof of theorem \ref{thm:treewidth_triangulated_graph}.

	Our induction statement is a bit stronger. That is, for a perfect elimination order $[v_1, \cdots, v_n]$, there exists a tree decomposition of $G$ with treewidth equal to the maximum simplex number, and there is a bag containing exactly the vertex $v_1$ and all of its neighbours. This bag is also a leaf in the tree decomposition.

	The base case: when $G$ has only one vertex, the statement is trivially true.

	Let $T$ be our desired tree decomposition promised by the induction hypothesis on the subgraph induced by $[v_1, \cdots, v_n]$, and we want to show that the induction statement is also true when we add a new vertex $v_0$ to the front of the perfect elimination order, so that the new perfect elimination order is $[v_0, v_1, \cdots, v_n]$, and we create a new tree decomposition $T'$ of the new graph $G'$ induced by $[v_0, v_1, \cdots, v_n]$.

	Now, let $N = \{v_a, v_b, \cdots\}$ be the set of neighbours of $v_0$ in the subgraph induced by $[v_0, v_1, \cdots, v_n]$. 
	Without loss of any generality let $a < b < \cdots$. 
	Now, $v_a$ is simplicial too and it is ordered first among all the neighbours of $v_0$, so by induction hypothesis there is a bag (not necessarily a leaf) in $T$, which we label as $B$, such that $B$ contains the vertex $v_a$ and all of its neighbours, which includes all of the elements in $N$. 
	Make a new bag $\{v_0\} \bigcup N$ and attach it to $B$.
	This gives us the desired tree decomposition.
\end{proof}

\begin{figure}[ht]
	\centering
	\resizebox{0.9\textwidth}{!}{
		\input{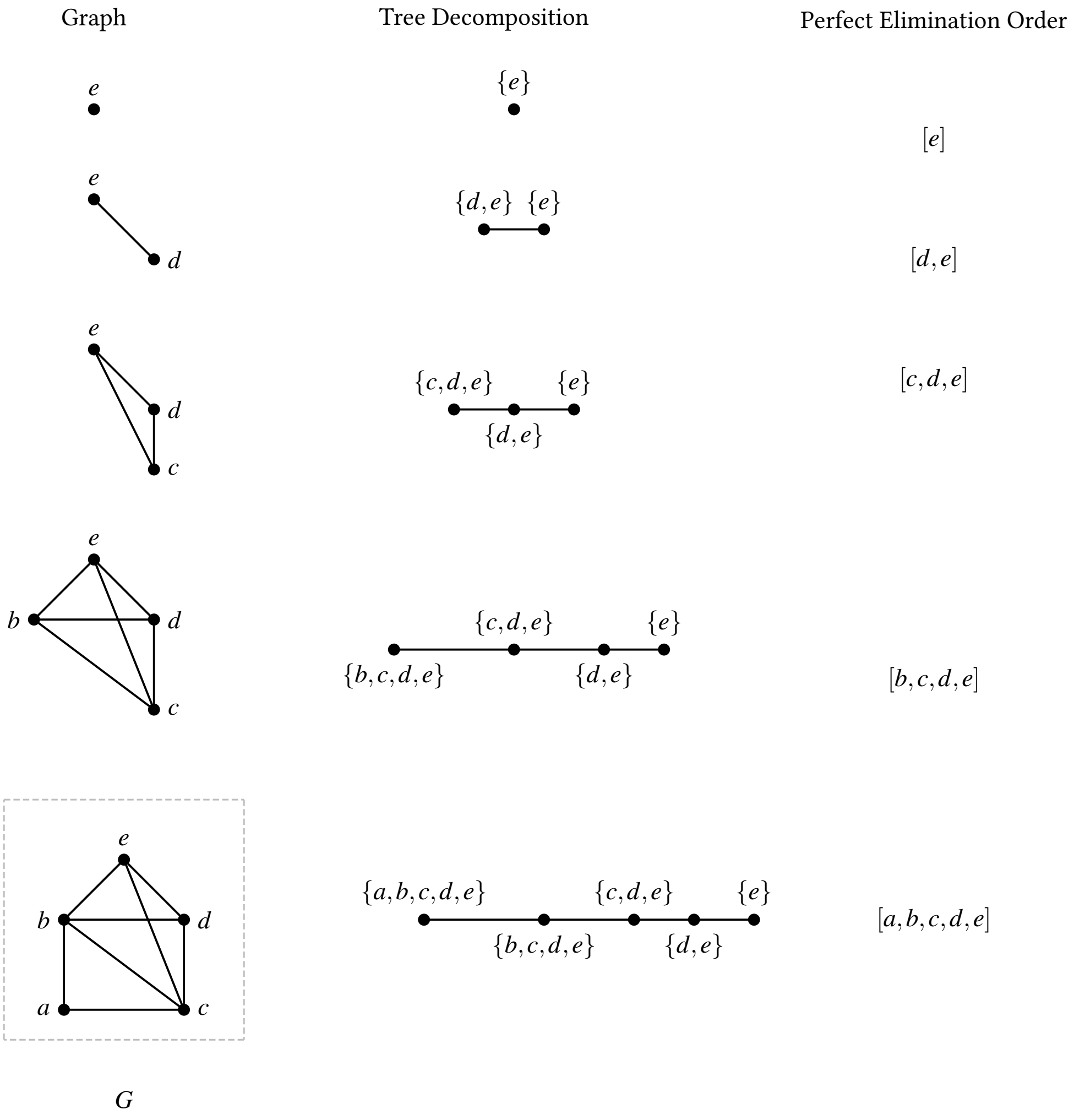}
	}
	\caption{Demonstration of constructing a tree decomposition from a perfect elimination ordering using algorithm \ref{alg:tree_decomposition_from_peo}. The perfect elimination ordering is $[a,b,c,d,e]$ on the graph $G$ shown on the lower left and boxed in dashed lines.} 
	\label{fig:tree_decomposition_from_perfect_elimination_ordering}
\end{figure}

The constructive steps of theorem \ref{thm:tree_decomposition_from_perfect_elimination_ordering} are summarised in Algorithm \ref{alg:tree_decomposition_from_peo}.
Figure \ref{fig:tree_decomposition_from_perfect_elimination_ordering} is a demonstration of algorithm \ref{alg:tree_decomposition_from_peo} to construct a tree decomposition from a perfect elimination ordering.

\begin{algorithm}[ht]
	\caption{Tree decomposition from a perfect elimination ordering}\label{alg:tree_decomposition_from_peo}
	\begin{algorithmic}[1]
		\Require A triangulated graph $G$ and a perfect elimination ordering $[v_1, v_2, \dots, v_n]$ of $G$.
		\Ensure The final output is a tree decomposition $(T, \{X_t\})$ of $G$ whose width equals the maximum simplex number of $G$.
		\State Initialise $T$ as a single node $t_1$ with bag $X_{t_1} \gets \{v_n\}$.
		\For{$i \gets n-1, n-2, \dots, 1$}
			\State $N \gets \{ v \in \{v_{i+1}, \dots, v_n\} \mid v \text{ is adjacent to } v_i \}$. \Comment{a clique, since $v_i$ is simplicial}
			\State Let $v_a$ be the vertex in $N$ with the smallest index.
			\State Let $B$ be a bag of $T$ that contains $\{v_a\} \cup N$. \Comment{exists, since $\{v_a\} \cup N$ is a clique}
			\State Add a new node $t$ with bag $X_t \gets \{v_i\} \cup N$, and attach $t$ to $B$.
		\EndFor
		\State \Return $(T, \{X_t\})$.
	\end{algorithmic}
\end{algorithm}

Using theorem \ref{thm:tree_decomposition_from_perfect_elimination_ordering} and theorem \ref{thm:treewidth_triangulated_graph}, we arrive at the following important theorem.

Theorem \ref{thm:tree_decomposition_from_perfect_elimination_ordering} ties well with lemma \ref{lem:contracting-makes-neighbours}, and together they give us the following lemma.

\begin{lemma}[Contraction Order Gives a Tree Decomposition]\label{lem:contraction_order_tree_decomposition}
  Let $G$ be a tensor network with no loose bonds, regarded as a graph.
	Given a contraction order of $G$, it is possible to find a tree decomposition of $G^*$ (which is the line graph of $G$), such that the complexity of the contraction order lower than or equal to $O(n \cdot d^{s})$, where $n$ is the number of edges in $G$ and $s$ is the size of the maximal bag in the tree decomposition.
\end{lemma}

A tensor with no loose bond denote as constant after full contraction. We use this just for convenince.

\begin{proof}
	We know by lemma \ref{lem:contracting-makes-neighbours} that contracting an edge is equivalent to simplicial elimination of the corresponding vertex in the line graph.
	Therefore, given a contraction order of $G$, $[e_1, e_2, e_3, \cdots]$, the corresponding vertex order $[v_{e_1}, v_{e_2}, v_{e_3}, \cdots]$ is a perfect elimination order of some triangulation of $G^*$.
	Now, Theorem \ref{thm:tree_decomposition_from_perfect_elimination_ordering} states that from this perfect elimination ordering we can find a tree decomposition of $G^*$ with the size of the maximal bag minus one equal to the maximum degree of any vertex in the simplicial elimination ordering.
	This lemma follows from lemma \ref{lem:contraction-complexity-vertex-order}.
\end{proof}

We now arrive at the following important theorem.

\begin{theorem}[Treewidth is the Complexity of Tensor Contraction]\label{th:ultimate_complexity_tensor_contraction}
	Let $G$ be a tensor network with no loose bonds and let $G^*$ be its line graph.
	Let all bonds be of dimension $d$ and let $G^*$ have $n$ vertices.
	The best possible complexity of contracting $G$, bond by bond, is $O(n \cdot d^{\text{tw}(G^*) + 1})$, where $\text{tw}(G^*)$ is the treewidth of $G^*$.
\end{theorem}

\begin{proof}
	The proof consists of two parts. First, there exists a contraction order with complexity $O(n \cdot d^{\text{tw}(G^*) + 1})$.
	Second, no contraction order can have complexity better than $O(n \cdot d^{\text{tw}(G^*) + 1})$.

	The existence of a contraction order with complexity $O(n \cdot d^{\text{tw}(G^*) + 1})$ is guaranteed by lemma \ref{lem:treewidth_contraction_order}.

	The other direction is guaranteed by lemma \ref{lem:contraction_order_tree_decomposition}, which states that given any contraction order, we can find a tree decomposition of $G^*$ with the size of maximal bag equal to the exponent of the complexity of the contraction order.
\end{proof}

Theorem \ref{th:ultimate_complexity_tensor_contraction} is an equivalent statement of proposition 4.2 in Markov and Shi's paper \cite{Markov_2008}.
Our proof, however, is completely original.

Theorem \ref{th:ultimate_complexity_tensor_contraction} basically states that finding the best contraction order of a tensor network is equivalent to finding the treewidth of its line graph. 
This is, of course, not surprising, as we know both problems are NP-hard \cite{treewidthNPComplete} \cite{NPcomplete_tensor_network_contraction}.

It is important to note that theorem \ref{th:ultimate_complexity_tensor_contraction} only governs the complexity of a general contraction algorithm where in each step one bond of the tensor network is contracted.
It is possible that a more efficient algorithm can exist that exploits the special structure of the tensor network and contracts multiple bonds at once.

An extreme example is that if we somehow know that a certain tensor is zero, then the whole tensor network must be zero no matter how complicated it is.

\begin{figure}[ht]
	\centering
	\input{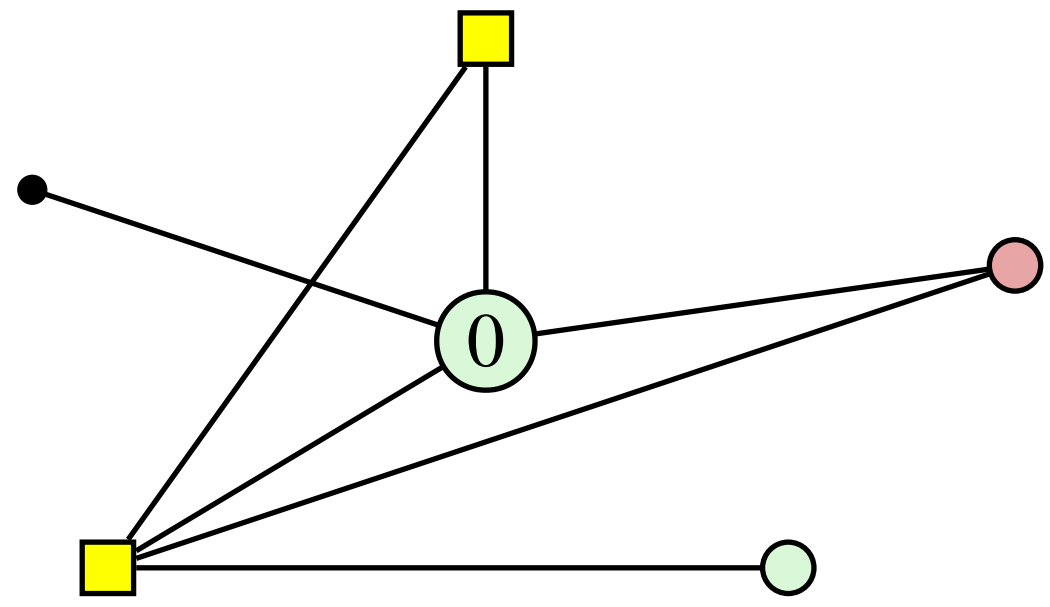}
	\caption{If the central tensor labelled with $0$ is always evaluated to 0, then the whole tensor network must be zero regardless of how complicated it is.}
	\label{fig:zero_tensor_network} 
\end{figure}

Our next chapter will also discuss a more efficient algorithm that exploits the special structure of the tensor network using ZX calculus.

\chapter{ZX Calculus} \label{ch:zx_calculus}

This chapter is a quick introduction to ZX calculus, a graphical calculus for quantum circuits.
Most parts of this chapter are based on the book \cite{KissingerWetering2024Book} and the lecture notes \cite{vandeweteringNotes}.

\section{Definition of ZX Calculus}

The theory of ZX calculus consists of two parts: the tensor networks, which are called ZX diagrams, and certain ways they can transform into each other, which are called the rewrite rules.
As ZX was invented to represent quantum circuits, all bonds in ZX diagrams are 2-dimensional.

ZX diagrams are generated by the following three classes of generators

\begin{definition}[Generators of ZX diagrams]\label{def:zx_generators}
ZX diagrams have two classes of generators, $x$ and $z$ spiders, and one special tensor called the Hadamard gate.
\begin{enumerate}
	\item A rank-$k$ $z$-spider with phase $\phi$, which we denote as $z_{\phi}$, is a symmetric tensor such that $z_{\phi}(\underbrace{0, \cdots, 0}_{k}) = 1$, $z_{\phi}(1, 1, \cdots, 1) = \exp(i \phi)$, and all other entries are $0$.
	\item The Hadamard gate, denoted as $H$, is a 2-legged symmetric tensor that equals 
		\begin{equation}
			\frac{1}{\sqrt{2}}\begin{pmatrix}
				1 & 1 \\ 1 & -1
			\end{pmatrix}
		\end{equation} \label{eq:hadamard}
		 As a matrix, it is unitary and self-adjoint.
	\item A rank-$k$ $x$-spider with phase $\phi$, denoted as $x_{\phi}$, is a symmetric tensor that is the conjugation of $z_{\phi}$ of the same rank by the Hadamard gate.
\end{enumerate}
\end{definition}

\begin{figure}[ht]
	\centering
	\input{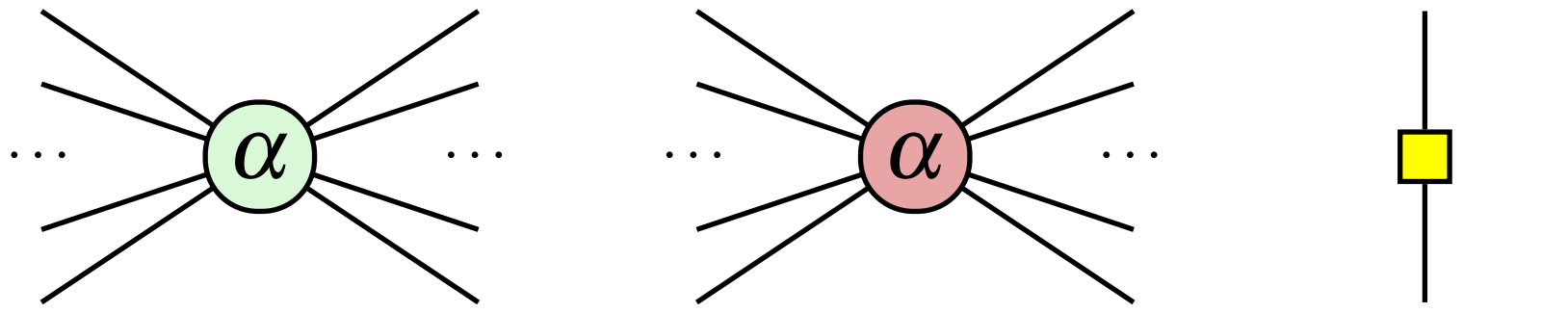}
	\caption{Generators of ZX diagrams.}
	\label{fig:zx_generators}
\end{figure}

These generators are shown in figure \ref{fig:zx_generators}. 
$z$-spiders are shown as green nodes, $x$-spiders red nodes, and their phases are shown as labels on the nodes. 
The Hadamard gate is shown as a yellow square.

Since ZX calculus was invented to represent quantum circuits, most authors treat ZX diagrams as linear maps with an input and an output instead of a generic tensor.
Following the conventional bra-ket notation where $\ket{a}$ denotes a state (a vector, usually written as a column) and $\bra{a}$ denotes its dual (co-vector, also called an effect, written as a row), $z$, $x$ and $H$ nodes can be represented in the following ways
\begin{align*}
	z_{\phi} &= \ket{0}^{\otimes n}\bra{0}^{\otimes m} + e^{i \phi} \ket{1}^{\otimes n}\bra{1}^{\otimes m} \\
	H &=\frac{1}{\sqrt{2}}(\ket{0}\bra{0} + \ket{0}\bra{1} + \ket{1}\bra{0} - \ket{1}\bra{1}) \\
	x_{\phi} &=  \ket{+}^{\otimes n}\bra{+}^{\otimes m} + e^{i \phi} \ket{-}^{\otimes n}\bra{-}^{\otimes m}
\end{align*}
where $\ket{+}$ and $\ket{-}$ denote the orthonormal basis $\frac{1}{\sqrt{2}}(\ket{0} + \ket{1})$ and $\frac{1}{\sqrt{2}}(\ket{0} - \ket{1})$ respectively.
Note that $H\ket{+} = \ket{0}$, $H\ket{-} = \ket{1}$, and $HH = I$.

The second part of ZX calculus is its rewrite rules,
which are the statements that certain combinations of generators are equivalent to each other.
Two ZX diagrams are equivalent if they represent the same tensor or they represent the same tensor up to a global constant. (This is because in quantum computation a global constant does not change the result of a quantum measurement.)
Throughout this dissertation, a strict equality will be denoted by a $=$ sign, while an equality up to a global constant will be denoted by a $\propto$ sign.\footnote{In the literature of ZX calculus, the equal $=$ sign often denotes both strict equality and equality up to a global constant.}
There are infinitely many rewrite rules, but the following are the most crucial ones.
\begin{definition}[Rewrite rules of ZX diagrams]
	Eight of the most important rewrite rules are shown in figure \ref{fig:zx_rewrites} and explained as follows:
	\begin{enumerate}
	\item Spider fusion, denoted as $\color{red} \mathbf{f}$. 
		A ZX diagram consisting only of $z$-spiders is equivalent to a single $z$-spider with the sum of the phases of the original spiders as its phase.
	\item Color change, denoted as $\color{red} \mathbf{c}$.
		A $z$-spider with phase $\phi$ is equivalent to an $x$-spider with phase $\phi$ conjugated by Hadamard gates on all its legs.
	\item Hadamard cancellation, denoted as $\color{red} \mathbf{hh}$.
		Two Hadamard gates are equivalent to the identity.
	\item Identity removal, denoted as $\color{red} \mathbf{id}$.
		A single rank-2 $z$-spider with phase $0$ and one input and one output leg is equivalent to the identity (a single bond).
	\item $\pi$-copy, denoted as $\color{red} \mathbf{p}$. A $\pi$-phase $x$-spider can be copied through a $z$-spider, and vice versa.
	\item Copy rule, denoted as $\color{red} \mathbf{cp}$. A $0$-phase $x$-spider can be copied through a $z$-spider.
	\item Strong complementarity, denoted as $\color{red} \mathbf{sc}$. 
		Two sets of 0-phase $z$ and $x$ spiders that form a complete bipartite graph can be transformed into a simpler graph where the set of $z$ spiders is replaced with the same-ranked $x$-spider, and the set of $x$-spiders is replaced with the same-ranked $z$-spider.
		This is the most \emph{crucial} one for our later discussion. 
	\item Euler decomposition, denoted as $\color{red} \mathbf{e}$. 
		The Hadamard gate can be decomposed into a sequence of $z$ and $x$ spiders with specific phases following the Euler decomposition of single qubit unitaries.
		We only showed one of the possible decompositions in figure \ref{fig:zx_rewrites}, but all the possible decompositions are included in the rule set.
\end{enumerate}

\begin{figure}[ht]
	\centering
	\resizebox{\linewidth}{!}{
		\input{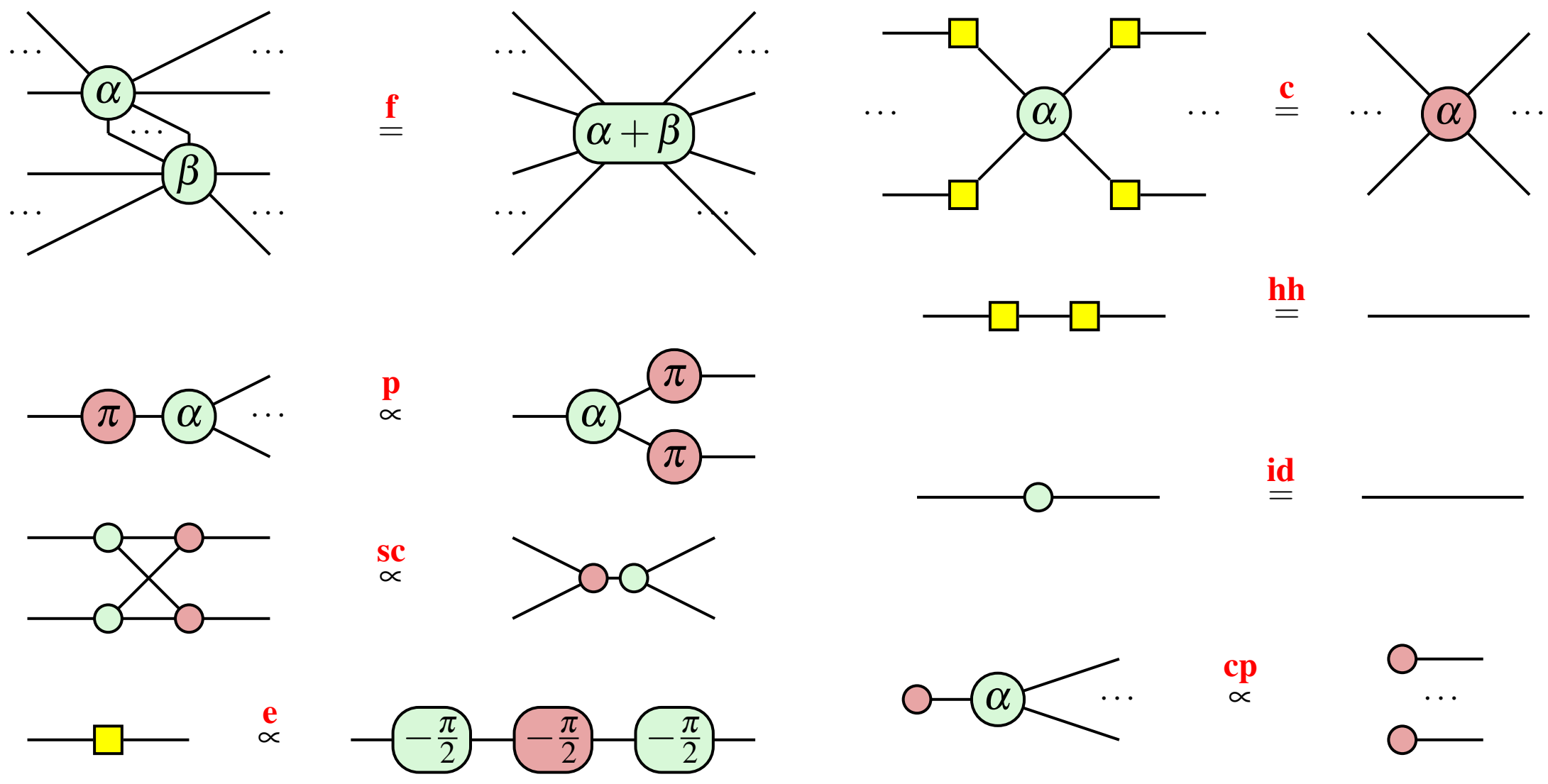}
	}
	\caption{Rewrite rules of ZX diagrams. }
	\label{fig:zx_rewrites}
\end{figure}

As this thesis is focused on tensor contraction, we need to keep track of the constant phase term for the rewriting rules. 
Fortunately, this is simple.

\begin{lemma}[Constant in Rewriting Rules]\label{lem:zx-constants}
	For the rule strong complementarity ($\color{red} \mathbf{sc}$), $\pi$-copy ($\color{red} \mathbf{p}$), and copy rule ($\color{red} \mathbf{cp}$), the left-hand side and right-hand side of the rewriting rules are equal up to a global constant.
	If there are $n$ $x$-spiders on the left-hand side of the rewriting rules, and $m$ $x$-spiders on the right-hand side, then there is a constant factor of $(\frac{1}{\sqrt{2}})^{n-m}$ on the right-hand side.
\end{lemma}

There is also a meta rule.
\begin{enumerate}
	\item Since a $z$-spider represents the same linear transformation as an $x$-spider but in a different basis, there exists a dual rewrite rule for each rule, where $z$ and $x$ spiders are swapped.
\end{enumerate}

As ZX calculus is indeed one particular kind of tensor network, all of the properties of tensor networks apply to ZX diagrams, in particular
\begin{enumerate}
	\item As all of the ZX generators are symmetric tensors, \emph{only connectivity matters} (lemma \ref{lem:only-connectivity-matters}) applies.
  \item All of the tensor operations defined in chapter \ref{ch:tensors} apply to ZX diagrams.
\end{enumerate}
\end{definition}\label{def:zx_rewrites}

The surprising property of ZX rewrite rules is that they are \emph{complete} in the following sense. 

\begin{theorem}[Completeness of ZX rewrite rules]\label{thm:zx-completeness}
	For any two ZX diagrams $D_1$ and $D_2$, if they represent the same tensor, then there exists a sequence of rewrites from definition \ref{def:zx_rewrites} that transforms $D_1$ into $D_2$.
\end{theorem}

As the completeness of ZX rewrite rules is not relevant to our later discussion of the tensor contraction algorithm, the reader is encouraged to refer to the original paper for the proof \cite{completenessZXCalculus}.

\section{Universality and Quantum Circuits}

Let us demonstrate the power of ZX calculus in this section by proving the universality theorem of ZX diagrams.

The following quantum gates can be represented as ZX diagrams

\begin{figure}[ht]
	\centering
	\input{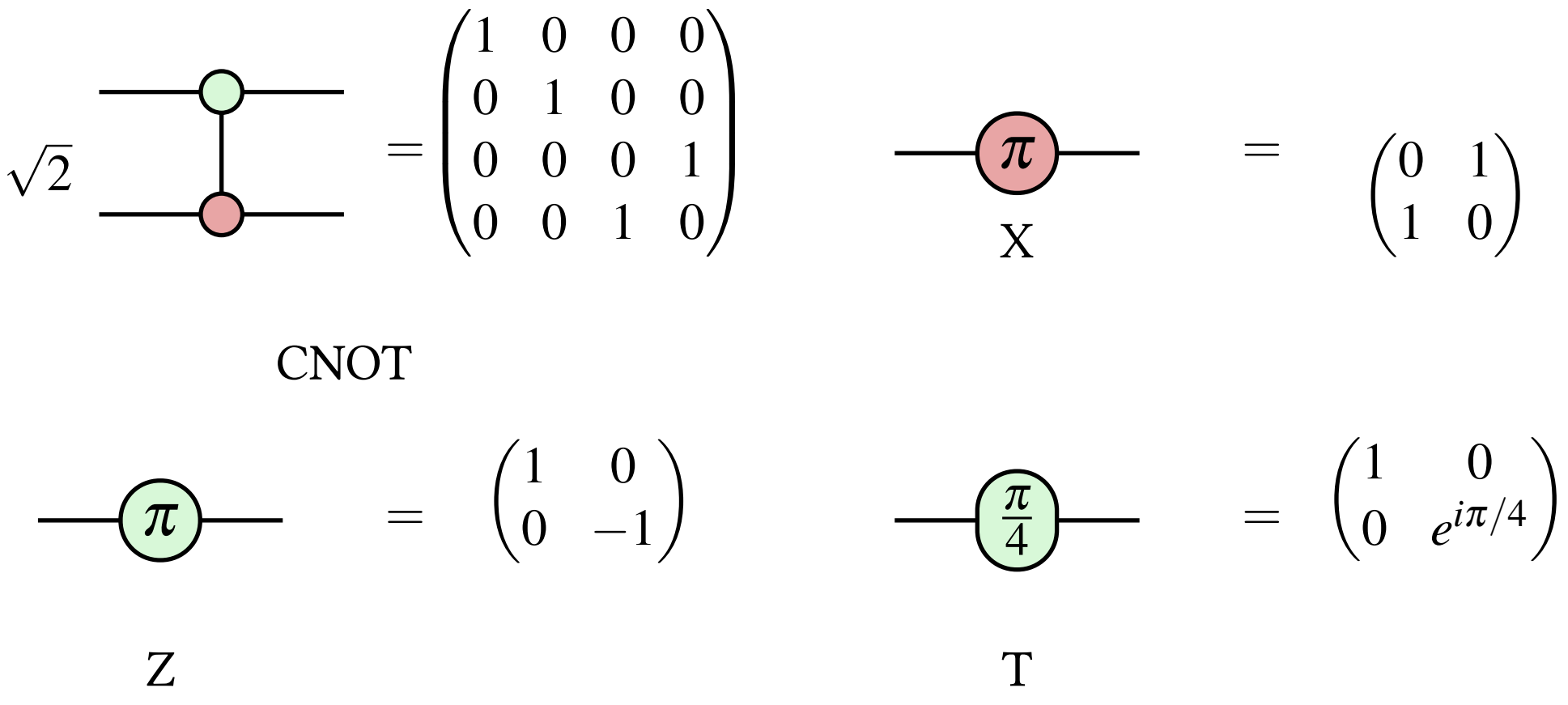}
	\caption{Common quantum gates represented as ZX diagrams.}
	\label{fig:zx_quantum_gates}
\end{figure}

In general, $z$-spiders and $x$-spiders are the rotation gates around the $z$ and $x$ axes of the Bloch sphere, respectively.
Therefore, by Euler decomposition, any single qubit unitary can be represented as a ZX diagram.
Since all unitary gates can be decomposed into single qubit unitaries and CNOT gates\footnote{See Chapter 4.2 of \cite{Nielsen_Chuang_2010}.}, we have arrived at the following important lemma.

\begin{lemma}[ZX diagrams are universal for quantum circuits]
	ZX diagrams can represent any unitary transformation up to a global phase, and therefore can represent any quantum circuit.
\end{lemma}

\begin{figure}[ht]
	\centering
	\input{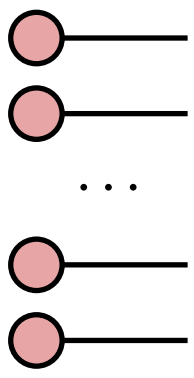}
	\caption{The state $\ket{0 \cdots 0}$ represented as a ZX diagram.} 
	\label{fig:zero_state}
\end{figure}

Now, since ZX diagrams can represent the state $\ket{0 \cdots 0}$ as shown in figure \ref{fig:zero_state}, and for any state in $\C^{2^n}$, there exists a unitary transformation that transforms $\ket{0 \cdots 0}$ to that state, we have the following corollary.

\begin{corollary}
	ZX diagrams can represent any state in $\C^{2^n}$.
\end{corollary}

Using the Choi–Jamiołkowski isomorphism, which states that any linear transformation can be represented as a state in a larger Hilbert space, we have demonstrated the universality of ZX diagrams.

\begin{theorem}[Universality of ZX diagrams]\label{thm:zx-universality}
	Let $V = \C^{2^n}$ for some $n \in \N$. 
	A ZX diagram can represent any linear transformation $T: V \to V$ and any state $\ket{\psi} \in V$.\end{theorem}

\subsection{Random Circuit of Google Sycamore}

Let us demonstrate the universality of ZX calculus by expressing the random circuit used by Google's Sycamore quantum processor \cite{GoogleSycamore} as a ZX diagram.

Google's Sycamore quantum processor simulates a random quantum circuit that is generated in the following way: 
\begin{enumerate}
	\item Start with the state $\ket{0 \cdots 0}$.
	\item Apply a layer of single-qubit gates on each qubit selected from the set $\{X^{1/2}, Y^{1/2}, W^{1/2}\}$, where $W = \frac{1}{\sqrt{2}}(X + Y)$. 
	\item Choose two qubits at random and apply an iSwap gate on them. 
\end{enumerate}

\begin{figure}[ht]
	\centering
	\input{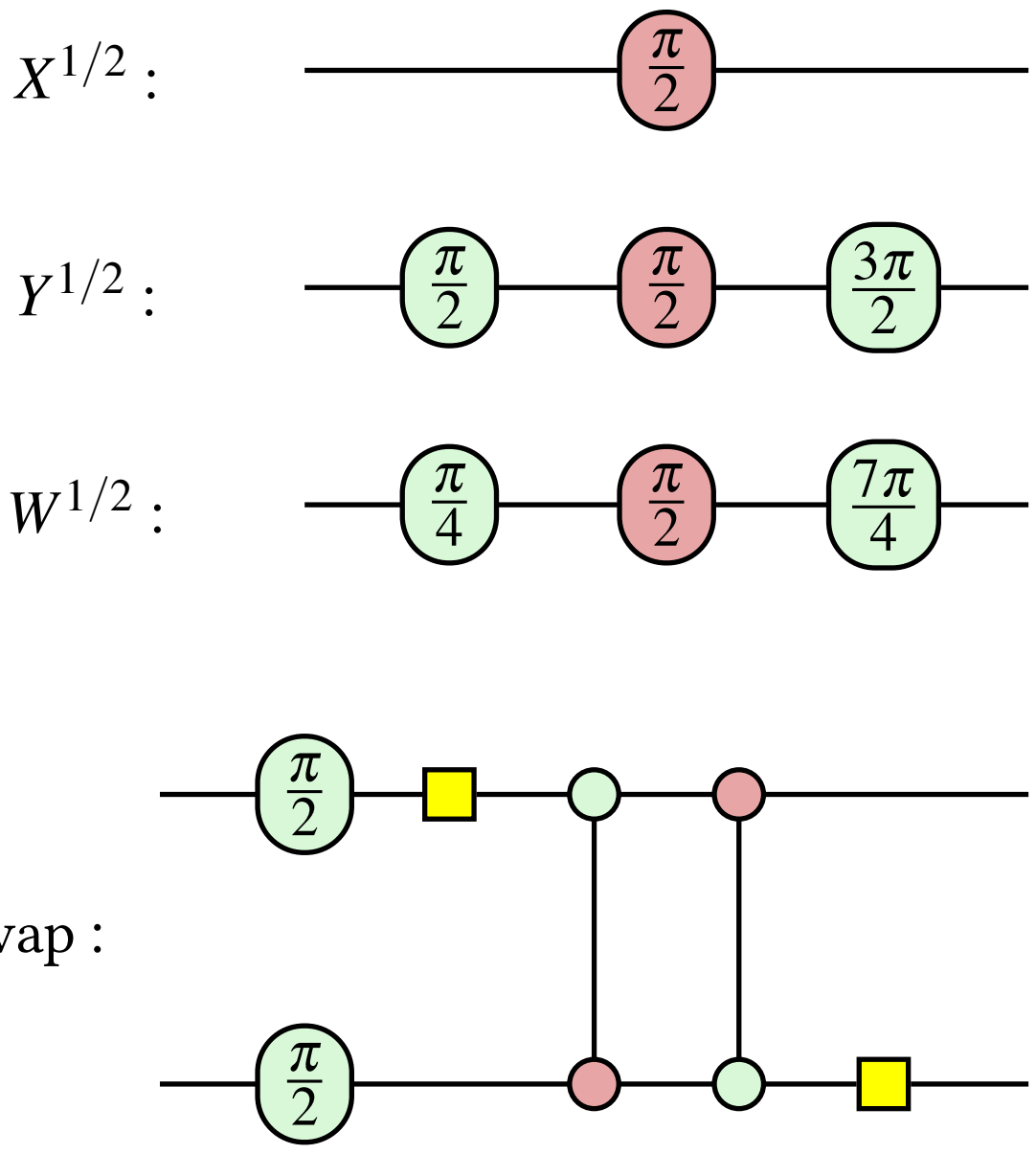}
	\caption{Generators of the Sycamore random circuit expressed as ZX diagrams.}
	\label{fig:sycamore_generators}
\end{figure}

The generators of the Sycamore random circuit can be expressed in ZX diagrams as shown in figure \ref{fig:sycamore_generators}, as can be verified by direct computations.

\section{Graph-like ZX diagrams}

Let us first introduce a new notation, dashed-blue edges, that denotes a Hadamard gate between two spiders.

\begin{figure}[ht]
	\centering
	\input{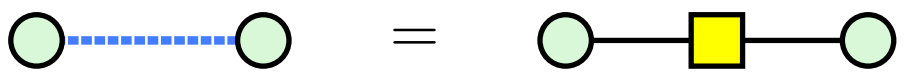}
	\caption{Dashed-blue edges denote a Hadamard gate between two spiders.}
	\label{fig:dashed_blue_edge}
\end{figure}

\begin{definition}[Dashed-blue edges]
	A dashed-blue edge between two spiders denotes a Hadamard gate between them, as shown in figure \ref{fig:dashed_blue_edge}.
\end{definition}

We now prove several identities using the ZX rewrite rules.

\begin{lemma}[Double Edges between $z$- and $x$-spiders Cancel]\label{lem:double_edge_cancellation}
	As shown in figure \ref{fig:double_edge_cancellation}, two edges between a $z$-spider and an $x$-spider can be removed without changing the tensor represented by the ZX diagram.
	Note that this identity only asserts that the two diagrams are proportional.
	\begin{figure}
		\centering 
		\caption{Double edges between $z$- and $x$-spiders cancel.} 
   	    \input{figures/cancellation.tikz}
		\label{fig:double_edge_cancellation}
	\end{figure}
\end{lemma}

\begin{proof}
	The proof exploits strong complementarity and the copy rule, and is shown in figure \ref{eq:proof-cancellation}.

	\begin{figure}
		\centering
		\resizebox{\linewidth}{!}{
			\input{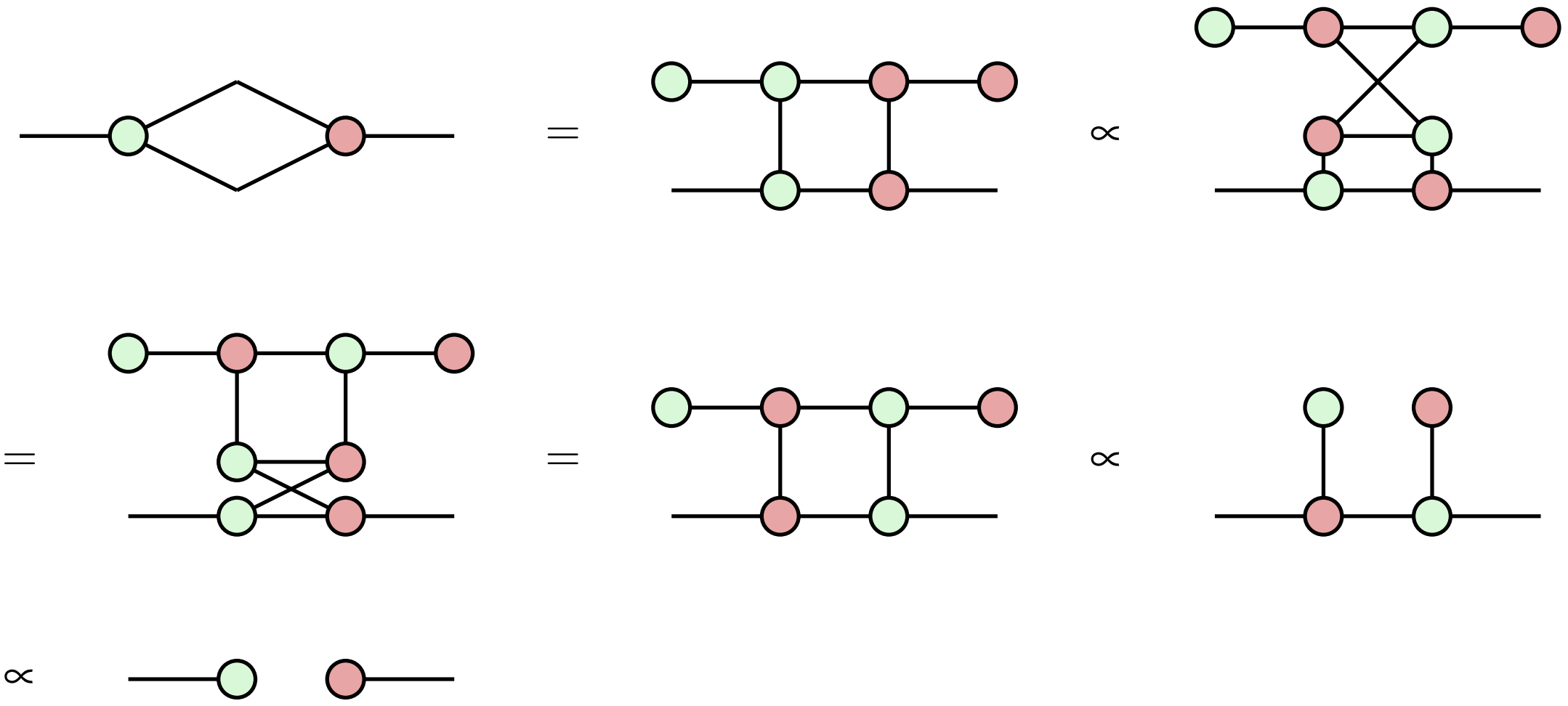}
		}
		\caption{Proof of the double edge cancellation identity.}
		\label{eq:proof-cancellation}
	\end{figure}
\end{proof}

\begin{figure}[ht]
	\centering
	\resizebox{0.8\linewidth}{!}{
		\input{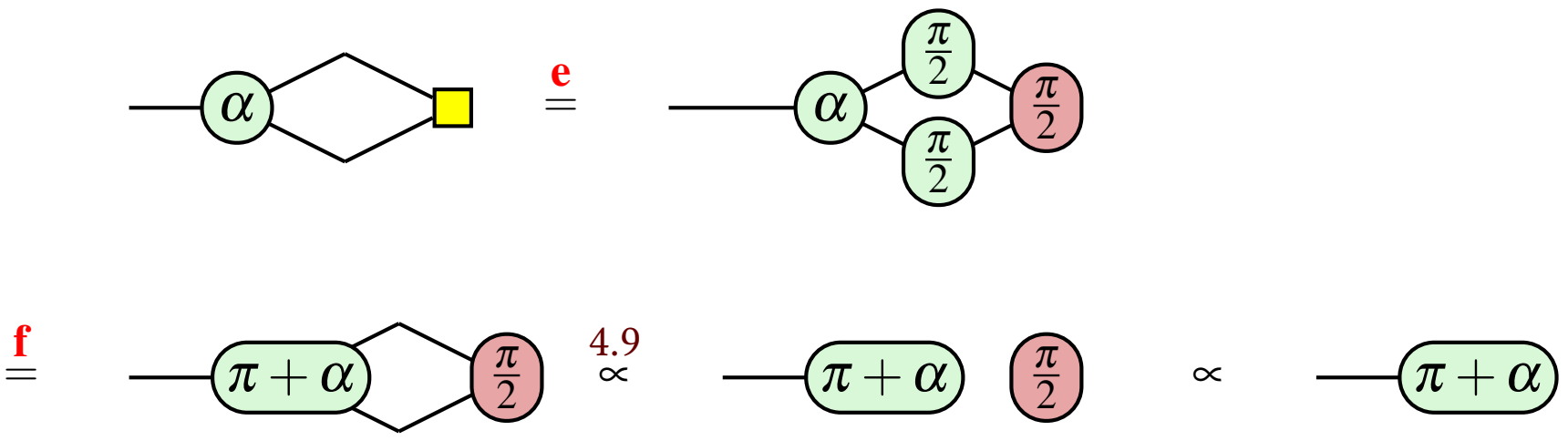}
	}
	\caption{Proof of the Hadamard self-loop cancellation identity.}
	\label{fig:hadamard_self_loop_cancellation}
\end{figure}

\begin{lemma}[Hadamard self-loop cancels]\label{lem:hadamard_self_loop_cancellation}
  A Hadamard self loop on a spider can be removed and only change the tensor represented by the ZX diagram up to a global constant.
	This identity is proved using Euler decomposition, spider fusion, and lemma \ref{lem:double_edge_cancellation}. The proof is shown in figure \ref{fig:hadamard_self_loop_cancellation}.
\end{lemma}

\begin{figure}[ht]
	\centering
	\resizebox{0.8\linewidth}{!}{
		\input{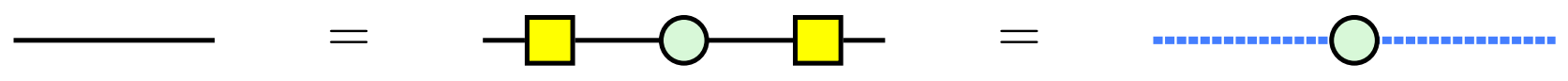}
	}
	\caption{Turning an edge into a Hadamard edge joined by a 0-phase $z$-spider.}
	\label{fig:turning_edge_to_hadamard_edge}
\end{figure}

\begin{lemma}[Turning an Edge into a Hadamard Edge]\label{lem:turning_edge_to_hadamard_edge}
	Any edge can be turned into a Hadamard edge joined by a 0-phase $z$-spider, as shown in figure \ref{fig:turning_edge_to_hadamard_edge}.
\end{lemma}

We are now ready to define a specific type of ZX diagram that will be important for our later discussion on the tensor contraction algorithm.

\begin{definition}[Graph-like ZX diagrams]
	A ZX diagram is called \define{graph-like} if it satisfies the following conditions:
	\begin{enumerate}
		\item All spiders are $z$-spiders.
		\item All edges between spiders are Hadamard edges (dashed-blue edges).
		\item There are no parallel edges or self-loops.
		\item Every input and output is connected to a spider, and every spider is connected to at most one input or output. \end{enumerate}
\end{definition}

It is evident that, using lemma \ref{lem:double_edge_cancellation} and lemma \ref{lem:hadamard_self_loop_cancellation}, any ZX diagram can be transformed into a graph-like ZX diagram.

\begin{theorem}[Graph-like form of ZX diagrams]\label{th:all_to_graph_like}
	Any ZX diagram, in particular any quantum circuit, can be transformed into a graph-like ZX diagram.
\end{theorem}

\begin{figure}[ht]
	\centering
	\resizebox{0.8\linewidth}{!}{
		\input{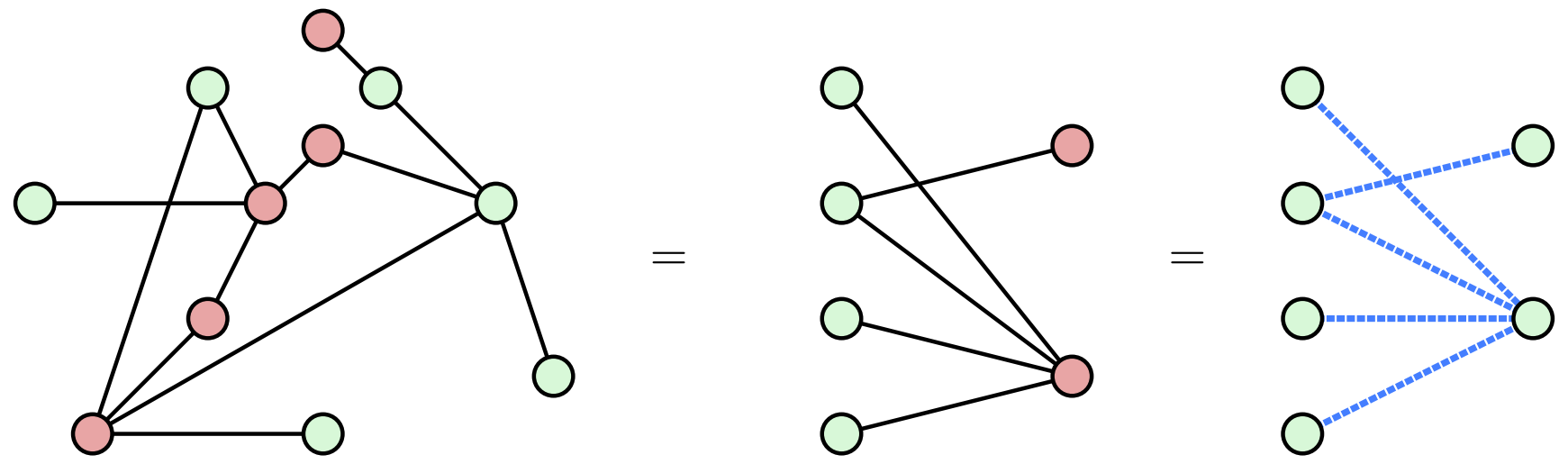}
	}
	\caption{Transformation of a ZX diagram into a graph-like ZX diagram.}
	\label{fig:graph_like_transformation}
\end{figure}

\begin{proof}
	The transformation procedure is thus: 
	\begin{enumerate}
		\item Fuse as many $z$- and $x$-spiders as possible so that there are no interconnections between $z$- and $x$-spiders. 
		\item Use color change to transform all $x$-spiders into $z$-spiders by adding Hadamard gates on all their legs. 
		\item Use lemma \ref{lem:double_edge_cancellation} and lemma \ref{lem:hadamard_self_loop_cancellation} to remove all parallel edges and self-loops.	
		\item If there are non-Hadamard edges, use lemma \ref{lem:turning_edge_to_hadamard_edge} to add 0-phase $z$-spiders to turn them into Hadamard edges.
	\end{enumerate}
	
	See figure \ref{fig:graph_like_transformation} for an example of the transformation.
\end{proof}

\begin{lemma}[Quantum Circuits in Graph-like form]
	Any quantum circuit can be transformed into a graph-like ZX diagram.
\end{lemma}

\section{Local Complementation}

This section explains the local complementation identity, which is crucial for our later discussion on the tensor contraction algorithm.

\begin{lemma}[$K_3$ graph identity]
	As shown in figure \ref{fig:k3_local_complementation}, a complete graph $K_{3}$ with 
	$z$-spiders connected by Hadamard gates can be restructured so that the $K_{3}$ loop is opened.
	\begin{figure}
		\centering
		\resizebox{0.8\linewidth}{!}{
			\input{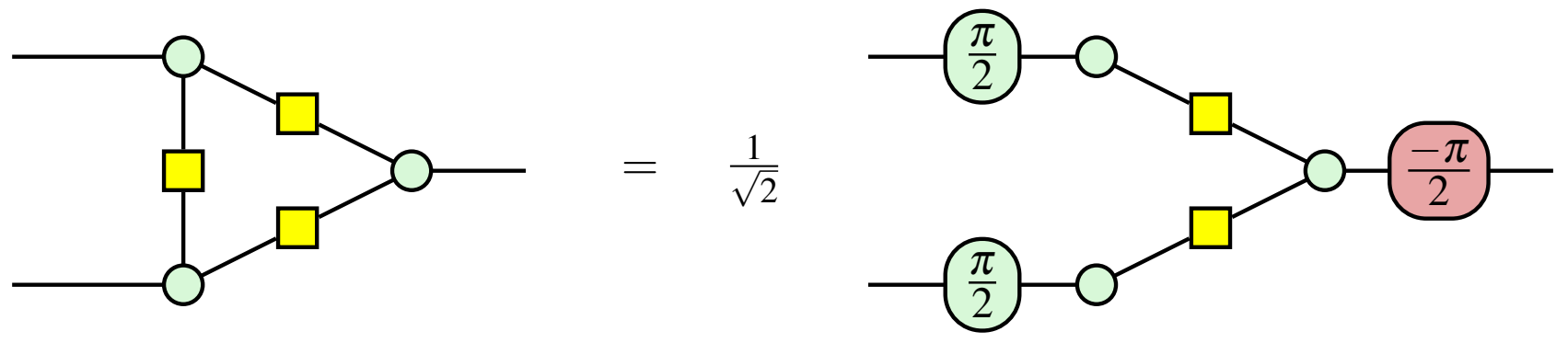}
		}
		\caption{The $K_3$ local complementation.}
		\label{fig:k3_local_complementation}
	\end{figure} 
\end{lemma}\label{lem:k3_local_complementation}

\begin{proof}
	Of course, we can prove it by brute force and evaluate the tensor that each side of the equation represents. Yet here we present a purely graphical proof.
	\begin{figure}
		\centering
		\resizebox{0.8\linewidth}{!}{
			\input{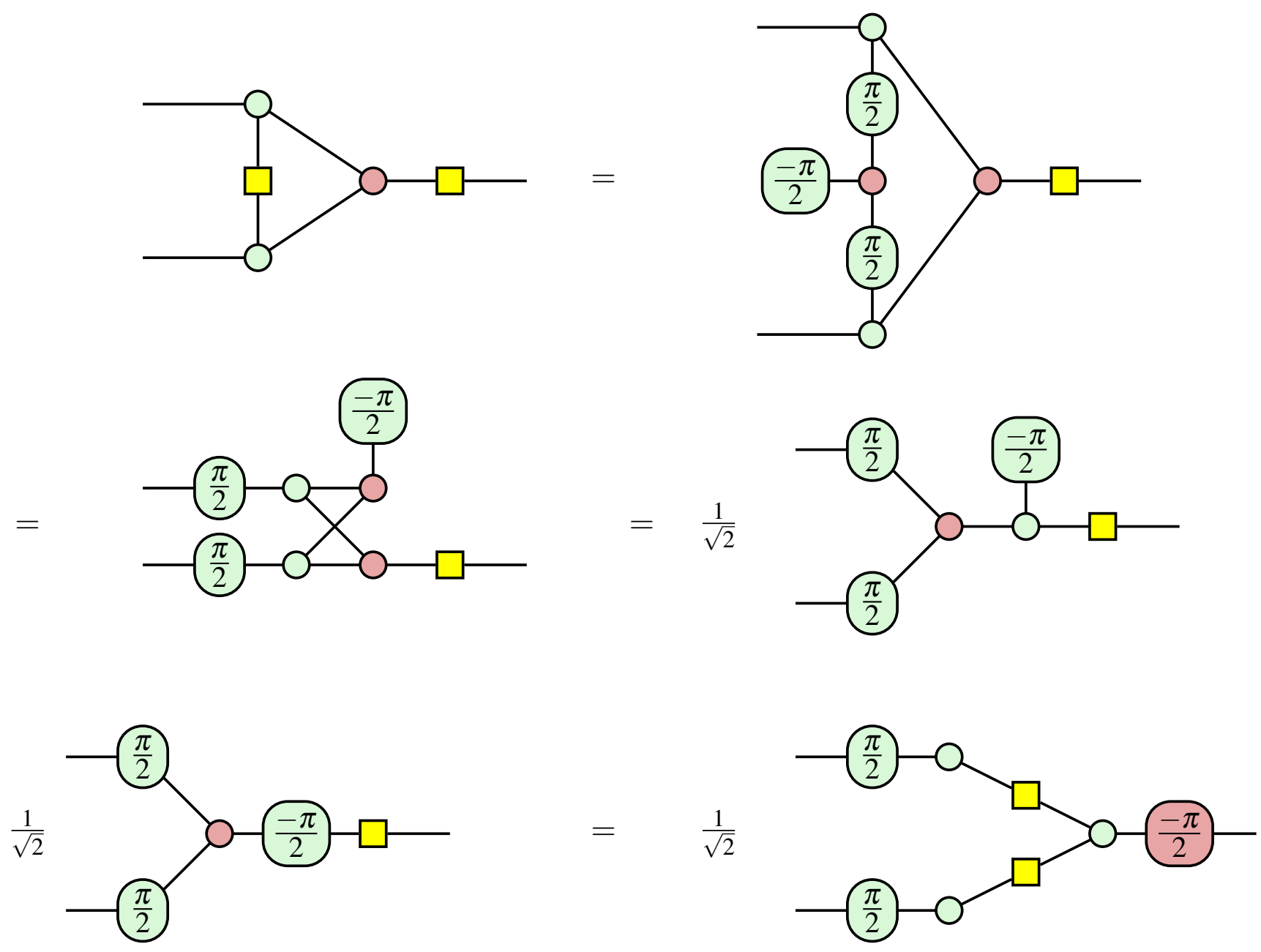}
		}
		\caption{Proof of the $K_3$ local complementation.}	
		\label{fig:k3_identity_proof}
	\end{figure}
\end{proof}

Using lemma \ref{lem:k3_local_complementation}, we can prove the following more general identity.

\begin{lemma}
	Let the box labelled with $K_m$ denote the ZX diagram of a complete graph with $m$ $z$-spiders connected by Hadamard gates. 
	Then, as shown in figure \ref{fig:k_m_local_complementation}, the $K_m$ graph can be restructured so that the $K_m$ clique is turned into a star graph.
\begin{figure}
	\centering
	\includegraphics[width=0.8\linewidth]{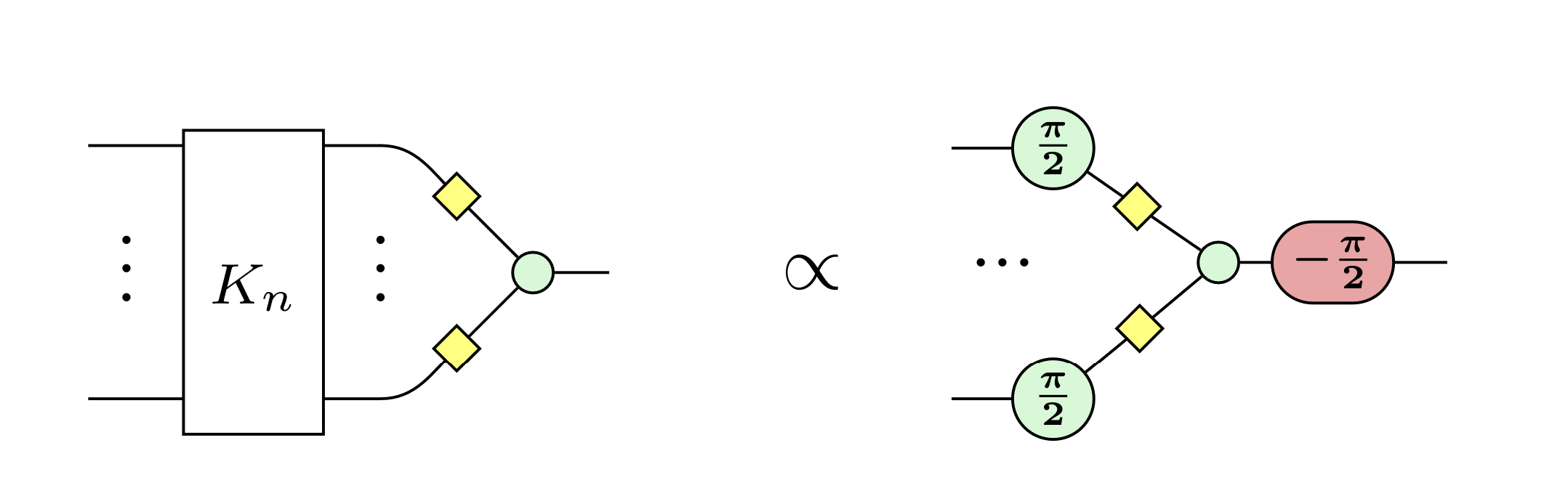}
	\caption{The $K_{m}$ local complementation.}
	\label{fig:k_m_local_complementation}
\end{figure}
\end{lemma}\label{lem:k_m_local_complementation}

The proof is by induction on $m$ and uses lemma \ref{lem:k3_local_complementation} as the base case. 
The reader is referred to lemma 5.2.4 of \cite{KissingerWetering2024Book} for the detailed proof.

Using lemma \ref{lem:k_m_local_complementation}, we can prove the general local complementation identity, which will be important for our later discussion on the tensor contraction algorithm.

Let us first define local complementation of a graph.

\begin{figure}[ht]
	\centering
    \input{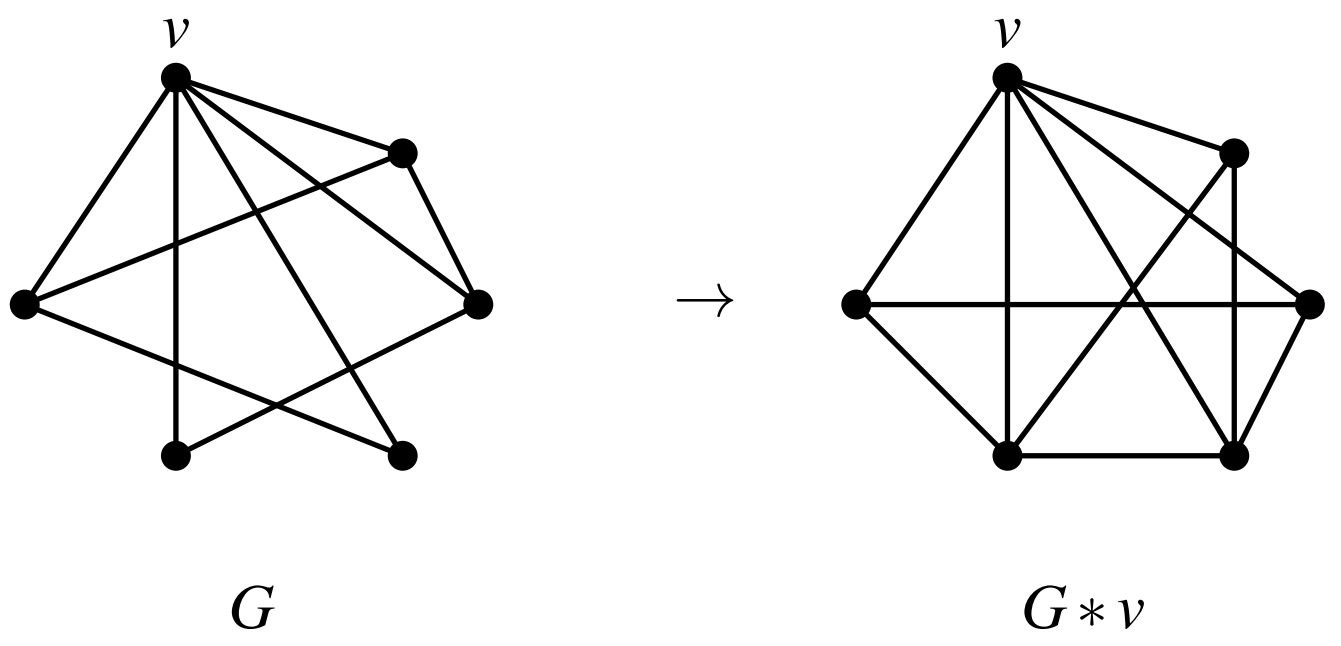}
	\caption{Local complementation of the graph $G$ on the vertex $v$. The subgraph induced by the neighbours of $v$ is replaced with its complement graph.}
	\label{fig:local_complementation}
\end{figure}

\begin{definition}[Local Complementation]
	The local complementation of a graph $G$ on a vertex $v$ is the operation that replaces the subgraph of $G$ induced by the neighbours of $v$ with its complement graph. 
	The resulting graph is denoted as $G*v$.
	Let $u, w$ be two neighbours of $v$ in $G$. $u$ is connected to $w$ in $G*v$ if and only if $u$ is not connected to $w$ in $G$. All other connections are left unchanged.
	An example of local complementation is shown in figure \ref{fig:local_complementation}.
\end{definition}

\begin{figure}[ht]
	\centering
	\includegraphics[width=0.8\linewidth]{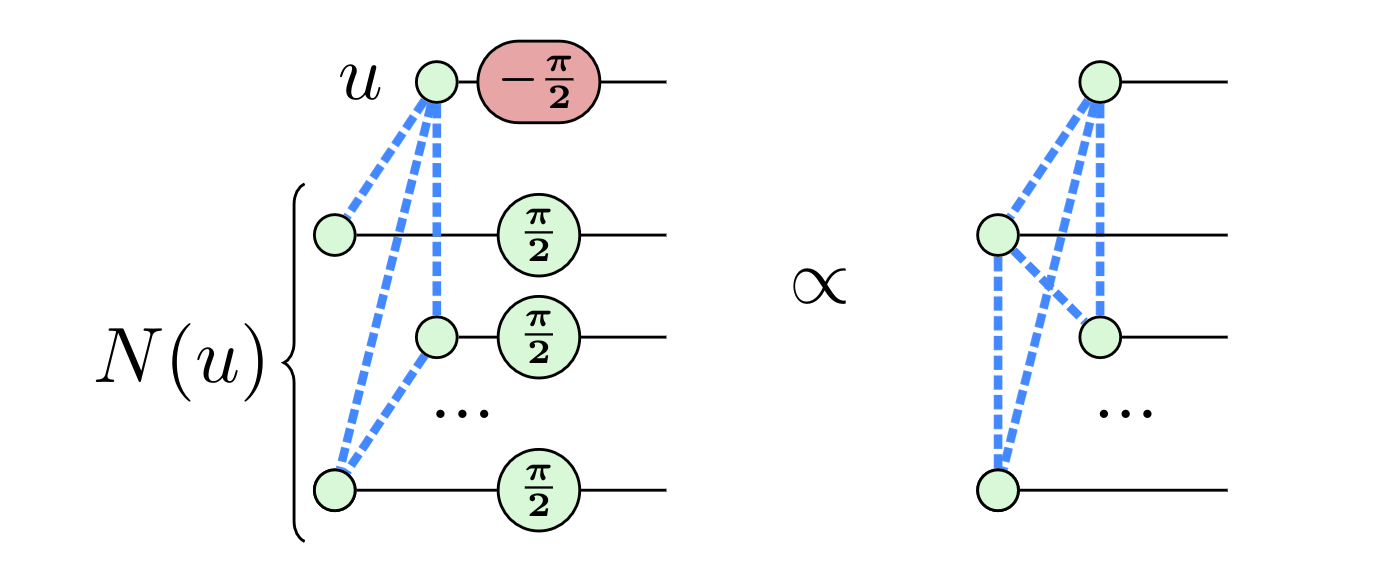}
	\caption{General Local complementation identity. The ZX diagram of the local complementation of $G$ on $v$ is equivalent to the original ZX diagram up to a global constant and some local unitaries applied to $v$ and its neighbours.}
	\label{fig:local_complementation_identity}
\end{figure}

\begin{theorem}[Local Complementation Identity]
	Let $G$ be a ZX diagram consisting of $z$-spiders connected by Hadamard gates, and let $v$ be a vertex in $G$. 
	This ZX diagram is equivalent, up to a global constant and up to some local unitaries applied to $v$ and its neighbours, to the ZX diagram of the local complementation of $G$ on $v$, denoted as $G*v$.
	See figure \ref{fig:local_complementation_identity}.
\end{theorem}

The concept of local complementation was introduced in \cite{Duncan_2020}, and is discussed in detail in the textbook \cite{KissingerWetering2024Book}.

\section{A Categorical Approach to ZX Calculus}\label{sec:categorical-zx}

In this section, we will use category theory to justify the claim we have made repeatedly in this dissertation: ZX calculus captures the underlying abstract structure of tensor networks.

The ZX calculus was born in the categorical quantum mechanics programme of Abramsky and Coecke \cite{abramsky2004categorical}, and much of its literature is written in the language of categories \cite{coecke2017picturing, vandeweteringNotes}.
Although nothing in this thesis depends on the categorical language, it is worth explaining this approach.
Fuller accounts are given by Heunen and Vicary \cite{heunen2019categories}; the exposition follows section 7 of \cite{vandeweteringNotes}.

\begin{definition}[The category $\mathbf{ZX}$]\label{def:cat-zx}
	The category $\mathbf{ZX}$ has the natural numbers as its objects, the object $n$ being read as $n$ parallel wires.
	A morphism $D: n \to m$ is a ZX diagram with $n$ inputs and $m$ outputs, the composition $D' \circ D$ plugs the outputs of $D$ into the inputs of $D'$, and $\id_n$ is the diagram of $n$ bare wires.
	On objects the monoidal product is $n \otimes m := n + m$, with the empty diagram as its unit; on morphisms it stacks one diagram above the other.
\end{definition}

Since wires may be crossed, $\mathbf{ZX}$ is \emph{symmetric} monoidal.
A symmetric monoidal category whose objects are the natural numbers and whose tensor product is their addition is called a \define{PROP}, short for \emph{product and permutation category}, and many authors accordingly speak of the PROP of ZX diagrams.
The coherence conditions imposed upon a monoidal category \cite{maclane1998categories} are what licenses drawing diagrams at all rather than writing nested expressions; the principle that \emph{only connectivity matters} (lemma \ref{lem:only-connectivity-matters}) is the diagrammatic shadow of their coherence theorem \cite{joyal1991geometry, selinger2011survey}.

In definition \ref{def:tensor} we declined to distinguish the inputs of a tensor from its outputs, whereas definition \ref{def:cat-zx} appears to insist upon the distinction; the two are reconciled by compact closure.
A symmetric monoidal category is \define{compact closed} when every object $A$ has a dual $A^{*}$ equipped with a cup $\eta: 0 \to A^{*} \otimes A$ and a cap $\varepsilon: A \otimes A^{*} \to 0$ satisfying the yanking equations \cite{kelly1980coherence}, which let an output be bent into an input and back again.
In $\mathbf{ZX}$ every object is its own dual, and both the cup and the cap are the phase-free rank-two $z$-spider: read as undirected tensors in the sense of chapter \ref{ch:tensors}, the identity wire, the cup and the cap are literally the same tensor, so that bending a leg is a redrawing rather than an operation, and nothing was lost by forgetting the distinction in the first place.
Reflecting a diagram horizontally and negating its phases gives a \define{dagger}: involutive, contravariant, the identity on objects, and on the side of tensors the adjoint, it makes $\mathbf{ZX}$ a dagger compact closed category \cite{selinger2007dagger}.

Until now we have deliberately confused a diagram with the tensor it denotes; to speak of completeness the two must be separated.

\begin{definition}[Interpretation]\label{def:interpretation}
	Let $\mathbf{Qubit}$ be the PROP whose object $n$ is the vector space $(\C^2)^{\otimes n} \cong \C^{2^n}$ and whose morphisms $n \to m$ are the linear maps between them, a full subcategory of finite dimensional complex vector spaces.
	The \define{interpretation} $\interp{D}$ of a ZX diagram $D$ is the linear map obtained by contracting the tensor network $D$ as in chapter \ref{ch:tensors}.
\end{definition}

As contraction respects both plugging and stacking, $\interp{\cdot}: \mathbf{ZX} \to \mathbf{Qubit}$ is a strong monoidal functor which is the identity on objects.
It is not injective on morphisms, nor could it be: the very point of the rewrite rules is that distinct diagrams denote the same map.
Let $\sim$ therefore be the smallest equivalence relation on diagrams such that $D_1 \sim D_2$ for every rule of figure \ref{fig:zx_rewrites}, and such that $D_1 \sim D_2$ implies both $E \circ D_1 \circ C \sim E \circ D_2 \circ C$ and $E \otimes D_1 \otimes C \sim E \otimes D_2 \otimes C$ for all diagrams $C$ and $E$; these two conditions are the licence, used freely above, to rewrite a subdiagram in place.
As $\sim$ is a congruence, its classes $\overline{D}$ are the morphisms of a PROP $\mathbf{ZX}/\!\sim$, and the assertion that $\interp{\overline{D}} := \interp{D}$ is well defined is exactly the \define{soundness} of the rewrite rules, which need be checked only on the rules of figure \ref{fig:zx_rewrites}, with the constants of lemma \ref{lem:zx-constants}.

Two theorems already stated now become statements about this single functor.
Theorem \ref{thm:zx-universality} says that every linear map is $\interp{D}$ for some diagram $D$, which for a functor is the analogue of surjectivity and is called \define{fullness}; theorem \ref{thm:zx-completeness} says that $\interp{D_1} = \interp{D_2}$ implies $D_1 \sim D_2$, the analogue of injectivity, called \define{faithfulness}.
Universality and completeness together therefore assert that $\interp{\cdot}: \mathbf{ZX}/\!\sim \; \to \mathbf{Qubit}$ is full and faithful, which is to say an equivalence of categories.

A \define{monoid} in a monoidal category is an object $M$ with morphisms $\mu: M \otimes M \to M$ and $\nu: 0 \to M$ satisfying the following identities.

\begin{align*}
	\mu \circ (\mu \otimes \id) &= \mu \circ (\id \otimes \mu) \\
	\mu \circ (\nu \otimes \id) &= \id = \mu \circ (\id \otimes \nu)
\end{align*}

\begin{figure}[ht]
	\centering
	\input{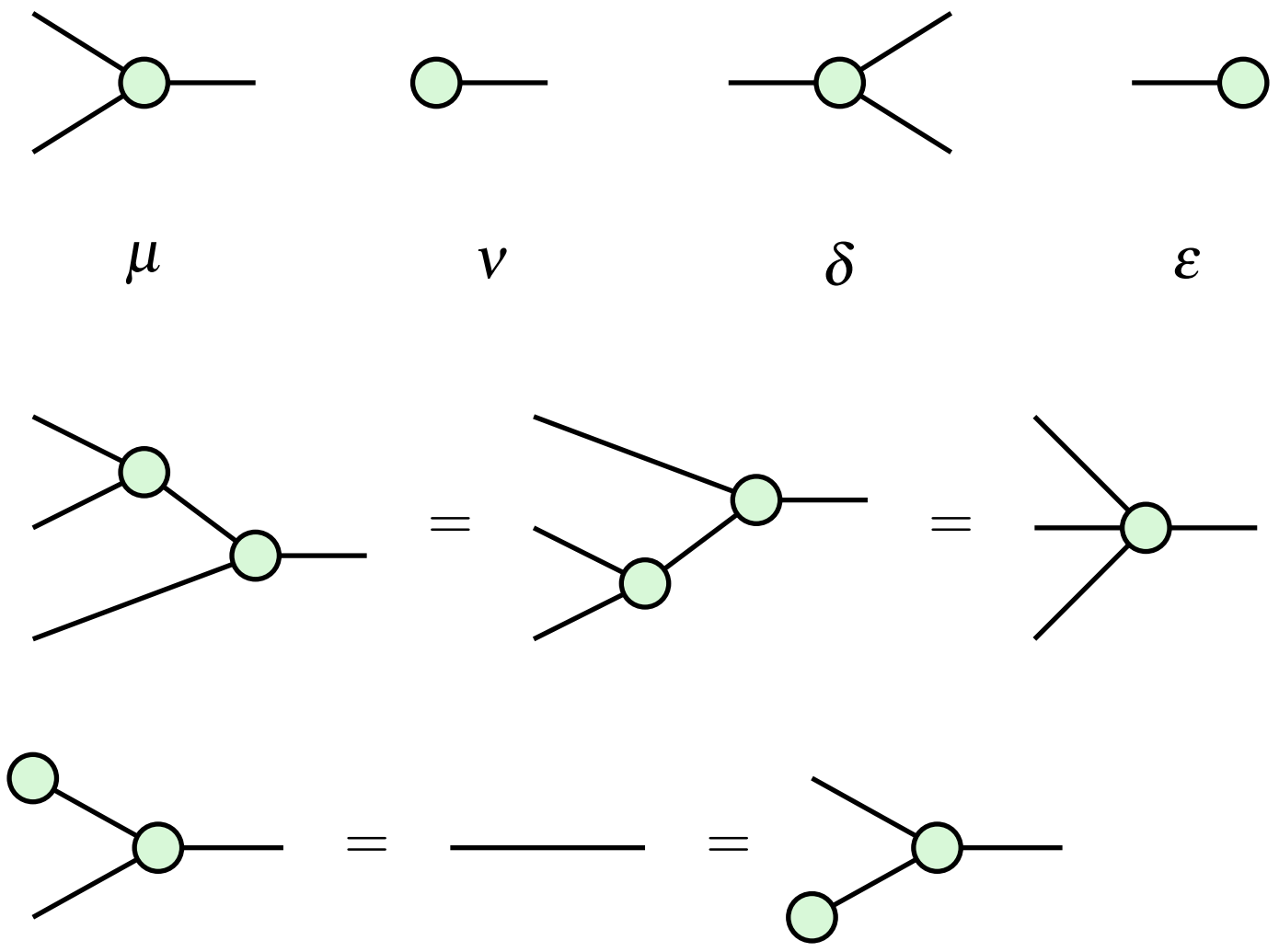}
	\caption{The $z$-spider as a monoid and a comonoid. Top: the multiplication $\mu$, the unit $\nu$, the comultiplication $\delta$ and the counit $\epsilon$. Middle: associativity, both sides reducing by spider fusion to the same rank-four spider. Bottom: the unit law. The comonoid laws are the mirror images of these. Redrawn after equations (110), (111) and (113) of \cite{vandeweteringNotes}.}
	\label{fig:frobenius_monoid}
\end{figure}

Take $M = 1$, let $\mu$ be the rank-three $z$-spider read with two inputs and one output, and let $\nu$ be the rank-one $z$-spider: associativity and unitality are then instances of spider fusion and identity removal, and reading the same two spiders backwards gives a \define{comonoid} $\delta = \mu^{\dagger}$, $\epsilon = \nu^{\dagger}$.
These four maps and the laws they obey are drawn in figure \ref{fig:frobenius_monoid}.

\begin{figure}[ht]
	\centering
	\input{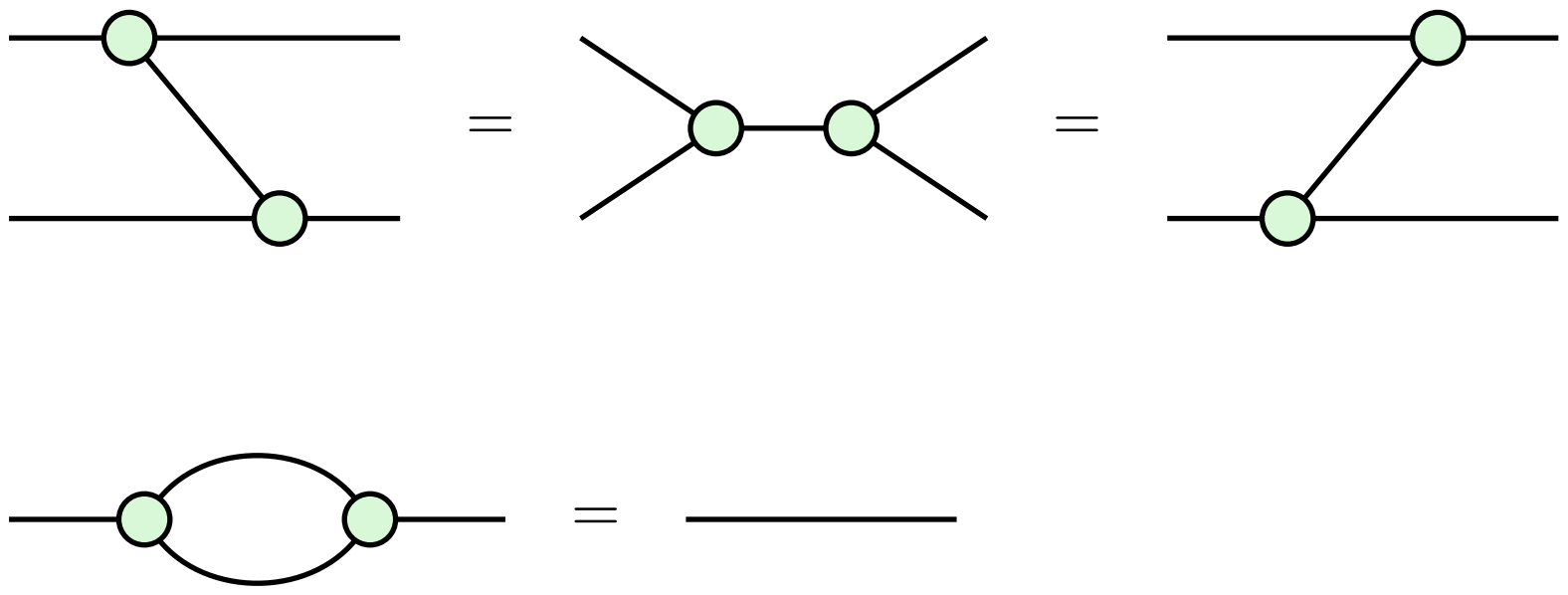}
	\caption{The Frobenius conditions (top) and speciality (bottom) for the $z$-spider. Redrawn after equations (115) and (116) of \cite{vandeweteringNotes}.}
	\label{fig:frobenius_conditions}
\end{figure}

A monoid and a comonoid on one object that satisfy $(\id \otimes \mu) \circ (\delta \otimes \id) = \delta \circ \mu = (\mu \otimes \id) \circ (\id \otimes \delta)$ form a \define{Frobenius algebra}, \emph{special} when $\mu \circ \delta = \id$ and a \emph{dagger}-Frobenius algebra when the two halves are related by $\dagger$ as above.
Both conditions are drawn in figure \ref{fig:frobenius_conditions}, and both are again instances of spider fusion, so that the $z$-spiders, and by the meta rule the $x$-spiders, form a commutative special dagger-Frobenius algebra.

The illuminating fact is the converse, due to Coecke, Pavlovi\'{c} and Vicary \cite{coecke2013orthogonal}: every commutative special dagger-Frobenius algebra on a finite dimensional Hilbert space $V$ arises from an orthonormal basis $\ket{v_1}, \cdots, \ket{v_n}$ of $V$, its spiders being $\sum_{i} \ket{v_i \cdots v_i} \bra{v_i \cdots v_i}$.
Spider fusion is thus no arbitrary axiom but a diagrammatic characterisation of the notion of an orthonormal basis, of which the $z$-spider is the instance for the computational basis.
Frobenius algebras long predate this use \cite{kock2004frobenius}, entering quantum theory through Coecke and Pavlovi\'{c}.

\begin{figure}[ht]
	\centering
	\input{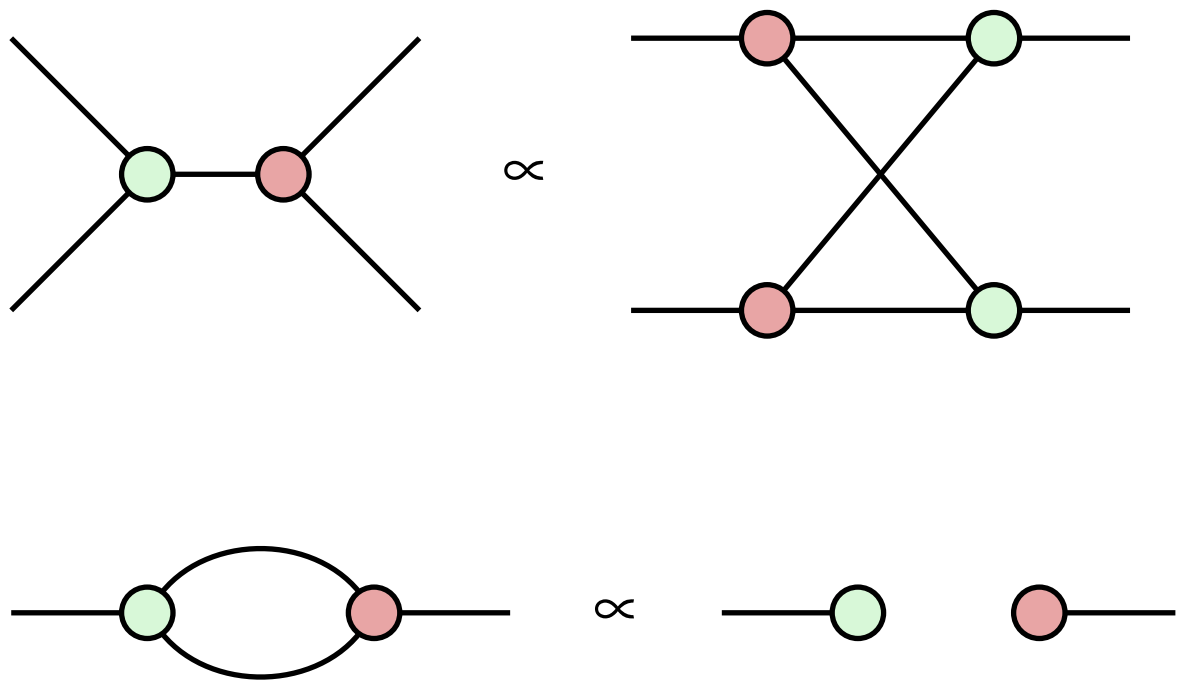}
	\caption{The bialgebra law (top) and the Hopf law (bottom) for the $z$- and $x$-spiders. The bialgebra law is the two-by-two case of strong complementarity, and the Hopf law, the antipode here being the identity, is the double edge cancellation of lemma \ref{lem:double_edge_cancellation}. Both hold only up to the scalar of lemma \ref{lem:zx-constants}, whence the $\propto$. Redrawn after equations (118) and (119) of \cite{vandeweteringNotes}.}
	\label{fig:bialgebra_hopf}
\end{figure}

The interaction of the two colours is likewise a familiar structure.
A \define{bialgebra} is a monoid together with a comonoid whose comultiplication is a homomorphism of the monoid; written diagrammatically, as in the upper half of figure \ref{fig:bialgebra_hopf}, its principal law is exactly the two-by-two case of strong complementarity ($\color{red} \mathbf{sc}$), with the copy rule ($\color{red} \mathbf{cp}$) supplying the unit conditions.
A bialgebra carrying an antipode that obeys the Hopf law is a \define{Hopf algebra}, and here the antipode is the identity, so that the Hopf law reduces to the lower half of figure \ref{fig:bialgebra_hopf}, which the reader will recognise as lemma \ref{lem:double_edge_cancellation}.

Either colour may supply the monoid, and a pair of Frobenius algebras so related is called a pair of interacting Frobenius algebras \cite{duncan2016interacting}; in the qubit model the law holds only up to the scalar of lemma \ref{lem:zx-constants}, so what we have is a scaled bialgebra.
The rule that does the most work in this dissertation is therefore the abstract statement that two orthonormal bases are complementary, which is how Coecke and Duncan introduced the calculus \cite{coecke2011interacting}.

The ellipses of figure \ref{fig:zx_rewrites} deserve a remark, since each rule containing them is strictly a family of rules indexed by the number of legs.
Such families are made precise by the \define{!-box} of Dixon and Duncan \cite{dixon2009graphical}, a subdiagram which may be duplicated any number of times, zero included, and which allows the rules to be given to automated tools such as Quantomatic \cite{kissinger2015quantomatic}; not every family is so expressible, the complete graph $K_m$ of lemma \ref{lem:k_m_local_complementation} being the standard counterexample \cite{millerbakewell2019finite}.

We are now in a position to state the claim of this section properly.
A ZX diagram is a morphism in a free dagger compact closed PROP presented by two commutative special dagger-Frobenius algebras that interact as a Hopf algebra, and no part of that description mentions $\C^2$.
Every equation derived by the rewrite rules therefore holds in each model of the theory, under any monoidal functor out of $\mathbf{ZX}/\!\sim$, and not merely under the interpretation of definition \ref{def:interpretation}.
This is what we mean when we say that ZX calculus captures the abstract structure of tensor networks: the qubit tensors of definition \ref{def:zx_generators} are one interpretation of that structure, the qudit tensors and the ZXW calculus of chapter \ref{ch:generic_diagram} are others, and the graphical reasoning, local complementation included, is the same in all of them.

\chapter{An Algorithm using ZX Calculus} \label{ch:zx-algorithms}

Theorem \ref{th:ultimate_complexity_tensor_contraction} is a final verdict on any bond-by-bond contraction algorithm of tensor networks.
However, it is still possible to find a more efficient algorithm if we can exploit the structure of the tensor network.
This chapter attempts to present such an algorithm for quantum circuits using ZX calculus.

\section{Graph Operations and Their Effects on Contraction Complexity}

Let us first look at the effect of some of the common graph operations on the complexity of tensor contraction and treewidth.

\begin{figure}[ht] \centering 
	\input{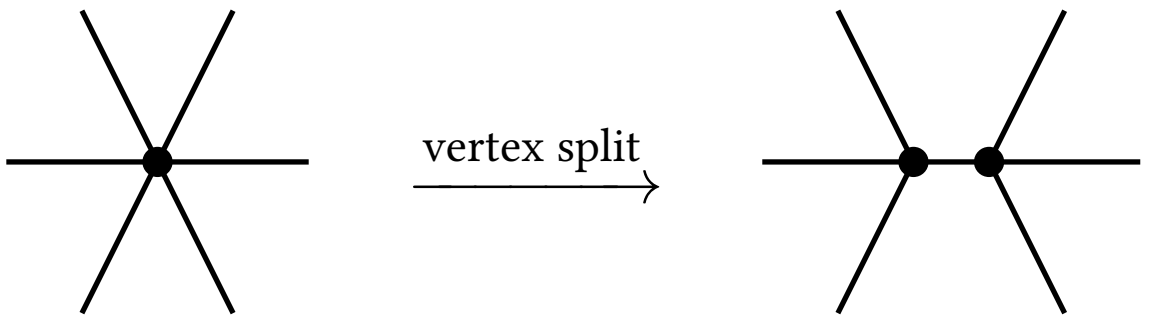}
	\caption{The operation of splitting a vertex}
	\label{fig:vertex_split}
\end{figure}

\begin{lemma}[Splitting a Vertex Does not Increase Complexity]\label{lem:split_vertex}
  Let $G$ be a graph of a tensor network (we assume $G$ is connected and contains at least 2 edges.) , $G'$ be another graph obtained by the operation of splitting a vertex, which is the opposite of spider fusion (f in figure \ref{fig:zx_rewrites}, and in figure \ref{fig:vertex_split}).
	The best bond-by-bond tensor contraction algorithm on $G'$ is not more complex than that of $G$.
\end{lemma}

Splitting a vertex may create a different tensor.
This lemma only regards the complexity of bond-by-bond contraction, and is not concerned with the equality of the resulting tensor.

\begin{figure}[ht]
    \centering 
    \input{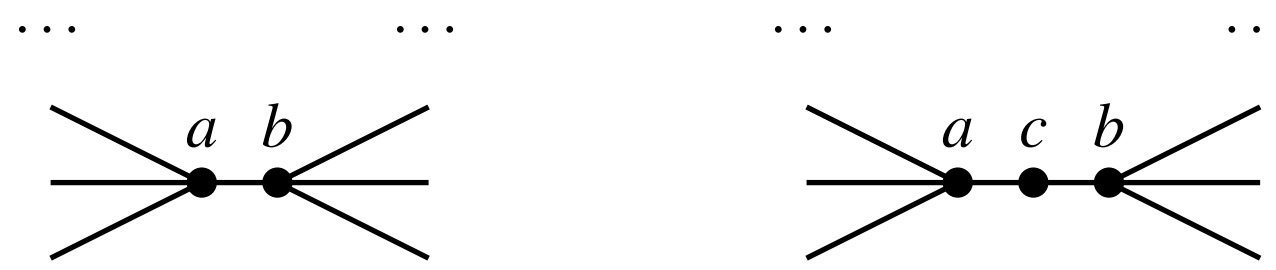}
	\caption{Dividing an edge does not increase the complexity of contraction. 
			Left: the original graph. Right: the graph after dividing an edge.}
    \label{fig:divide_edge}
\end{figure}

\begin{lemma}[Dividing an Edge Does not Increase Complexity]\label{lem:split_edge}
  Let $G$ be a graph of a tensor network (which we assume to be connected and contains at least two edges.), $G'$ be another graph obtained by the operation of dividing an edge (as shown in figure \ref{fig:divide_edge}).

	The best bond-by-bond tensor contraction algorithm on $G'$ is not more complex than that of $G$.
\end{lemma}

Again, similar to lemma \ref{lem:split_vertex}, this lemma only regards the complexity of bond-by-bond contraction, and is not concerned with the equality of the resulting tensor.
The proof of lemma \ref{lem:split_edge} is similar to that of lemma \ref{lem:split_vertex}.
Detailed proofs of both lemmas can be found in the appendix.

It is worth noting that the condition of the tensor network $G$ being a connected non-trivial graph is necessary for both lemmata.
This is because, if $G$ is the graph with two vertices and one edge, its line graph $G$ is a single vertex graph, which has treewdith 0, and both operations will make it into a two-vertex tree which has teeewidth 1.

Let us define a graph operation based on the K3 local complementation (lemma \ref{lem:k3_local_complementation}) in ZX calculus.

\begin{definition}[K3 remove edge]
	Let the graph $G$ denote a tensor network. Let $v$ be a vertex of degree two with both of its neighbours connected.
	Let $G'$ be the diagram with K3 local complementation (lemma \ref{lem:k3_local_complementation}) applied to $v$.
	The transformation from $G$ to $G'$ is called K3 remove edge.
\end{definition}

\begin{figure}[ht]
	\centering
	\input{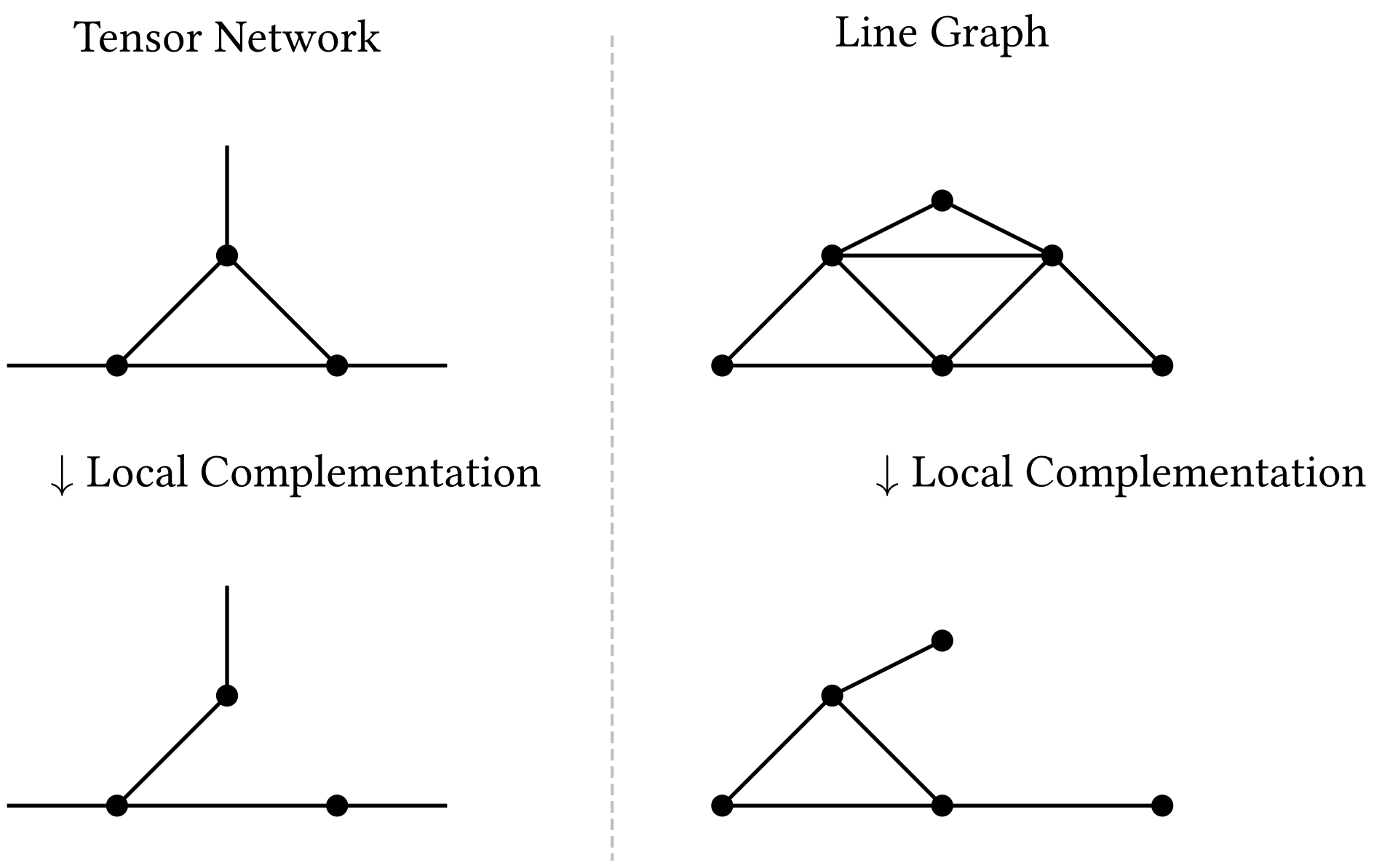}
	\caption{K3 complementation on $G$ is equivalent to removing a vertex in the line graph $G^*$.}
	\label{fig:k3_line_graph}
\end{figure}

K3 remove edge is demonstrated in figure \ref{fig:k3_line_graph}. K3 local complementation removes a bond and therefore a vertex in the line graph of $G$.
Therefore the complexity cannot be greater.

In practice, removing one vertex in the line graph may reduce the treewidth by one.
This is a significant reduction considering that the complexity of tensor contraction is exponential in the treewidth.

We now define another graph operation, called \define{Unfuse and K3 Complementation}, specifically for graph-like ZX diagrams.

\begin{figure}[ht]
	\centering
	\input{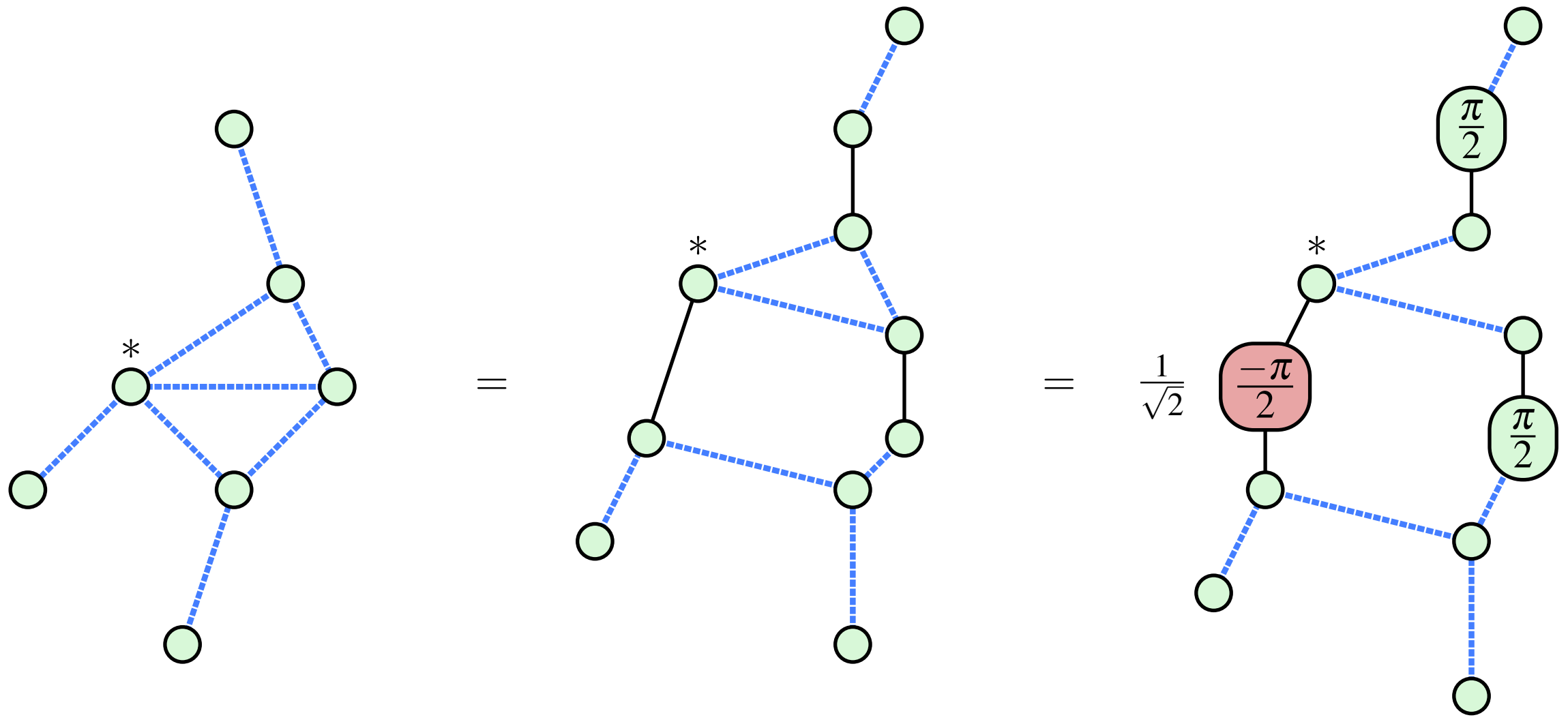}
	\caption{Unfuse a vertex to create a valid vertex for K3 local complementation. The vertex labelled with $*$ is the one on which we want to apply K3 local complementation.}
	\label{fig:unfuse_local_comp}
\end{figure}

\begin{definition}[Unfuse and K3 Complementation]\label{def:unfuse_k3_complementation}
	Let $G$ be a graph-like ZX diagram.
	Let $v$ be a vertex in $G$ such that $v$ has degree 3 or more and two of its neighbours $n_1$ and $n_2$ are connected by a Hadamard edge.
	We may unfuse $v$ to form $v_1, v_2$ such that $v_1$ is connected to $n_1$ and $n_2$, while the rest of the neighbours of $v$ are connected to $v_2$.
	By applying K3 local complementation to $v_1$, we can remove the edge $(n_1, n_2)$. 
	We label the resulting graph as $G'$.
	We do not require $G'$ to be graph-like.

	This process is shown in figure \ref{fig:unfuse_local_comp}.
\end{definition}

Unfuse and K3 complementation is a combination of dividing an edge, splitting a vertex, and K3 removing edge, therefore the contraction complexity of the resulting graph is no more than that of the original graph.

Unfuse and K3 complementation can be applied as long as there is a Hadamard triangle (a Hadamard triangle is a triangle whose edges are all Hadamard edges) among three $z$-spiders.
Since this operation does not create any triangles, we can apply this operation repeatedly until there are no more Hadamard triangles in the graph.

\begin{remark}
	Although we know that both of the two operations we defined above (unfuse and K3 complementation, K3 remove edge) do not increase the complexity of tensor contraction, because treewidth is a complicated graph property, we do not know how effectively they can reduce treewidth.
	
	The theoretical study of the effectiveness of these operations is extremely complicated, and, unfortunately, due to time constraints, is not covered in this thesis.
\end{remark}

\section{The Algorithm}

If our tensor network is a ZX diagram, we can apply the two operations (unfuse and K3 complementation, K3 remove edge) to preprocess the tensor network to potentially reduce the complexity of tensor contraction.

\begin{definition}[Rewrite Preprocessing]\label{def:rewrite_preprocessing}
	Given a graph-like ZX diagram $G$, we can apply two operations (unfuse and K3 complementation, K3 remove edge) iteratively until they can no longer produce any changes. 
	We call the operation the Rewrite Preprocessing of $G$.
\end{definition}

We propose a straightforward algorithm, shown in algorithm \ref{alg:zx_graph_state_contraction}.

\begin{algorithm}[ht]
	\caption{Tensor Contraction with Graph Operation}
	\label{alg:zx_graph_state_contraction}
\begin{algorithmic}[1]
	\State \textbf{Input:} A graph-like ZX diagram $G$.\footnote{Note that any quantum circuit can be converted to a graph-like diagram, as shown in theorem \ref{th:all_to_graph_like}.}
	\State Perform Rewrite Preprocessing on $G$ to obtain a new graph $G'$.
	\State Perform regular tensor contraction algorithm on $G'$.
\end{algorithmic}
\end{algorithm}

Both operations of K3 remove edge and unfuse and K3 complementation can be performed in constant time if a valid vertex is found.
Finding a valid vertex can be done, in the worst case, by checking all vertices and their neighbours, which is $O(n^2)$ time.
Note that contraction can be done in $O(n d^2)$ only for a tree tensor network, where $d$ is the dimension of the bond (in our case, $d=2$).
This gives reason to expect that our algorithm can be more efficient than the best bond-by-bond contraction algorithm applied to the original graph.

\begin{lemma}[Complexity of Algorithm \ref{alg:zx_graph_state_contraction}]
	Let $G$ be a graph-like ZX diagram with $n$ vertices.
	The complexity of algorithm \ref{alg:zx_graph_state_contraction} is $O(n^2)$ plus the complexity of the best bond-by-bond contraction algorithm on the resulting graph.
\end{lemma}

It is very difficult to carry out the rewrite preprocessing by hand even for simple graph-like diagrams.
To solve this issue, the entire rewrite preprocessing algorithm has been implemented in the Rust program cartographer available at \url{https://github.com/harryhanYuhao/cartographer}.
More on the details of implementation can be found in the appendix.
Check figure \ref{fig:preprocess_workflow}.

\section{Benchmarking Rewrite Preprocessing}

Our rewrite preprocessing algorithm has been benchmarked in the following way. 
First, we generate a random graph-like ZX diagram $G$ with $n$ vertices via the Erd\H{o}s--R\'{e}nyi algorithm \cite{erdos1959random}.
Its line graph is computed, which we label as $G^*$. We also compute the treewidth of $G^*$. 
The new graph $F$ is obtained by applying the rewrite preprocessing to $G$.
The treewidth of $F^*$ is also computed.
The treewidths are found via Tamaki's positive-instance driven dynamic programming algorithm \cite{tamaki2017positive}.

\begin{figure}[ht]
	\centering
	\includegraphics[width=0.8\textwidth]{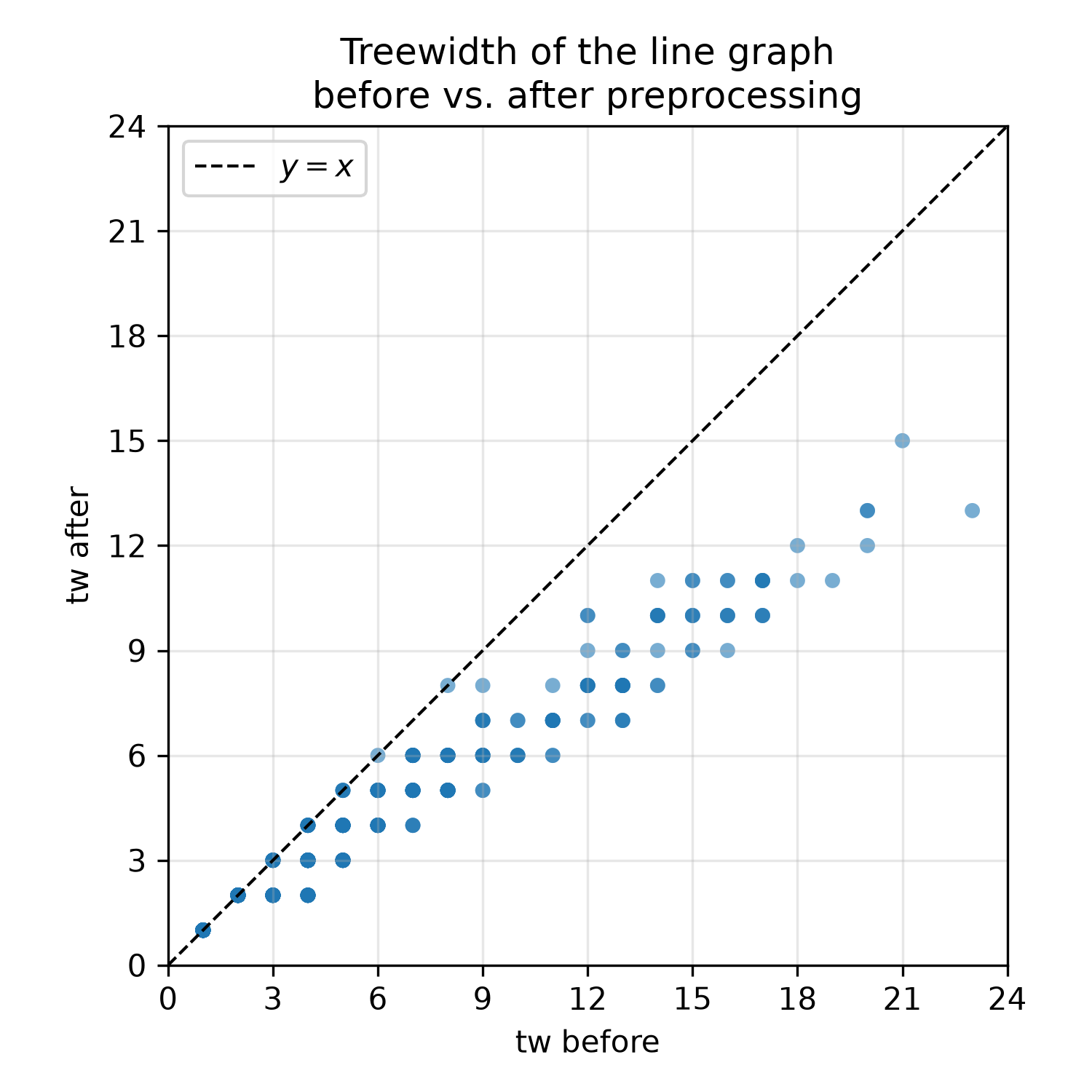}
	\caption{The scatter plot of treewidth before and after rewrite preprocessing. The x-axis is the treewidth of the line graph of the original graph, and the y-axis is the treewidth of the line graph of the graph after rewrite preprocessing. The red line is the line $y=x$. The points below the red line are the graphs whose treewidth is reduced by rewrite preprocessing.}
	\label{fig:tw_scatter}
\end{figure}

These are our key findings.
\begin{enumerate}
  \item Of the 441 random graphs generated, the treewidth of 254 decreased, that of 187 was unchanged, and none increased. \footnote{The computation takes more than 60 hours of cpu run time on the author's personal computer with Ryzen 7945 HX.}
	The comparison of treewidth is shown in figure \ref{fig:tw_scatter} and table \ref{tab:tw_decrease}.
	\item The average decrease of treewidth is $1.62$, out of the mean treewidth $5.56$.
  \item Our algorithm also increases the number of vertices in the line graph. The average node count before and after the preprocessing is $13.35$ and $15.56$, with mean increase of $2.21$.
	\item The treewidth decrease is observed on the original line graphs of both high (23) and low (3) treewidth, as shown in figure \ref{fig:tw_scatter}.
\end{enumerate}

\begin{table}
	\centering
\begin{tabular}{rrrrrr}
\toprule
$|V(G)|$ & $|V(G^*)|$ & $|V(F^*)|$ & $\mathrm{tw}(G^*)$ & $\mathrm{tw}(F^*)$ & decrease \\
\midrule
10 & 33 & 42 & 17 & 10 & 7 \\
11 & 35 & 44 & 16 & 10 & 6 \\
10 & 26 & 32 & 13 & 8 & 5 \\
20 & 43 & 50 & 15 & 11 & 4 \\
19 & 36 & 42 & 12 & 8 & 4 \\
9 & 23 & 28 & 12 & 8 & 4 \\
8 & 20 & 25 & 11 & 7 & 4 \\
7 & 16 & 19 & 9 & 6 & 3 \\
14 & 18 & 21 & 6 & 4 & 2 \\
7 & 15 & 18 & 8 & 6 & 2 \\
6 & 11 & 14 & 6 & 4 & 2 \\
6 & 11 & 14 & 6 & 4 & 2 \\
5 & 7 & 9 & 4 & 2 & 2 \\
5 & 8 & 10 & 5 & 3 & 2 \\
16 & 27 & 30 & 7 & 6 & 1 \\
14 & 20 & 22 & 7 & 6 & 1 \\
14 & 18 & 20 & 5 & 4 & 1 \\
11 & 11 & 12 & 4 & 3 & 1 \\
6 & 8 & 9 & 4 & 3 & 1 \\
6 & 10 & 12 & 5 & 4 & 1 \\
\bottomrule
\end{tabular}
\caption{The table of treewidth before and after rewrite preprocessing on 20 random graphs. 
$|V(G)|$ is the number of vertices in the original graph, $|V(G^*)|$ is the number of vertices in the line graph of the original graph, $|V(F^*)|$ is the number of vertices in the line graph of the graph after rewrite preprocessing, $\mathrm{tw}(G^*)$ is the treewidth of the line graph of the original graph, $\mathrm{tw}(F^*)$ is the treewidth of the line graph of the graph after rewrite preprocessing, and decrease is $\mathrm{tw}(G^*) - \mathrm{tw}(F^*)$.}
\label{tab:tw_decrease}
\end{table}

\begin{table}
	\centering
\begin{tabular}{rrrrrr}
\toprule
$|V(G)|$ & $|V(G^*)|$ & $|V(F^*)|$ & $\mathrm{tw}(G^*)$ & $\mathrm{tw}(F^*)$ & decrease \\
\midrule
10 & 36 & 45 & 20 & 12 & 8 \\
9 & 29 & 36 & 17 & 10 & 7 \\
9 & 31 & 39 & 18 & 11 & 7 \\
9 & 33 & 41 & 20 & 13 & 7 \\
12 & 41 & 50 & 20 & 13 & 7 \\
10 & 33 & 42 & 17 & 10 & 7 \\
10 & 32 & 40 & 17 & 11 & 6 \\
10 & 32 & 39 & 17 & 11 & 6 \\
10 & 32 & 40 & 17 & 11 & 6 \\
10 & 27 & 34 & 13 & 7 & 6 \\
\bottomrule
\end{tabular}
	\caption{Treewidth before and after rewrite preprocessing for the selected graphs with the largest decrease.}
\label{tab:tw_decrease_largest}
\end{table}

It is worth explaining why we did not choose to evaluate the run time of the tensor contraction algorithm before and after rewrite preprocessing.
This is because the run time is extremely large on high treewidth graphs, to the extent that it is almost impossible to run on personal computers.

\section{Limitations of Rewrite Preprocessing}

The benchmarks show very promising results. 
However, it is also important to realise the limitations of our rewrite preprocessing algorithm.

\begin{enumerate}
	\item Our algorithm so far only works on quantum circuits. That is, each bond is of dimension 2. The next chapter discusses the possibility of extending to general tensor networks.
	\item The generators of ZX diagrams are sparse, and often there are many different ways to exploit the sparsity to create a very efficient tensor contraction algorithm.
\end{enumerate}

I shall in particular explain the second point. 
Our rewrite rules only apply to the generators of ZX diagrams, which are sparse tensors (in particular the $z$-spider).
However, the same rewrite rule also applies to any other set of generators that are 
conjugate to the ZX generators up to a self-adjoint matrix, and conjugation may make sparse matrices no longer sparse.\footnote{
	Conjugation, however, does not change the rank of a matrix. Therefore rank may be a better measurement of the `sparsity' condition.
}

\chapter{Generic Diagrammatic Calculus} \label{ch:generic_diagram}

This chapter discusses more generic diagrammatic calculus that can represent any tensor.
Due to constraints of time, however, we are not able to deduce the analogue of local complementation rules for these generic diagrammatic calculi. 
This question is left for future work.

\section{ZXW Calculus}

So far, we have only discussed ZX calculus with two dimensional bonds. 
It is possible to extend ZX calculus to more generic tensors. 
ZXW calculus, pioneered by Wang et al.~\cite{wang2025penrosetensordiagramszx}, is one approach we describe in this section.

The generators and rewrite rules of ZX calculus are a subset of those in ZXW calculus.
There are many different flavours of ZXW calculus; here we will introduce the mixed dimensional qufinite ZXW calculus, where each bond can have arbitrary dimensions.

\begin{figure}[ht]
	\centering
	\resizebox{0.8\textwidth}{!}{
		\input{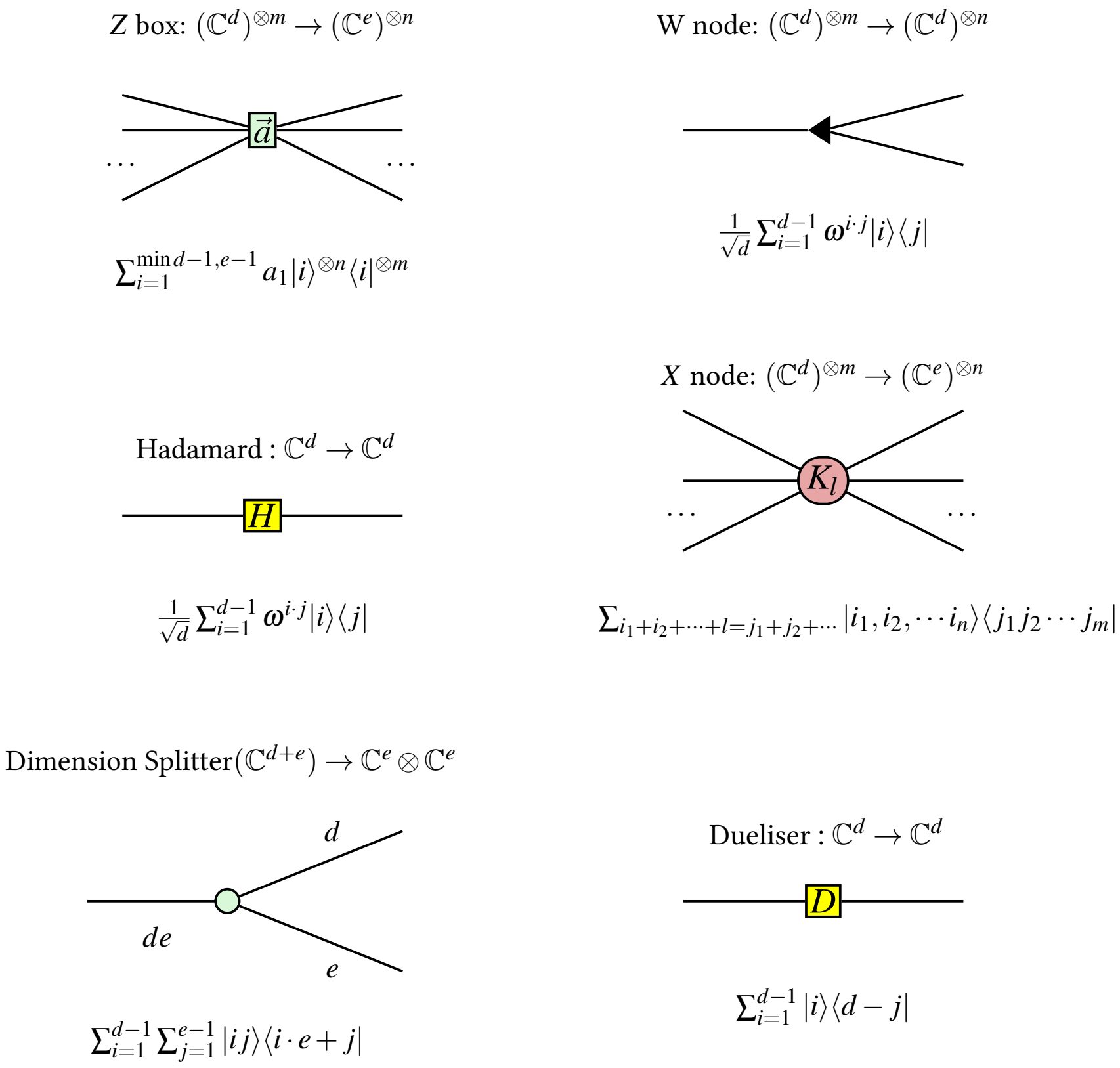}	
	}
	\caption{Generators of ZXW calculus. For the equation in Hadamard generator, $\omega$ denotes the $d$-th root of unity, where $d$ is the dimension of the bond.}
	\label{fig:zxw_generators}
\end{figure}

\begin{definition}[Generators of ZXW calculus]
	ZXW calculus has generators $Z$-box, Hadamard box, $W$-box, $X$-node, and dimension splitter as shown in figure \ref{fig:zxw_generators}.
	
	We have assumed that all wires have dimension $d$ unless otherwise labelled, and wires on the left side of a box are inputs, while wires on the right side are outputs.
\end{definition}

ZXW calculus is different from ZX calculus in two key aspects:
\begin{enumerate}
	\item All bonds in ZX are two dimensional, while bonds in ZXW can have arbitrary dimensions. The dimension of the bonds is labelled when necessary as a number near the bond.
	\item All generators in ZX calculus are symmetric tensors. The $x$-node and $W$-box in ZXW are not symmetric.
\end{enumerate}

\begin{figure}[ht]
	\centering
	\resizebox{0.8\textwidth}{!}{
		\input{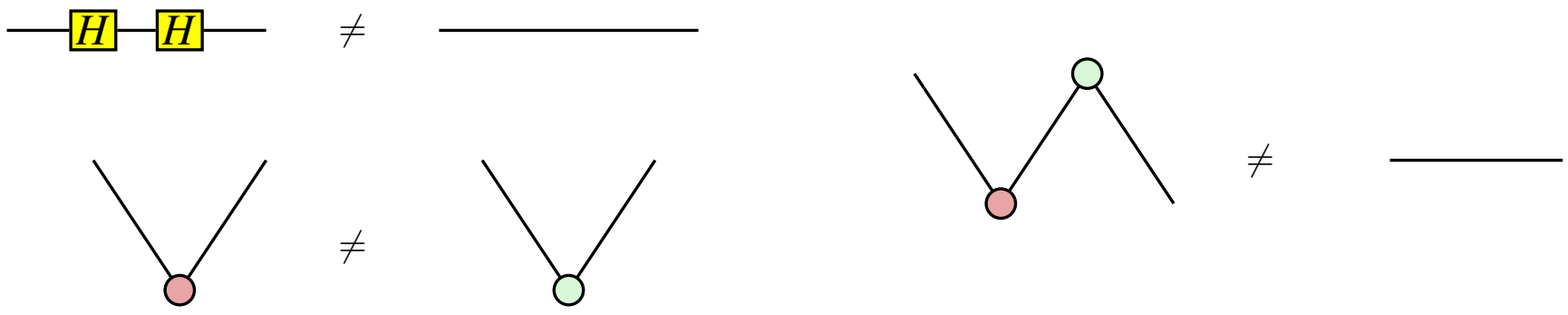}	
	}
	\caption{Some rules in ZX calculus do not hold in ZXW calculus.}
	\label{fig:zxw_differences}
\end{figure}

Because of these differences, some rules in ZX calculus do not hold in ZXW calculus, as shown in figure \ref{fig:zxw_differences}.

ZXW has a large number of rewrite rules.
We only list a few of them here.
For a comprehensive list of rewrite rules, please refer to the papers by Wang et al.~\cite{wang2025penrosetensordiagramszx, Po_r_2023, wang2026completenessqufinitezxwcalculus}.

\begin{lemma}[ZXW Rewrite Rules]
	Five of the interesting ZXW rewrite rules are summarised in the following figure.
	We still have identity rules for the $Z$-box and $X$-node (although $X$ is no longer symmetric, and other identities, like those shown in figure \ref{fig:zxw_differences}, do not hold).
	Hadamard gate $H$ is still unitary, but it is no longer self-adjoint.
	$$H^{\dagger}  = H^3,  \quad HH^{\dagger} = I$$
	$X$-nodes are still similar to $Z$-nodes (they are conjugate to each other), but now $HHHZH = X$.

	Most importantly, we still have strong complementarity between $Z$-box and $X$-node as well as another bialgebra rule between $Z$-box and $W$-box.
\end{lemma}

\begin{figure}[ht]
	\centering
	\resizebox{\textwidth}{!}{
		\input{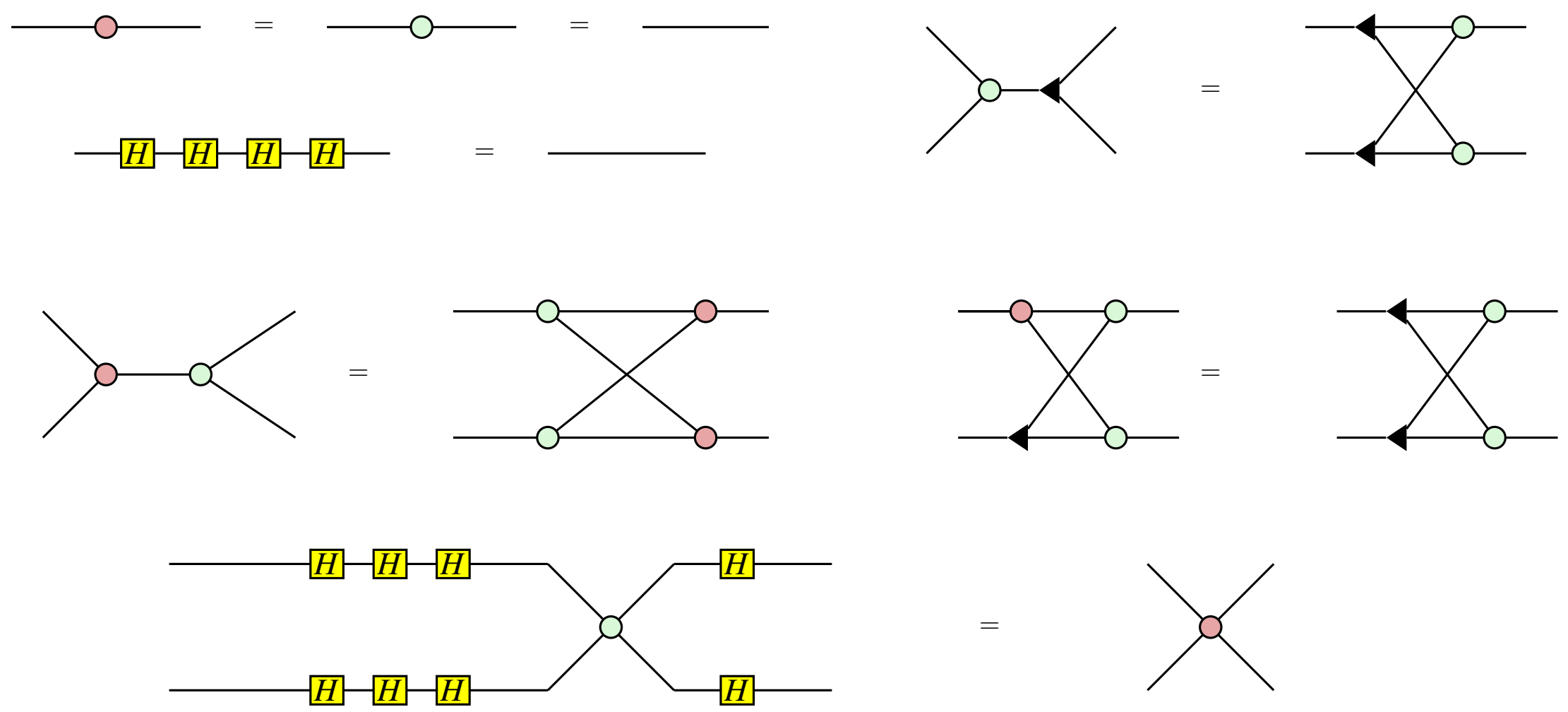}	
	}
	\caption{Some rewrite rules of ZXW calculus. The empty $z$ node is defined as $\sum \ket{i} \ket{j}$, and the empty $x$ node is defined as $x K_0$.}
	\label{fig:zxw_rewrite_rules}
\end{figure}

Because of the differences in bond-dimension and more types of generators, in general, doing rewrites in ZXW calculus is a lot more complicated than in ZX calculus.

Fortunately, the universality of ZXW calculus still holds.

\begin{theorem}[Universality of ZXW calculus]\label{thm:zxw_universality}
	Any tensor can be represented as a qufinite mixed-dimensional ZXW diagram.
	That is, using the generators of ZXW calculus and any tensor operation we have defined in chapter \ref{ch:tensors}, we can represent any tensor as a ZXW diagram.

	In particular, they can be represented with only two generators: $Z$-box and $X$-node.
\end{theorem}

The reader is referred to the original paper by Wang et al.~\cite{wang2025penrosetensordiagramszx} for a proof of this theorem.

\section{Generic Diagrams with Symmetric Generators}

As ZXW calculus also has strong complementarity rules, does it have an analogue of local complementation rules?
The answer is affirmative, but may not be of much use to us.

We cannot readily apply our graph theoretical operations to ZXW diagrams because the $X$-node and $W$-box are not symmetric, and therefore their diagrams cannot be regarded as graphs.

In general, a graph, as in graph theory, is a very versatile structure that allows 
\begin{enumerate}
	\item multiple edges between two vertices, 
	\item loops, 
	\item directed edges,
	\item coloured vertices, therefore different types of vertices
\end{enumerate}
However, in none of the graph theoretical formulations do we specify the orders of edges.
The order of the edges to a non-symmetric node does matter as we have shown in chapter \ref{ch:tensors}.
This means none of the graph theoretical operations we have are applicable to ZXW diagrams.

The natural question is, is there another diagrammatic language with symmetric generators that has the universality property of ZXW calculus? 
That is, is there a set of symmetric generators that can represent any tensor?

The answer is yes. 

\begin{theorem}[Existence of Symmetric Generators with Universality Property]
	There exists a finite set of symmetric generators that can represent any tensor via the tensor operations defined in chapter \ref{ch:tensors}.
\end{theorem}

\begin{figure}[ht]
	\centering
	\resizebox{0.8\textwidth}{!}{
		\input{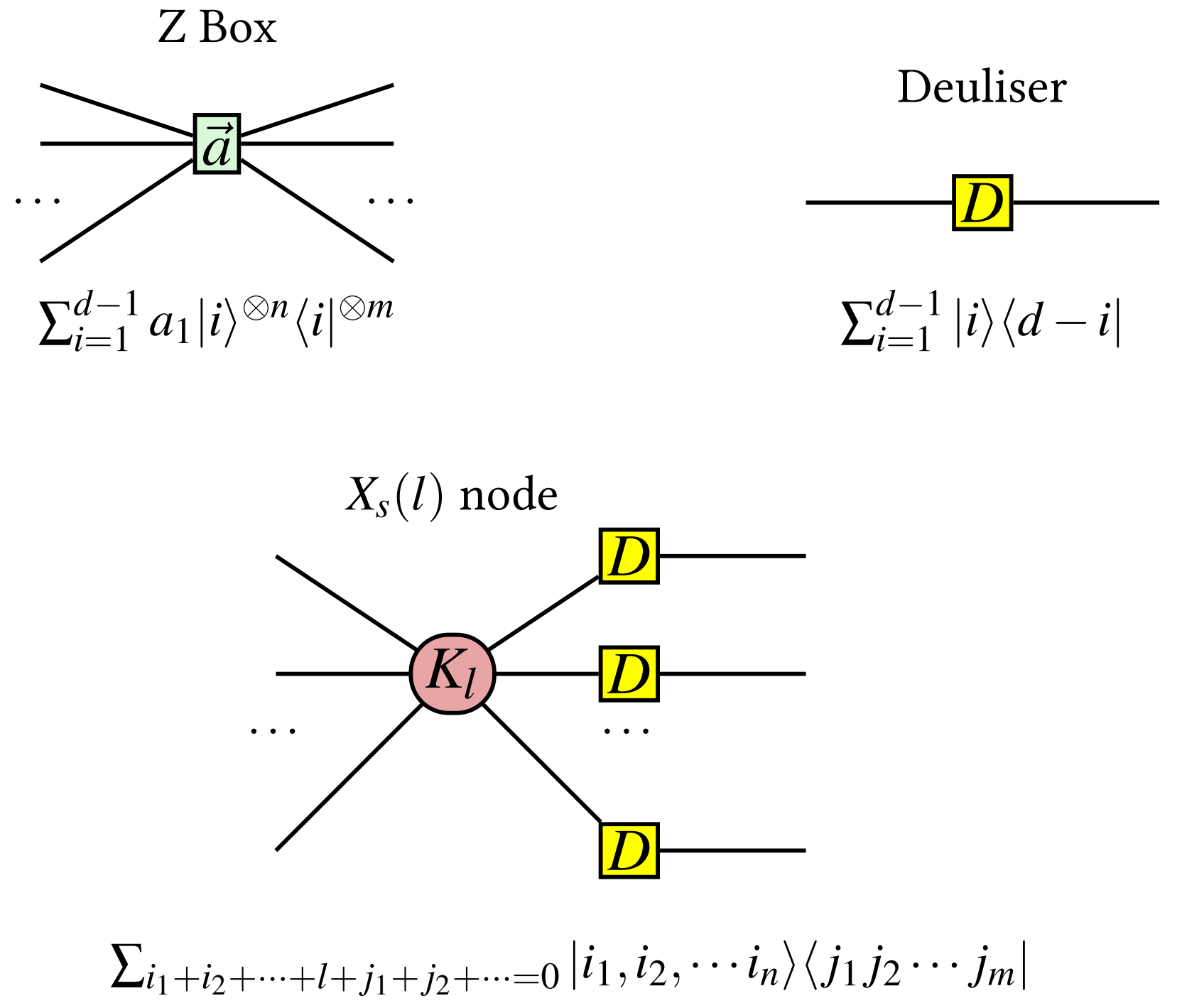}	
	}
	\caption{Symmetric generators with the universality property.}
	\label{fig:symmetric_generators}
\end{figure}

\begin{proof}
	In theorem \ref{thm:zxw_universality}, we have shown that any tensor can be represented as a ZXW diagram using only two generators: $Z$-box and $X$-node.

	$Z$-box is symmetric, but $X$-node is not.

	Applying the dualiser on each bond of the input of the $X$-node will make a symmetric tensor, which we denote as $X_s$-node.
	As the dualiser is symmetric, and two dualisers cancel, we have arrived at our finite set of generators that has the universality property: $Z$-box,  $X_s$-node, and dualiser.

	See figure \ref{fig:symmetric_generators} for a demonstration.
\end{proof}

We believe that this theorem opens the prospect of proposing many other diagrammatic calculi that exploit the graph theoretical operations, as we have done in chapter \ref{ch:zx-algorithms}.
Constrained by time, we were not able to explore this direction and leave it for future work.

\chapter{Conclusion and Future Work} \label{ch:conclusion}

Finding the optimal contraction order of a tensor network is NP-hard.
This dissertation makes the statement precise: the best possible bond-by-bond contraction complexity is $O(n \cdot d^{\mathrm{tw}(G^*)+1})$, where $G^*$ is the line graph of the network (theorem \ref{th:ultimate_complexity_tensor_contraction}) and $n$ is number of edges in $G$.
The proof depends on one observation: contracting a bond is exactly a simplicial elimination in the line graph, so tree decompositions of $G^*$ and contraction orders determine one another.

This correspondence motivates the central innovation of the thesis: rather than searching for a good contraction order on a fixed network, we may try to transform the network itself into an equivalent one whose line graph has smaller treewidth --- since the cost is exponential in treewidth, even a small reduction pays off.
Chapter \ref{ch:zx-algorithms} develops this into the paradigm of \emph{Tensor Contraction with Graph Operation}: two ZX rewrite operations, K3 remove edge and unfuse and K3 complementation, are proved never to increase contraction complexity, and their exhaustive application as a preprocessing step (algorithm \ref{alg:zx_graph_state_contraction}) carries only quadratic overhead.
In 441 random graph-like diagrams benchmarked with our Rust implementation \emph{cartographer}, the line-graph treewidth fell in  cases, never rose, and fell by $1.62$ on average out of the mean treewidth 5.56 .

\section{Future Work}

While we have proved that our preprocessing steps in chapter \ref{ch:zx-algorithms} never increase treewidth, we only have empirical evidence that they do decrease the treewidth of line graphs. 
Investigating the theoretical conditions under which the treewidth of line graphs decreases is an interesting direction for future work.

Our graph-theoretical rewrites currently only apply to tensor networks with dimension-2 bonds.
We believe it is possible to extend our rewrite rules to more general tensor networks using more generic diagrammatic languages such as ZXW calculus.

Our powerful computer program has generated many interesting data, which are shown in the appendix. 
These data may reveal some deep theoretical properties of treewidth and random graphs such as those shown in figure \ref{fig:tw_vs_degree} and \ref{fig:tw_fit}, which will also be an interesting direction for future work.


\begin{spacing}{0.9}


\bibliographystyle{apalike}
\cleardoublepage
\bibliography{References/references} 



\end{spacing}


\begin{appendices} 

\chapter{More on Graph Theory} \label{appendix:graph_theory}

\section{More on Treewidth}\label{sec:treewidth}

Let us first define several more concepts and prove several useful lemmata.

\begin{definition}[Vertex Subtree]
    Let $G$ be a graph and $T$ its tree decomposition. 
    For any vertex $v \in G$, the set of bags in $T$ that contain $v$ is the vertex subtree of $v$ in $T$.
\end{definition}

The reason why it is called a subtree is that the vertex subtree of $v$ must be connected by property 3 of tree decomposition.

\begin{lemma}[Clique Bag]\label{lem:clique_bag}
	If a graph $G$ has a clique of size $k$, then any tree decomposition of $G$ must have a bag containing all the vertices of the clique. Therefore, the treewidth of $G$ is at least $k-1$.
\end{lemma}

\begin{proof}
    Let the clique consist of vertices $v_1, v_2, \ldots, v_k$.

	Now, the vertex subtree of $v_1$ intersects with the vertex subtree of $v_2, v_3 , \ldots, v_k$, because all of them are adjacent. 
	The vertex subtree of $v_2$ intersects with the vertex subtree of $v_3, \ldots, v_k$, and so on.
	By simple set theory we see that the intersection of all the vertex subtrees of $v_1, v_2, \ldots, v_k$ is non-empty.
	All bags in the intersection contain all the vertices of the clique, and therefore the treewidth of $G$ is at least $k-1$.
\end{proof}

\section{Miscellaneous Proofs}\label{sec:proofs}

\begin{proof}[Proof of Lemma \ref{lem:minor_treewidth}]
	We only need to prove that if a graph $G$ can be obtained from $H$ by performing one of the three graph-minor operations, then the treewidth of $G$ is less than or equal to the treewidth of $H$. The lemma follows as any minor can be obtained by a finite sequence of these three operations.

	Let $T$ be a tree decomposition of $H$.
	We have three cases to consider.
	\begin{enumerate}
		\item If $G$ is obtained from $H$ by deleting an edge, then $T$ is also a tree decomposition of $G$. 
		\item If $G$ is obtained from $H$ by deleting a vertex $v$, let $T'$ be the tree decomposition of $G$ isomorphic to $T$ with $v$ removed from all bags. (We allow empty bags). It is clear that $T'$ is a tree decomposition of $G$ and the width of $T'$ is less than or equal to the width of $T$.
		\item If $G$ is obtained from $H$ by contracting an edge $(u, v)$, let $w$ denote the new vertex after contraction. 
		Let $T'$ be the tree decomposition of $G$ isomorphic to $T$ with $u$ and $v$ replaced by a new vertex $w$ in all bags that contain $u$ or $v$. 
	\end{enumerate}

	In all cases, we can construct a tree decomposition of $G$ from a tree decomposition of $H$ with width less than or equal to the width of $T$. Therefore, the treewidth of $G$ is less than or equal to the treewidth of $H$.
\end{proof}

\begin{lemma}[Connection between Triangulated Graph and Cut Vertices]\label{lem:triangulated_cut_vertices}
	Let $G$ be a triangulated graph. Let $P$ be a cycle in $G$ and let vertices $u, v \in G$ be a pair of non-adjacent vertices in $P$. 
	$u, v$ are cut vertices of $P$ (removing $u, v$ makes $P$ disconnected). 
	Let $C_1, C_2$ be the two components of $P$ after removing $u, v$.
	If $C_1, C_2$ are not adjacent in $G$, then $u$ must be adjacent to $v$ in $G$.
\end{lemma}

\begin{proof}
	Provided $C_1, C_2$ are not adjacent in $G$, there must be a cycle of length four.
	If not, let $\mathcal{C}$ be the smallest cycle of $G$ that contains $u, v$.
	Since $G$ is triangulated, $\mathcal{C}$ must have a chord.
	This chord makes the cycle smaller, until the length of the cycle is four.
	See figure \ref{fig:triangulated_cycle} for an example of this process.

	\begin{figure}[ht]
		\centering
		\input{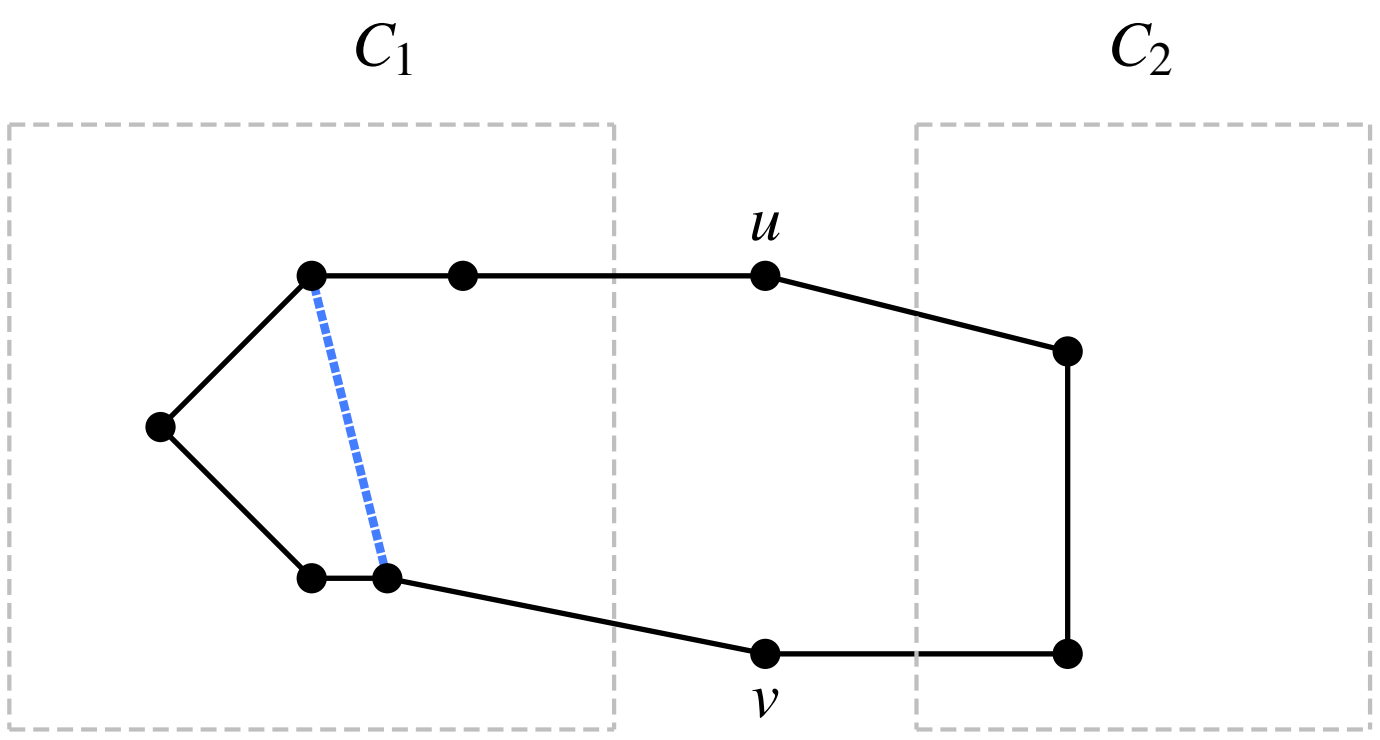}
		\caption{Provided $C_1$ and $C_2$ are not adjacent, any chord in the cycle of length greater than 4, such as the one in dashed blue, will produce a smaller cycle that still contains $u$ and $v$.}
		\label{fig:triangulated_cycle}
	\end{figure}
	
	Let $\mathcal{C}$ be the length four cycle. There are two cases: 
	\begin{enumerate}
		\item $u, v$ are adjacent in $\mathcal{C}$. Then we are done.
		\item $u, v$ are not adjacent in $\mathcal{C}$. Then the cycle must be of the form
		$[a, u, b, v]$, where $a \in C_1$ and $b \in C_2$. Since $a$ cannot connect to $b$, $u$ must be adjacent to $v$.
	\end{enumerate}
\end{proof}

\begin{proof}[Proof of Lemma \ref{lem:simplicial_vertex}]
	We shall prove the following stronger statement. 

	\emph{
	For any triangulated graph $G$, given any vertex $v \in G$, let $H$ be the set of vertices in $G$ that are farthest away from $v$.
	Then there exists a simplicial vertex in $H$.
	}
	
	Recall that the distance between two vertices $u$ and $v$ in a graph is the length of the shortest path connecting them.

	The proof is by induction. 
	The base case is when $G$ has only one vertex $v$. This means $H = \{v\}$ and $v$ is simplicial.

	For a generic triangulated graph $G$ with a chosen vertex $v$, let $H_1$ be one of the connected components of $H$. Let $S$ be the set of vertices that are adjacent to $H_1$ and are not in $H_1$.
	See figure \ref{fig:simplicial_vertex} for a demonstration of this setup.

	\begin{figure}[ht]
		\centering
		\input{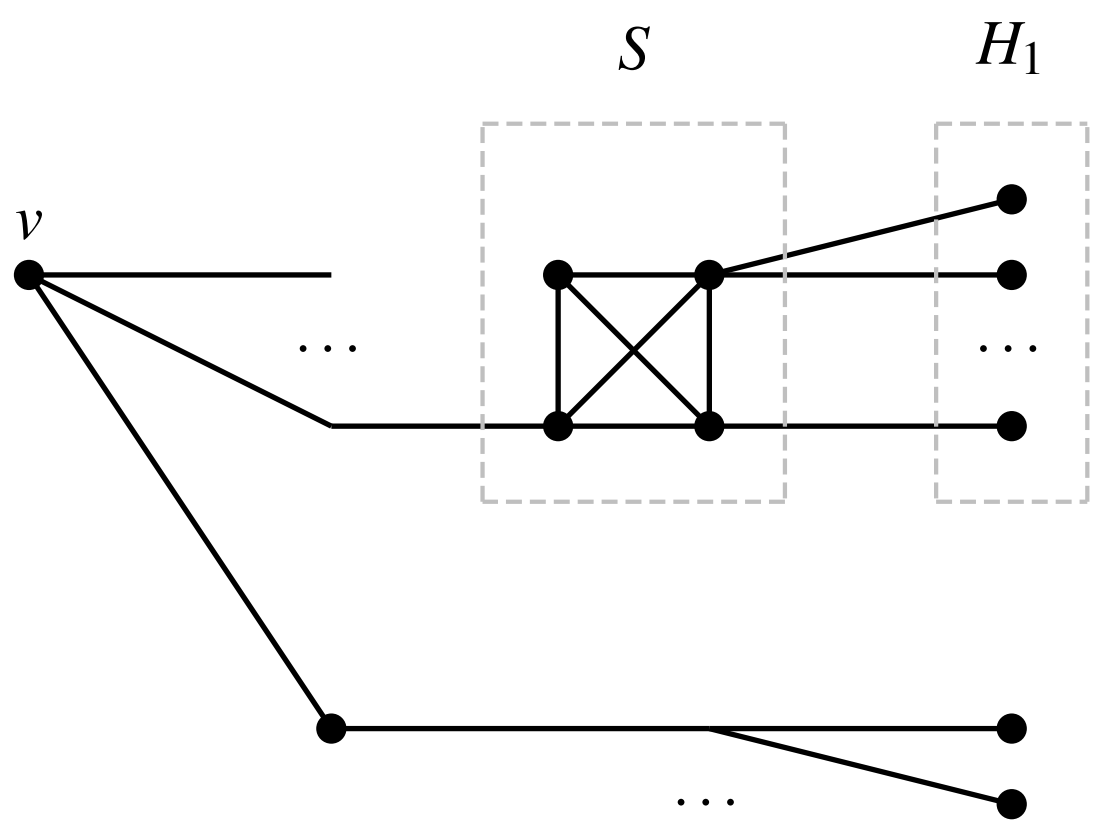}
		\caption{Demonstration of the proof of lemma \ref{lem:simplicial_vertex}. $H_1$ is a connected component of farthest vertices from $v$. $S$ is the set of vertices that are adjacent to $H_1$ and are not in $H_1$.
		}
		\label{fig:simplicial_vertex}
	\end{figure}
	We claim that $S$ is a clique.

	If $S$ contains only one vertex, there is nothing to prove.
	Assume $S$ contains at least two vertices. Let $u, w\in S$ be any distinct vertices.
	Both of $u, w$ are connected to $v$ via some vertices not in $H$, as they are closer to $v$. Both of them are also connected to some vertices in the connected component $H_1$.
	Note that no vertices in $H_1$ shall connect to any other vertices closer to $v$ except those in $S$, as this would make the distance from $v$ to $H_1$ smaller.
	Therefore we are in a position to apply lemma \ref{lem:triangulated_cut_vertices} to the cycle formed by $u, w$ and any two vertices in $H_1$.
	In such a way we have proved that any $u$ and $w \in S$ are adjacent, and therefore $S$ is a clique.

	Now apply the induction hypothesis to the subgraph induced by $H_1 \bigcup S$.
	If $H_1 \bigcup S$ is not a clique, then we can find a vertex $w \in H_1$ and $u \in S$ such that $w$ is at distance 2 from $u$, and applying the induction hypothesis to the subgraph induced by $H_1 \bigcup S$ with $u$ as the chosen vertex shows that there is a simplicial vertex in $H_1$. 
	This vertex is also simplicial in $G$ as it is not adjacent to any vertices outside of $H_1 \bigcup S$.
	If $H_1 \bigcup S$ is a clique, then any vertex in $H_1$ is simplicial in $G$.
\end{proof}

\section{Various Graph Operations and Treewidth}\label{sec:graph_operations}

\begin{lemma}[Splitting a Vertex Does not Increase the Treewidth of the Line Graph]\label{lem:appendix_split_vertex}\end{lemma}

\begin{proof}
  Let our vertex $a$ have $n+m$ incident edges. After splitting the vertex $a$ into two vertices $b$ and $c$, $b$ has $n$ incident edges and $c$ has $m$ incident edges. (We may safely assume that both $n$ and $m$ are greater than $1$, because otherwise the case is trivial and we have ruled out the case when the graph $G$ has treewidth $0$.)
	Splitting a vertex creates one more edge, whose corresponding vertex in the line graph we label as $\alpha$.
	The line graph of the graph $G$ before and after splitting a vertex is shown in figure \ref{fig:vertex_split_line}.

	\begin{figure}[ht]
		\centering 
		\input{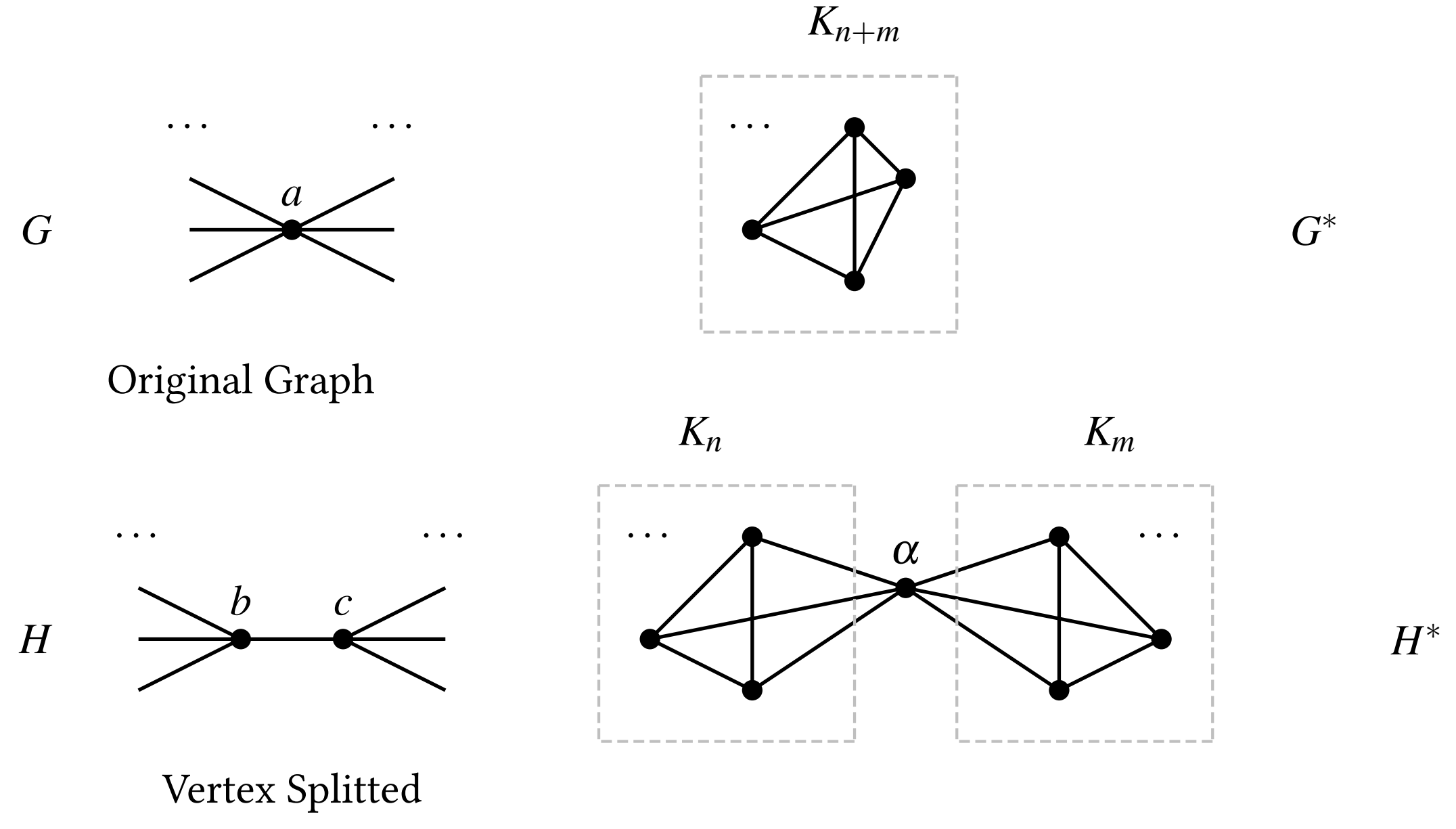}
		\caption{The line graph of $G$ before and after splitting a vertex.}
		\label{fig:vertex_split_line}
	\end{figure}

	Let us construct a tree decomposition of $H^*$ with treewidth no more than that of $G^*$.

	Since $G^*$ has a clique of size $n+m$, there must be a bag in the tree decomposition of $G^*$ that contains all of the vertices in this clique. 
	Denote this bag as $A$.
	We label these vertices as $v_1, \cdots, v_n, u_1, \cdots, u_m$.
	The treewidth of $G^*$ must be greater than or equal to $n+m -1$.

	If the treewidth of $G^*$ is greater than $n+m -1$, adding $\alpha$ to the bag $A$ gives a tree decomposition of $H^*$ with the same treewidth as $G^*$.

	Let us now consider the case when the treewidth of $G^*$ is $n+m-1$. 
	Again, here we have two cases to consider.
	\begin{enumerate}
		\item $A$ is a leaf in the tree decomposition of $G^*$. In this case there must be a vertex $w \in A$ such that $w$ is not in any other bag. We can remove $w$ from $A$ and add $\alpha$ to $A$, and attach the original bag that contains $w$ to $A$. This gives a tree decomposition of $H^*$ with the same treewidth as $G^*$.
		\item Removing $A$ from the tree decomposition of $G^*$ leaves two or more connected components. 
		Again, in this case we claim that there must be a vertex $w \in A$ such that there exist two components of the tree decomposition of $G^*$ that do not contain $w$.
		If it is not the case, it means that all the vertices in $A$ are connected to some components. Reducing that component to a single vertex makes a clique of size $n+m+1$, which is impossible as the treewidth of $G^*$ is $n+m-1$.

      So we have found at least one component $C_1$ such that it does not contain $w$. ($C_1$ can contain several connected components.)
		In this case we can create a new bag $B$ that contains $\alpha$ and all the vertices in $A$ except $w$, attach $B$ to $C_1$, connect $B$ to $A$, and remove the connections between $A$ and $C_1$
	\end{enumerate}
\end{proof}

\begin{lemma}[Dividing an Edge Does not Increase the Treewidth of the Line Graph]\label{lem:appendix_split_edge}\end{lemma}

\begin{proof}
	The line graph of the graph $G$ before and after dividing an edge is shown in figure \ref{fig:divide_edge_line_graph}.

  Again, we may safely assume that both $n$ and $m$ are greater than $1$, because otherwise the case is trivial and we have ruled out the case when the graph $G$ has treewidth $0$
	\begin{figure}[ht]
		\centering 
		\input{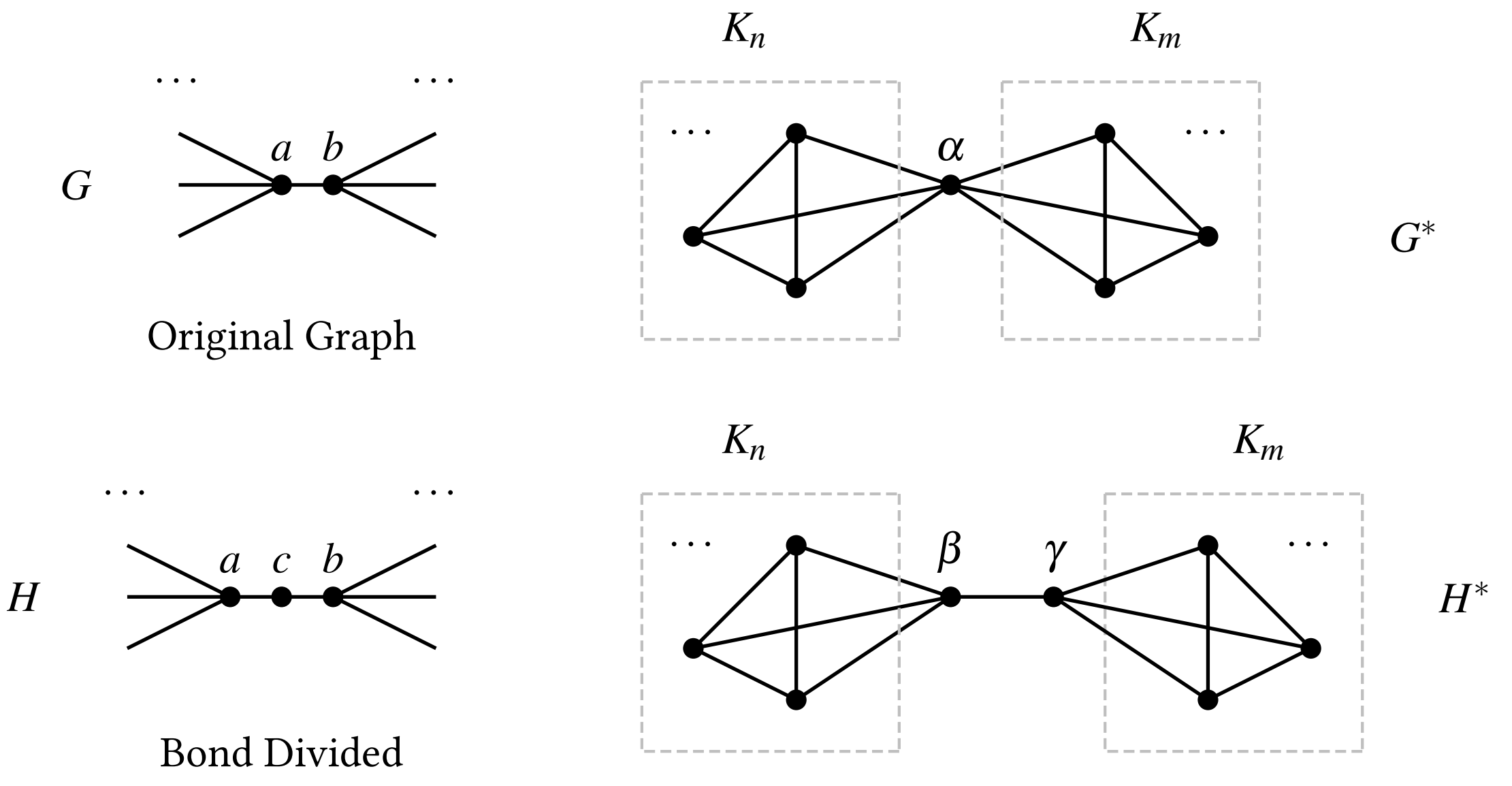}
		\caption{The line graph of $G$ before and after dividing an edge.}
		\label{fig:divide_edge_line_graph}
	\end{figure}

	Let us first consider the tree decomposition of $G^*$.
	By lemma \ref{lem:clique_bag}, there exist two bags, $A, B$, such that $A$ contains all of the vertices in $K_n$ and $\alpha$, while $B$ contains all of the vertices in $K_m$ and $\alpha$.

	There must be a path connecting $A$ and $B$ in the tree decomposition of $G^*$, and every bag on this path must contain $\alpha$ by the third property of tree decomposition. (This path may contain only $A$ and $B$, in which case $\alpha$ is a cut vertex of $G^*$.)

	Since $\alpha$ is not connected to any other vertices in $G^*$, we can remove $\alpha$ from all other bags that are not on the path connecting $A$ and $B$ without violating any of the properties of tree decomposition.
	Label this tree decomposition as $T(G^*)$.

	We will construct a tree decomposition of $H^*$, which we label as $T(H^*)$, and it shall have the same width as $T(G^*)$.
	Start with $T(H^*) = T(G^*)$ and replace $\alpha$ in $A$ with $\beta$ and $\alpha$ in $B$ with $\gamma$.
	All other $\alpha$ shall be removed.

	Now $T(H^*)$ is still a tree and all properties of tree decomposition are satisfied except that there is no bag specifying the connection between the vertex $\beta $ and $\gamma$.
	
	If the path from $A$ to $B$ does not contain any other vertices than the end point, we simply add another bag between $A$ and $B$ that contains $\beta$ and $\gamma$.
	If the path from $A$ to $B$ contains other vertices, what we need to do is add $\beta$ to the first $k$ bags, $\gamma$ to the rest of the bags, and $\gamma$ and $\beta$  shall both be added to one of the bags in the middle of the path.
	This is shown in figure \ref{fig:divide_edge_line_graph_tree_decomposition}.

	Changing $\alpha$ to $\beta$ or $\gamma$ in the new tree decomposition will not change the width of the tree decomposition. 
	The only issue is that we need to add both $\alpha$ and $\beta$ to one of the intermediate bags.
	This can also be done safely, as long as the orders of the bags on the path are not all the same. (Then there is one bag that has at least one fewer vertices than the maximum, and we can add two more vertices to it without increasing the width of the tree decomposition.)

	This is, of course, possible: if two adjacent bags on the path, $C, D$, contain different vertices and have the same order, we can add another bag $E$ that contains their intersection (which has fewer vertices) between them.
	
	This completes the proof.
	\begin{figure}[ht]
		\centering 
		\input{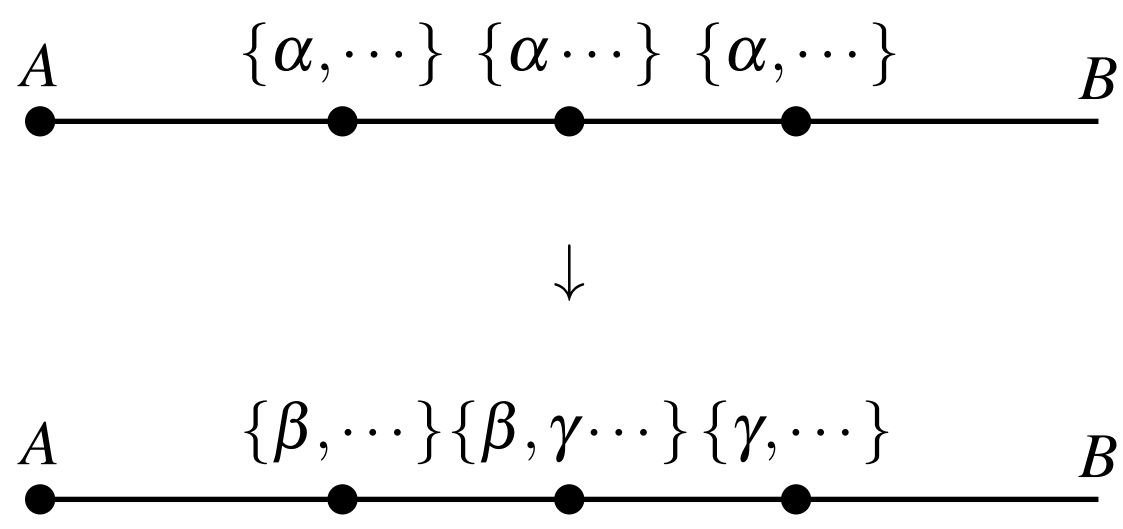}
		\caption{Up: the path $AB$ in the tree decomposition of $G^*$. Down: the path $AB$ in the tree decomposition of $H^*$.}
		\label{fig:divide_edge_line_graph_tree_decomposition}
	
	\end{figure}
\end{proof}

\begin{remark}
	The insight of the previous proof is that if there is a path $A_1, A_2, \cdots, A_k$ in the tree decomposition, then it is possible to create a bag $B$, connected only to $A_i$ and $A_{i+1}$, that contains the intersection of $A_i$ and $A_{i+1}$ without increasing the width of the tree decomposition.
	Moreover, the size of bag $B$ must be smaller than the size of $A_i$ and $A_{i+1}$, so we can add more vertices to $B$ without increasing the width of the tree decomposition.
\end{remark}

\begin{definition}[Graph Substitution]
	Let $G$ be a graph and $H$ be an induced subgraph of $G$. 
	$H$ is connected to the rest of $G$ via a set of edges $E$. 
	Regarding $H$ as a graph with loose edges, we can substitute $H$ with another graph $H'$ that has the same number of loose edges as $H$. 
	This is exactly the operation we perform for diagrammatic rewrites in ZX calculus.
\end{definition}

\begin{conjecture}[Graph Substitution may Reduce Treewidth]
	Let $G$ be a graph and $H$ be its induced subgraph. Let $S$ be another graph with the same number of loose edges as $H$. 
	Let $k$ be the number of loose edges in $H$ and $S$.
	Let $G'$ be the graph obtained by substituting $H$ with $S$ in $G$. 

	Regarding each loose edge as ending with a new vertex, we have the following inequality
	$$
	\text{tw}(G') \leq \text{tw}(G)  - \text{tw}(H) + \text{tw}(S) + k
	$$
\end{conjecture}

Several observations support this conjecture, but we were not able to prove it. 

\section{Treewidth of Common Graphs}
\begin{figure}[ht]
	\centering
	\begin{subfigure}{0.3\textwidth}
		\centering
		\resizebox{\linewidth}{!}{\input{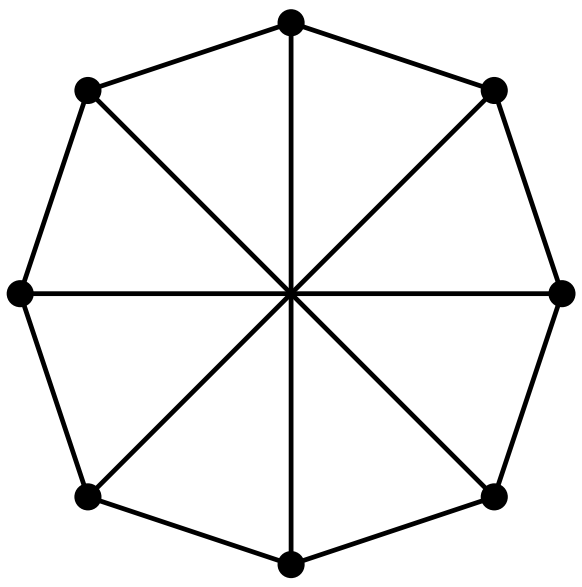}}
		\caption{The Wagner graph. 8 vertices, 12 edges, treewidth 4.}
		\label{fig:tw-wagner}
	\end{subfigure}\hfill
	\begin{subfigure}{0.3\textwidth}
		\centering
		\resizebox{\linewidth}{!}{\input{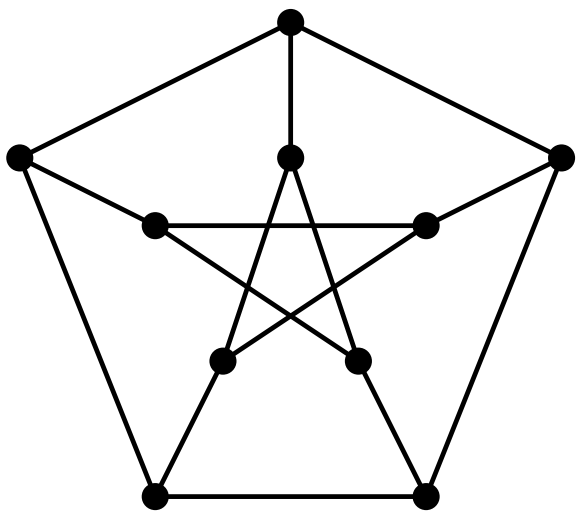}}
		\caption{The Petersen graph. 10 vertices, 15 edges, treewidth 4.}
		\label{fig:tw-petersen}
	\end{subfigure}\hfill
	\begin{subfigure}{0.3\textwidth}
		\centering
		\resizebox{\linewidth}{!}{\input{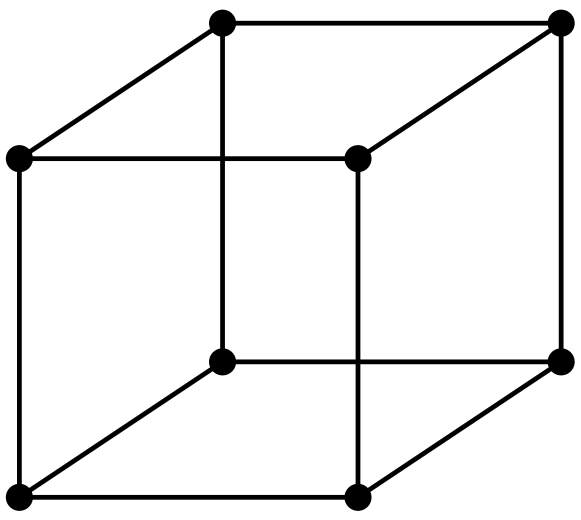}}
		\caption{The cubical graph (the graph of the cube). 8 vertices, 12 edges, treewidth 3.}
		\label{fig:tw-cubical}
	\end{subfigure}

	\medskip

	\begin{subfigure}{0.3\textwidth}
		\centering
		\resizebox{\linewidth}{!}{\input{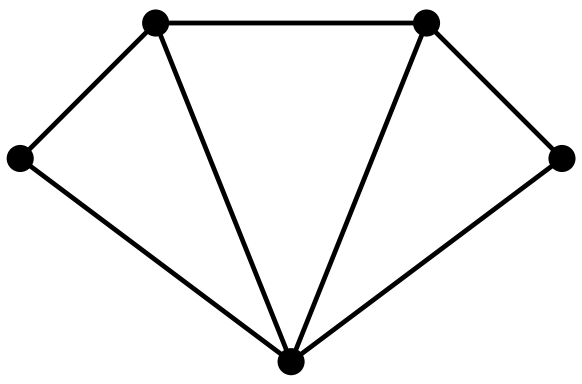}}
		\caption{The gem graph. 5 vertices, 7 edges, treewidth 2.}
		\label{fig:tw-gem}
	\end{subfigure}\hfill
	\begin{subfigure}{0.3\textwidth}
		\centering
		\resizebox{\linewidth}{!}{\input{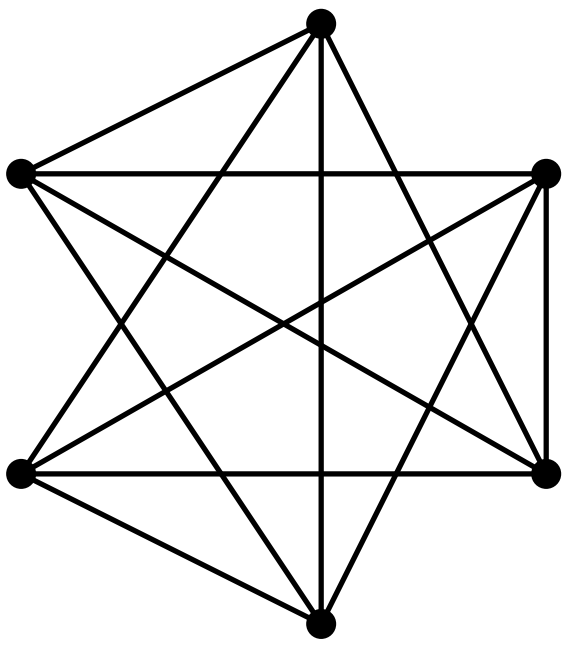}}
		\caption{The octahedron graph. 6 vertices, 12 edges, treewidth 4.}
		\label{fig:tw-octahedron}
	\end{subfigure}
	\caption{Several common graphs.}
	\label{fig:common-graph-treewidth}
\end{figure}

Here we list several common graphs and their treewidths in figure \ref{fig:common-graph-treewidth}.

\begin{figure}[ht]
	\centering
	\begin{subfigure}{0.3\textwidth}
		\centering
		\resizebox{\linewidth}{!}{\input{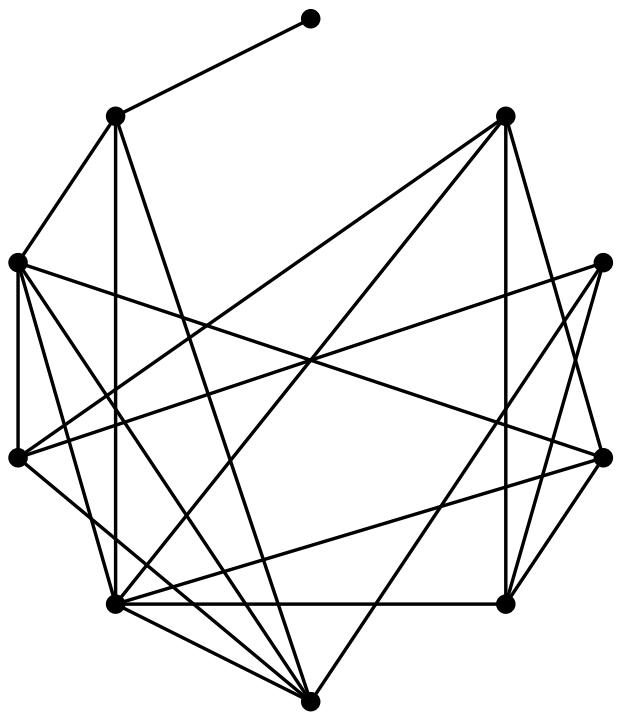}}
		\caption{10 vertices, 20 edges, treewidth 4}
	\end{subfigure}\hfill
	\begin{subfigure}{0.3\textwidth}
		\centering
		\resizebox{\linewidth}{!}{\input{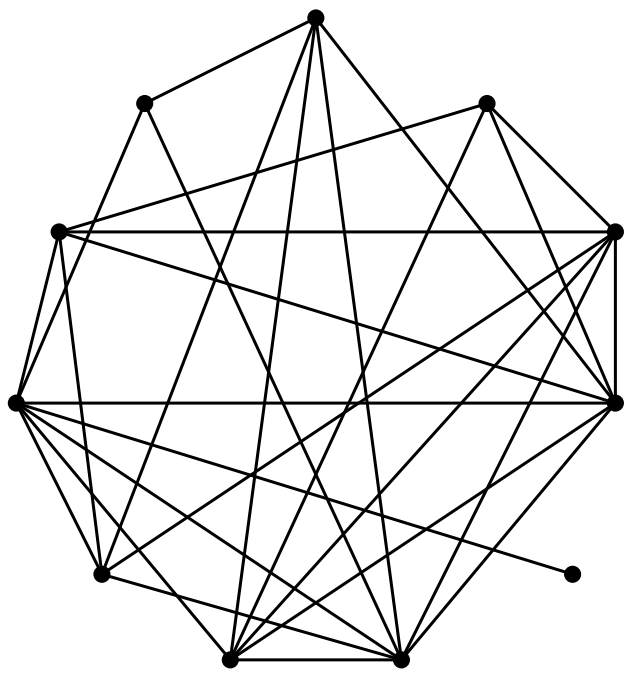}}
		\caption{11 vertices, 28 edges, treewidth 5}
	\end{subfigure}\hfill
	\begin{subfigure}{0.3\textwidth}
		\centering
		\resizebox{\linewidth}{!}{\input{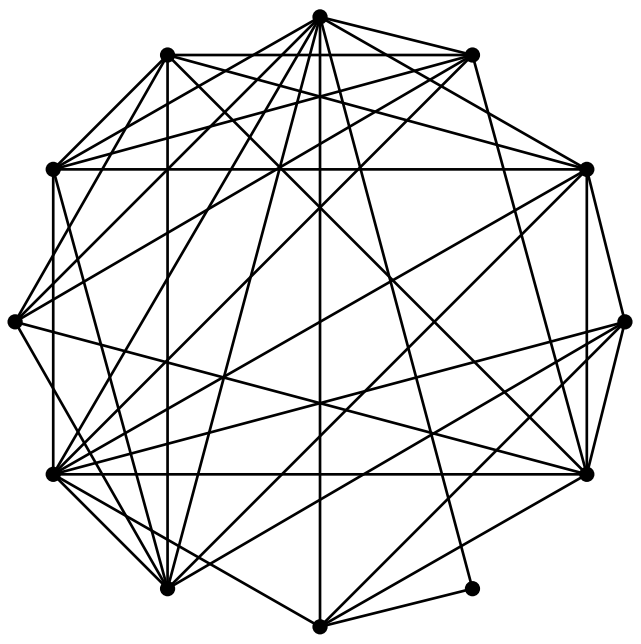}}
		\caption{12 vertices, 36 edges, treewidth 6}
	\end{subfigure}

	\medskip

	\begin{subfigure}{0.3\textwidth}
		\centering
		\resizebox{\linewidth}{!}{\input{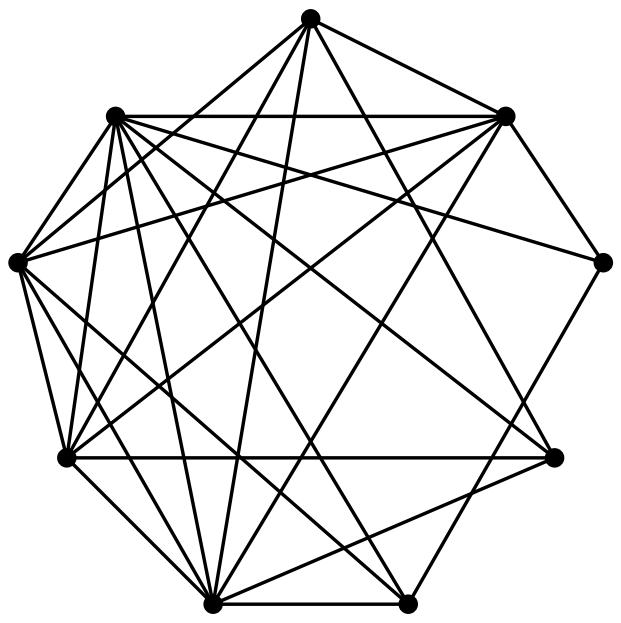}}
		\caption{9 vertices, 24 edges, treewidth 5}
	\end{subfigure}\hfill
	\begin{subfigure}{0.3\textwidth}
		\centering
		\resizebox{\linewidth}{!}{\input{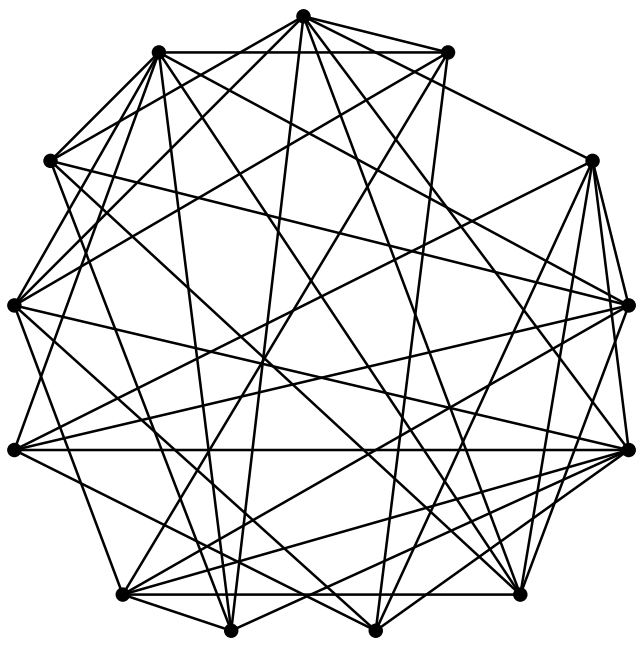}}
		\caption{13 vertices, 38 edges, treewidth 6}
	\end{subfigure}
	\caption{Several random graphs and their treewidths.}
	\label{fig:random-graph-treewidth}
\end{figure}

We have also listed the treewidth of some random graphs, as shown in figure \ref{fig:random-graph-treewidth}.
All of the treewidths and random graph generations are performed using the \emph{cartographer} package which I wrote in Rust for this thesis.
More information about the package can be found in section \ref{sec:cartographer}.

\chapter{Software Implementation}

There are two major pieces of software which I have used extensively for this project.
The first, \texttt{cartographer}, I wrote from scratch.
Available at \url{https://github.com/harryhanYuhao/cartographer}, it is a graph theory package that implements all of the graph operations mentioned in this thesis.
The second one, ZXW graphboard, available at \url{https://github.com/harryhanYuhao/graphboard}, is an open-source web-based graphing calculator for simple ZX diagrams.
Its graph utility in general is helpful for creating tikz diagrams.

\section{Cartographer}
\label{sec:cartographer}

Cartographer is the Rust compute crate behind all of the graph computations in this dissertation.
It implements the common graph algorithms used throughout the thesis.

\subsection{Functionality}

\begin{itemize}\setlength{\itemsep}{0pt}\setlength{\topsep}{1pt}\setlength{\parskip}{0pt}
  \item \textbf{Coloured multigraph core.} A graph is an undirected multigraph with per-vertex and per-edge colours.
  \item \textbf{Treewidth Algorithm.} Tamaki's positive-instance driven dynamic programming (PID-BT) \cite{tamaki2017positive} is implemented. 
  \item \textbf{Graph operations.} The crate also implements common graph operations such as vertex elimination, line graph construction, and random graph generations.
  \item \textbf{ZX algorithm.} This crate implements all of the ZX rewrite rules used in the dissertation.
\end{itemize}

The data for figure \ref{fig:tw_scatter} are created by the \texttt{example/benchmark.rs} file.

\section{ZXW Graphboard}
\label{sec:zxw_graphboard}

Although ZXW graphboard is open-source software, it has only recently been developed and therefore deserves a detailed description as well.
The ZXW graphboard is a browser-based application for drawing and reasoning about ZXW diagrams (as well as generic tikz diagrams): the user places the generators of the calculus on a canvas, connects them with wires, and evaluates the tensor the diagram represents with a Rust compute engine compiled to WebAssembly. It is deployed at \url{https://zxwgraphboard.netlify.app/} (documentation at \url{https://zxwgraphboard-doc.netlify.app/}) and runs client-side only; most of the ZXW TikZ figures in this dissertation were exported from it.

\subsection{Functionality}

\begin{itemize}\setlength{\itemsep}{0pt}\setlength{\topsep}{1pt}\setlength{\parskip}{0pt}
  \item \textbf{Editor.} All the generators used in this dissertation are built in ($Z$-boxes, $X$-nodes, Hadamard boxes, $W$-boxes, AND gates, boundary markers and an empty node), with phases typed as LaTeX expressions such as $\pi/4$ and rendered live with KaTeX. The canvas offers grid snapping, multi-tab editing, undo/redo and keyboard shortcuts, with autosave.
  \item \textbf{Tensor calculator.} A back end Rust WASM engine computes the tensor contraction of the diagram. All algorithms are implemented by hand from scratch.
  \item \textbf{TikZ and JSON export.} Diagrams export as self-contained TikZ figures or JSON documents for later editing; because the editor is a general graph canvas, the same export path produces TikZ for arbitrary graph diagrams.
\end{itemize}

\section{Graphs and Data}

I have made every effort to ensure that the software behaves as intended.
The figures in this section visualise the workflow carried out by the software and demonstrate its capabilities.

\subsection{A Complicated Graph}

\begin{figure}[ht]
	\centering
	\begin{subfigure}{0.4\textwidth}
		\centering
		\resizebox{\linewidth}{!}{\input{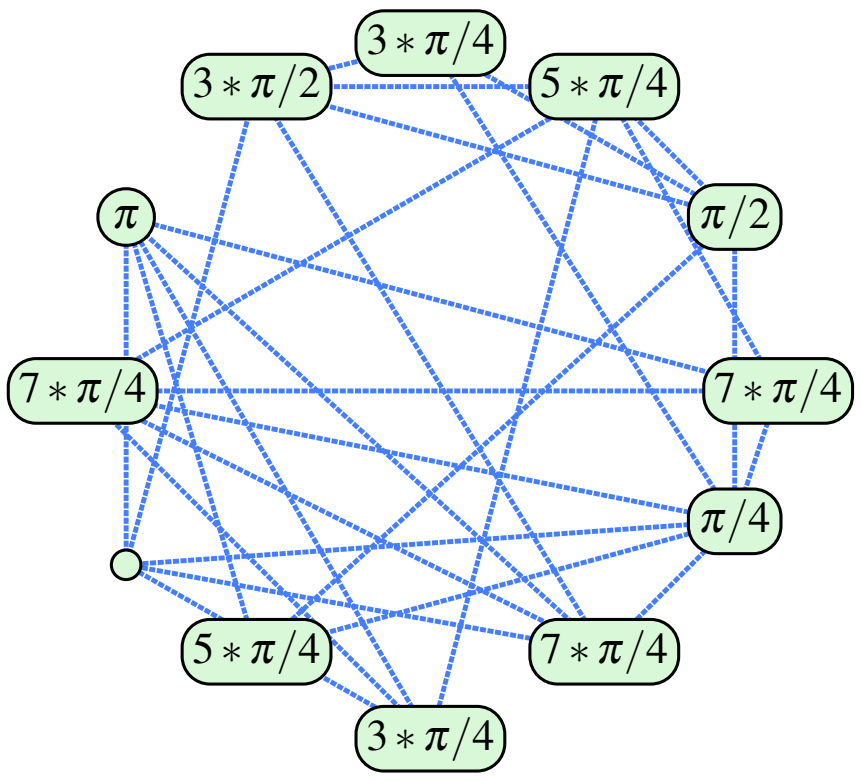}}
		\caption{A random graph-like ZX diagram $G$}
	\end{subfigure}\hfill
	\begin{subfigure}{0.5\textwidth}
		\centering
		\resizebox{\linewidth}{!}{\input{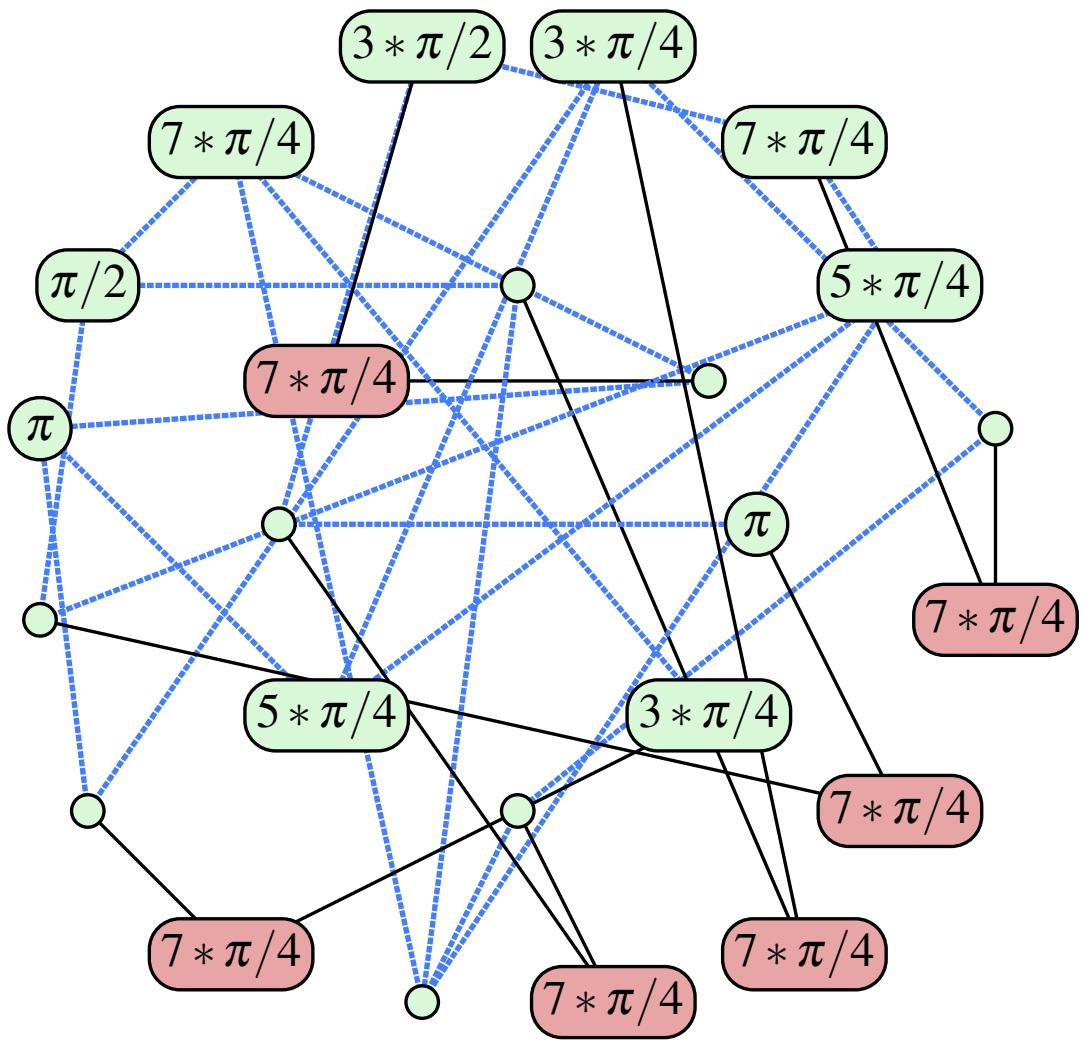}}
		\caption{Graph $F$, which is $G$ after the preprocessing algorithm (definition \ref{def:rewrite_preprocessing})}
	\end{subfigure}
	\medskip
	\begin{subfigure}{0.5\textwidth}
		\centering
		\resizebox{\linewidth}{!}{\input{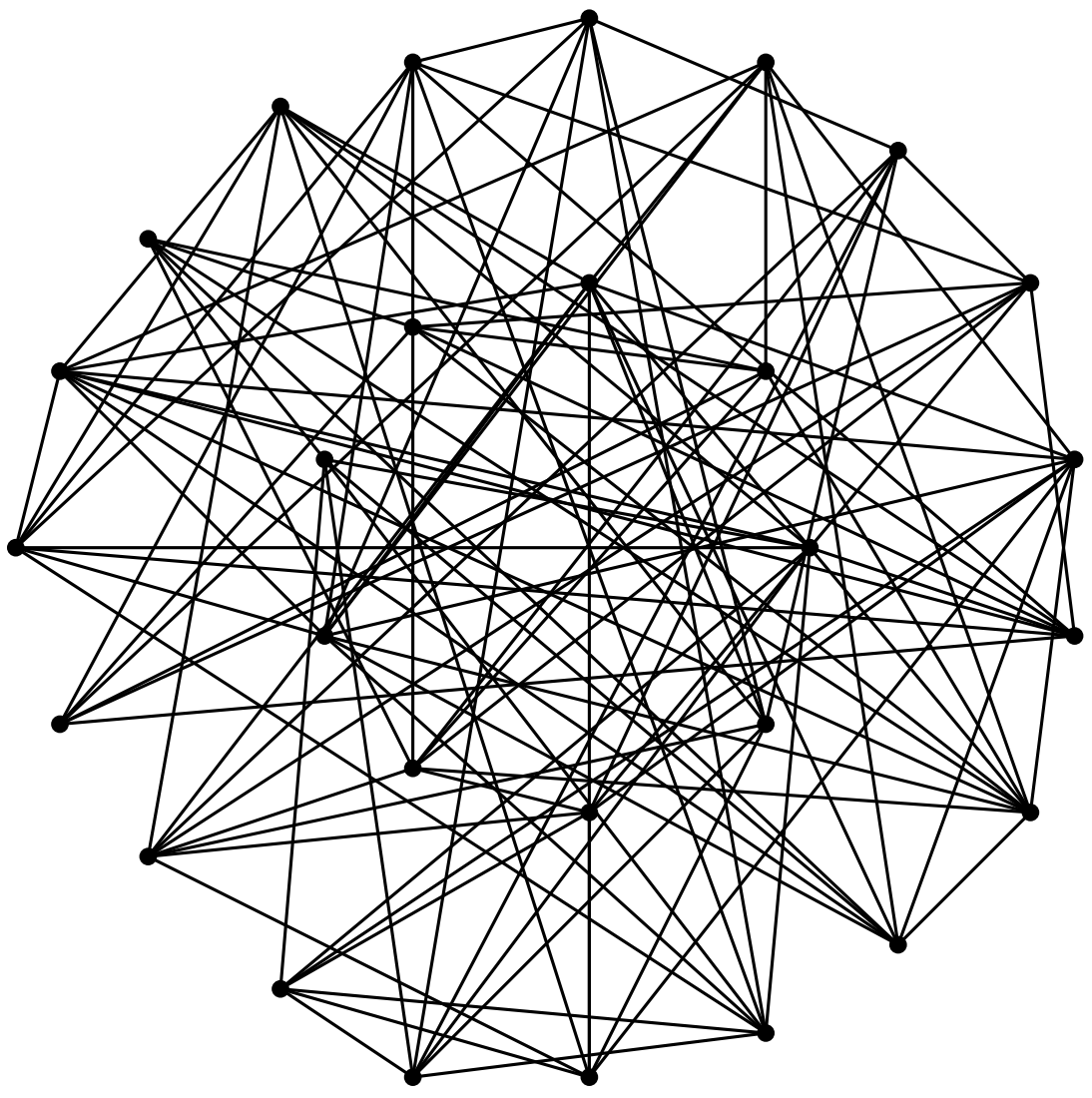}}
		\caption{Line graph of $G$, treewidth 11.}
	\end{subfigure}\hfill
	\begin{subfigure}{0.5\textwidth}
		\centering
		\resizebox{\linewidth}{!}{\input{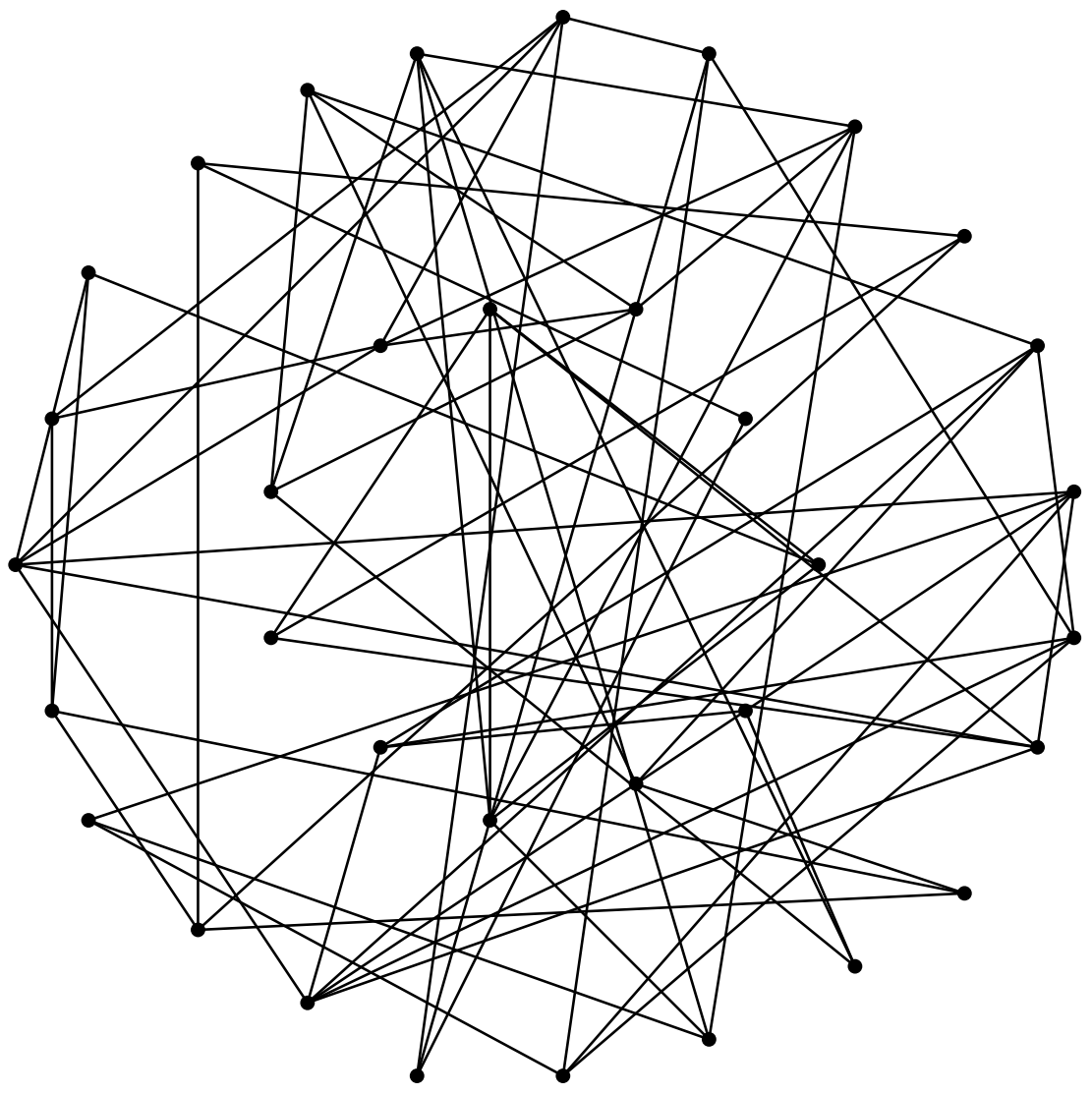}}
		\caption{Line graph of $F$, treewidth 7.}
	\end{subfigure}
	\caption{A random graph-like ZX diagram before and after preprocessing and their line graphs.}
	\label{fig:preprocess_workflow}
\end{figure}

In figure \ref{fig:preprocess_workflow}, we show a random graph-like ZX diagram $G$ and its line graph, as well as the graph $F$ after the preprocessing algorithm (definition \ref{def:rewrite_preprocessing}) and its line graph.

\subsection{Some Interesting Statistics}

\begin{figure}[ht]
	\centering
	\includegraphics[width=1\textwidth]{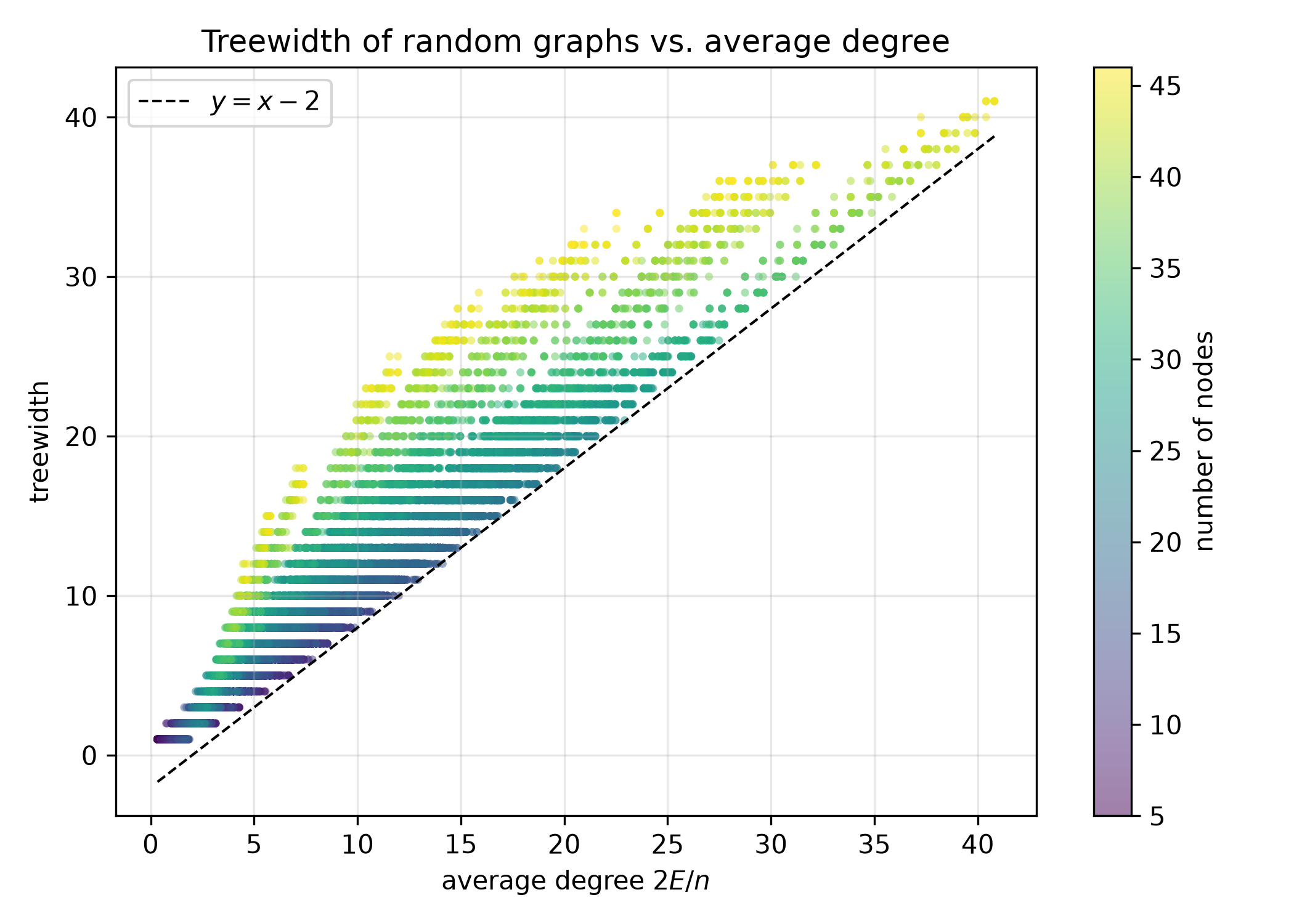}
	\caption{A scatter plot of treewidth vs average degree of random graphs.
		20244 graphs are randomly generated using the Erd\H{o}s--R\'{e}nyi algorithm and their treewidth and average degree are computed.
		The colour of the points indicates the number of vertices in the graph.
		The data is generated by the \texttt{example/treewidth\_hypo.rs} file in the cartographer crate.
	}
	\label{fig:tw_vs_degree}
\end{figure}

Since the package provides powerful graph theory and treewidth algorithms, I used it to generate some statistics of interest.
For example, figure \ref{fig:tw_vs_degree} shows a scatter plot of treewidth vs average degree of random graphs, based on 20244 graphs generated using the Erd\H{o}s--R\'{e}nyi algorithm.
From observing figure \ref{fig:tw_vs_degree}, I propose the following hypothesis:

\begin{conjecture}
	Let $G$ be a random graph generated by the Erd\H{o}s--R\'{e}nyi algorithm with a finite number of vertices.
	The probability that the treewidth of $G$ is smaller than $d-2$, where $d$ is the average degree of $G$, is zero.
\end{conjecture}

\begin{figure}[ht]
	\centering
	\includegraphics[width=1\textwidth]{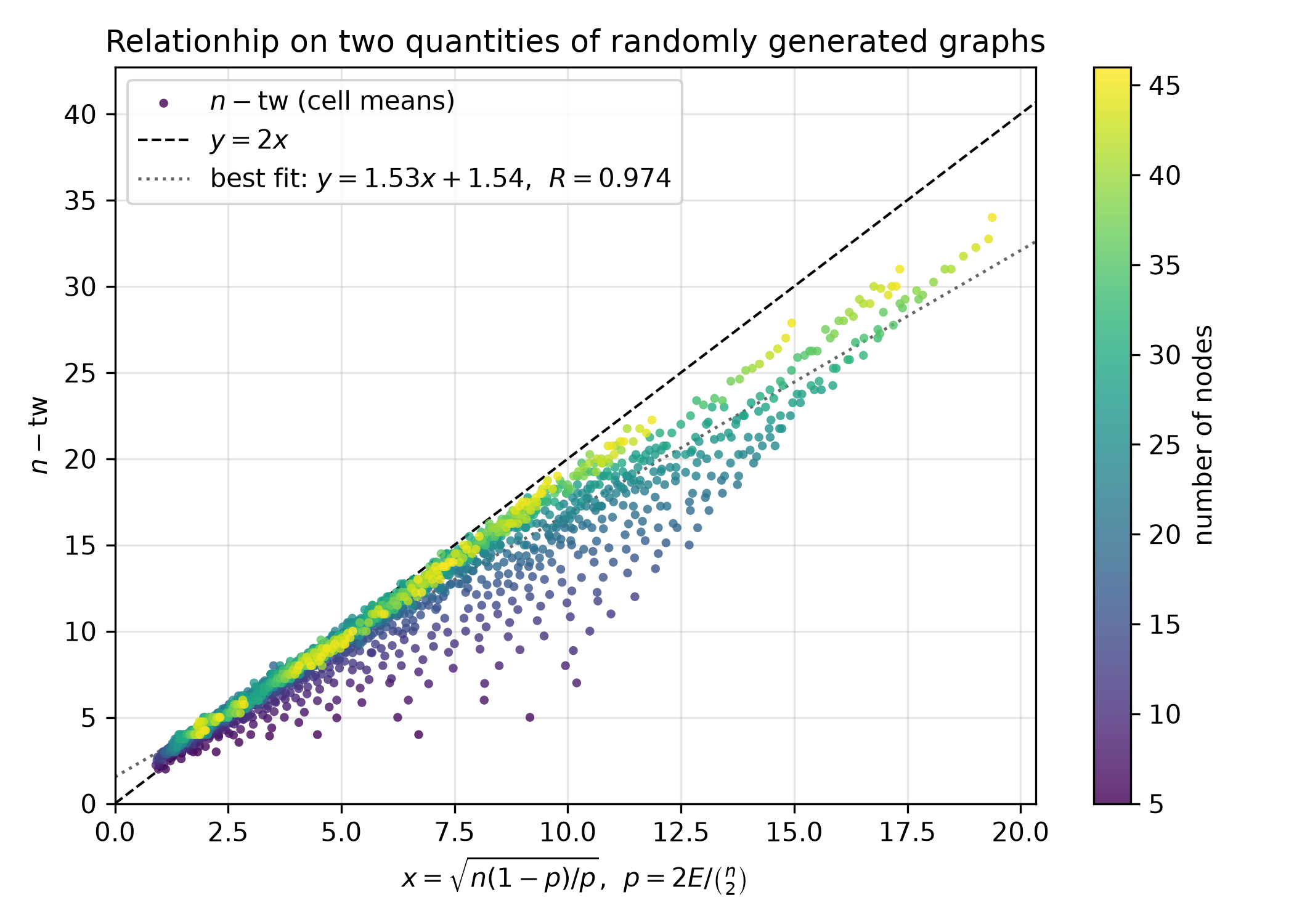}
	\caption{Using the same data as figure \ref{fig:tw_vs_degree}, we plotted $n-\mathrm{tw}$ versus $\sqrt{n(1-p)/p}$, where $n$ is the number of vertices and $p$ is the probability of edge generation in the Erd\H{o}s--R\'{e}nyi algorithm (so that the expected average degree is $p(n-1)$).
		The colour of each point denotes the number of vertices in the graph of that particular data point.
		Note that there is a strong linear correlation (especially when $n-\mathrm{tw}$ is low) between the two quantities.
	}
	\label{fig:tw_fit}
\end{figure}

Here I plotted another graph (Figure \ref{fig:tw_fit}), which shows a strong linear correlation between $n-\mathrm{tw}$ and $\sqrt{n(1-p)/p}$, where $n$ is the number of vertices and $p$ is the probability of edge generation in the Erd\H{o}s--R\'{e}nyi algorithm (so that the expected average degree is $p(n-1)$).

The treewidth of the Erd\H{o}s--R\'{e}nyi model $G(n,p)$ has in fact been well studied, and both the conjecture above and this correlation connect directly to that literature.
Kloks first showed that the treewidth of $G(n, d/n)$ is linear in $n$ with high probability for $d \geq 2.36$ \cite{kloks1994treewidth}; Gao improved the threshold to $m/n \geq 1.073$ in the uniform model $G(n,m)$ (where $m$ edges are placed uniformly at random among the $\binom{n}{2}$ possible positions) and conjectured that linearity holds throughout the supercritical regime \cite{gao2012treewidth}, which Lee, Lee, and Oum confirmed by way of rank-width \cite{lee2012rankwidth}.
In the dense regime, Perarnau and Serra proved that $\mathrm{tw}(G(n,p)) = n - O(\sqrt{n/p})$ asymptotically almost surely (that is, with probability tending to one as $n \to \infty$) \cite{perarnau2014treedepth}, so the deficit $n - \mathrm{tw}$ is of order $\sqrt{n/p}$ --- precisely the order of the horizontal axis $\sqrt{n(1-p)/p}$ of figure \ref{fig:tw_fit} when $p$ is fixed.
Their asymptotics also indicate how the conjecture should be read, since no event in a finite random graph has probability exactly zero: the intended statement is that $\mathrm{tw}(G) \geq d - 2$ holds with high probability.
This is immediate in the sparse regimes, where the treewidth is linear in $n$ while $d$ stays constant, and for fixed $p$ it follows from the dense asymptotics, since $d = p(n-1)$ and the deficit $O(\sqrt{n/p})$ is negligible compared with $n - d = n(1-p) + p$; the two sides are closest in the near-complete regime $1 - p = O(1/n)$.

\end{appendices}

\printthesisindex 

\end{document}